\documentclass[12pt,a4paper]{article}
\usepackage{jheppub}

\definecolor{klgreen}{rgb}{0.0, 0.5, 0.0}

\definecolor{color1}{rgb}{1,0,0}
\definecolor{color2}{rgb}{0,0,1}
\definecolor{color3}{rgb}{0,1,0}
\definecolor{color4}{rgb}{0.6,0.4,0.2}
\definecolor{color5}{rgb}{1,0,1}
\definecolor{color6}{rgb}{1,0.5,0}
\definecolor{color7}{rgb}{0.666667,0.666667,1}
\definecolor{color8}{rgb}{0.5,0,0.5}
\definecolor{color9}{rgb}{0.666667,0.666667,0}
\definecolor{green2}{rgb}{0.0, 0.5, 0.0}
\definecolor{gray2}{rgb}{0.5, 0.5, 0.5}

\newcommand{\exclude}[1]{}

\newcommand{\beq}{\begin{equation}}
\newcommand{\eeq}{\end{equation}}
\newcommand{\bea}{\begin{eqnarray}}
\newcommand{\eea}{\end{eqnarray}}
\long\def\/*#1*/{}

\newcommand{\junk}[1]{}

\renewcommand{\Im}{\mathrm{Im}}
\renewcommand{\Re}{\mathrm{Re}}

\title{\Large Holographic Non-Hermitian Lattices and Junctions and their RG Flows}
\author[1,2]{Daniel Are\'an,}
\author[1,2]{David Garcia-Fariña}
\affiliation[1]{Instituto de F\'isica Te\'orica UAM/CSIC, Calle Nicol\'as Cabrera 13-15, 28049 Madrid, Spain}
\affiliation[2]{Departamento de F\'isica Te\'orica, Universidad Aut{\'o}noma de Madrid, Campus de Cantoblanco, 28049 Madrid, Spain}

\preprint{IFT-UAM/CSIC-24-148}

\emailAdd{daniel.arean@uam.es}
\emailAdd{david.garciafarinna@estudiante.uam.es}

\usepackage{physics} 
\usepackage{tensor} 
\usepackage{subcaption}
\usepackage{graphicx}

\abstract{
We construct and study inhomogeneous non-Hermitian strongly coupled holographic field theories. We consider two models: a lattice where in each site there is some inflow/outflow of matter and a Hermitian/non-Hermitian/Hermitian junction.
By tuning a complex external gauge field, we find a non-Hermitian model which can be mapped to a Hermitian one via a complexified $U(1)$ gauge transformation. On the other hand, in the absence of the  
gauge field we find non-Hermitian solutions with a purely imaginary current. Despite this, all expectation values respect $\mathcal{PT}$ symmetry and thus we expect the system to feature unitary time evolution.
Nonetheless, we have not found a map bringing these solutions to 
a Hermitian description.
We study the RG flows of solutions with imaginary current
finding that in the IR they can be mapped to a Hermitian conformal fixed point via a complexified $U(1)$ transformation. Remarkably, we also obtain solutions where the null energy condition is violated near the boundary of the dual geometry.
}

\begin{document}

\maketitle

\section{Introduction}

One of the main axioms of Quantum Mechanics is that the Hamiltonian should be Hermitian. Underlying this assumption lies the need for unitary evolution, which is equivalent to demanding real eigenvalues of the Hamiltonian. However, when considering unitarity as the true physical restriction, one finds that Quantum Mechanics can also be consistently formulated with non-Hermitian Hamiltonians \cite{Bender:1998ke}. These non-Hermitian theories are instead required to have $\mathcal{PT}$ symmetry \cite{Bender:1998ke} or, more generally an antilinear symmetry as recently emphasized in \cite{Mannheim2018}. However, the existence of such symmetry is not a sufficient condition for unitary evolution as it can be spontaneously broken by the spectrum of the theory. This realization gave rise to the field of $\mathcal{PT}$-symmetric Quantum Mechanics, and later to that of $\mathcal{PT}$-symmetric Quantum Field Theory (see, for instance, \cite{Bender_2005} for an introduction to these topics).

Physically, one thinks of a non-Hermitian theory as a theory with influx and outflux of matter. Within this perspective, it is clear that $\mathcal{PT}$-symmetric theories correspond to the family of non-Hermitian theories where influx and outflux are balanced, ensuring that matter content is conserved. 

In practice, a $\mathcal{PT}$-symmetric theory can be mapped to a Hermitian one via a Dyson map \cite{Mostafazadeh:2001jk,Mostafazadeh:2001nr,Mostafazadeh:2002id}. Nonetheless, in many cases, the mapping can be highly non-trivial and give rise to non-local interactions \cite{Bender:2004sa}.\footnote{The construction of the Dyson map for a QFT in a perturbative regime may require resummation of series in the coupling constant and be therefore highly non-trivial~\cite{Alexandre:2022uns}. Also, it is worth noting that, even if the isospectral Hermitian theory is local, one expects the Dyson map to be non-local in the presence of interactions \cite{Li:2024xms}.}
This alone motivates us to study non-Hermitian theories for their own physical relevance. Moreover, once we rid ourselves of the constraints of Hermiticity and instead construct $\mathcal{PT}$-symmetric Hamiltonians, we can begin to study the vacua and the spontaneous breaking of $\mathcal{PT}$ as a function of the parameters of the model. This can then be viewed as tuning an open system to achieve a stationary state where influx and outflux are balanced.

\begin{figure}
\centering
\includegraphics[width=0.49\textwidth]{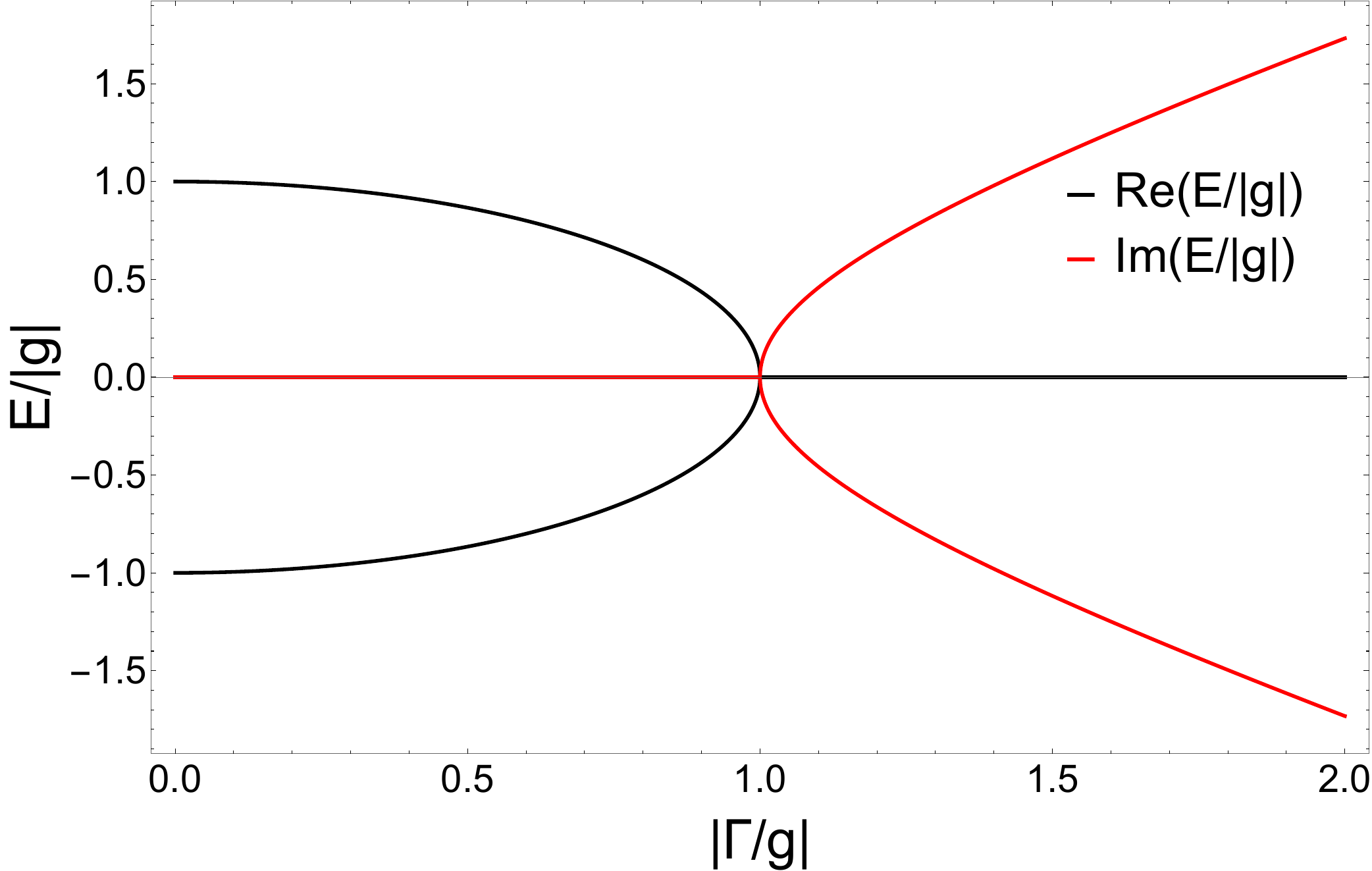}
\caption{Energy levels in the two-level system
\eqref{eq:H2level}. For $|\Gamma/g|>1$ the energies become complex conjugate and $\mathcal{PT}$-symmetry is broken.}
\label{fig:toymodel_2lvlEnergies}
\end{figure}

These features are nicely illustrated by the following
$\mathcal{PT}$-symmetric two-level Hamiltonian \cite{Bender_2005}
\begin{equation}
    H=\begin{pmatrix}-i \Gamma & g \\ g & i\Gamma\end{pmatrix}\,,
    \label{eq:H2level}
\end{equation}
as we review in~\cite{Arean:2024gks}. This system features both a $\mathcal{PT}$-symmetric and a $\mathcal{PT}$-broken regime. $\mathcal{PT}$-symmetry is intact when the hopping between the
two levels, set by $g$, is faster than the inflow (outflow) of matter from (to) the exterior, fixed by $\Gamma$; while it is broken in the regime where the inflow (outflow) is faster than the hopping and hence no stationary state is reached. As shown in Fig.~\ref{fig:toymodel_2lvlEnergies}, the energy levels of the system are real in the $\mathcal{PT}$-symmetric regime and complex in the $\mathcal{PT}$-broken phase. They meet at the so-called exceptional point $|\Gamma/g|=1$.

Non-Hermitian Quantum Mechanics has seen a wide range of applications to areas such as optics, cold atoms or condensed matters systems (see \cite{ElGanainy2018} for a review). Remarkably, applications to inhomogeneous condensed matter systems have been considered in~\cite{Ashida_2017}.

In the context of Quantum Field Theory, non-Hermitian models have garnished great attention in recent years starting with the work \cite{Bender_2004}. The effect of non-Hermitian couplings has been extensively studied 
(see references in \cite{Chernodub:2024lkr} for an overview), concluding that they can support real energy spectra, while also presenting novel phenomenology absent in Hermitian theories.
Spacetime-dependent non-Hermitian couplings were first introduced in \cite{Chernodub:2021waz} where the spectrum was found to become complex at high energies while remaining real in the IR, thus signaling spontaneous breaking of $\mathcal{PT}$ in the UV. This phenomenon was further explored in \cite{Chernodub:2024lkr} were the authors discussed the stability of the solutions. Focusing now on strongly coupled holographic field theories \cite{Maldacena:1997re,Aharony:1999ti,BigJ,zaanen2015holographic,Hartnoll:2018xxg}, non-Hermitian systems were introduced in \cite{Arean:2019pom} and have been further explored in \cite{Xian:2023zgu}, where the authors constructed the phase diagram and computed the electrical conductivity; and in \cite{Morales-Tejera:2022hyq}, where 
the effect of non-Hermitian quenches was studied.
A review of these developments can be found in \cite{Arean:2024gks}.
Non-Hermitian generalizations of the Sachdev-Ye-Kitaev model have also been studied in \cite{Cai:2022onu,Cao:2024xaw}.

In this paper we explore the effect of inhomogeneous non-Hermiticity.
In particular, we follow the approach pioneered in \cite{Arean:2019pom} and consider a theory where the non-Hermiticity is introduced via the sources of an operator charged under a $U(1)$ symmetry and its Hermitian conjugate. The inhomogeneity is then achieved by taking the sources to be functions of one of the spatial coordinates, say, $x^1$. Qualitatively, this setup describes a system where there is an $x^1$-dependent influx and outflux of matter. 
Explicitly we construct a lattice where in all sites there is some $x^1$-dependent inflow/outflow of matter, which we label \textit{non-Hermitian lattice}. We also consider a Hermitian/non-Hermitian/Hermitian junction which we dub \textit{non-Hermitian junction}.

Strikingly, we find a family of solutions presenting a purely imaginary current absent in the homogeneous setup. Nonetheless, $\mathcal{PT}$-symmetry is preserved and they still allow for unitary time evolution.
Despite this, we have only managed to identify the Dyson map to the Hermitian counterpart for the IR in the low-temperature limit. In this case, the solution becomes a complexified $U(1)$ rotation of the standard homogeneous Hermitian conformal fixed point and thus the Dyson map is the corresponding compensating transformation. This simplification seems to be related to the irrelevant character of the spatial deformation as in e.g. \cite{Donos:2012js,Donos:2014uba,Donos:2014yya}.

We explore the stability of the solutions with imaginary current in the non-Hermitian lattice through quasinormal mode (QNM) analysis and discuss the null-energy condition (NEC). We find stable solutions that violate the NEC, in contrast with the observations of \cite{Xian:2023zgu} for the homogeneous setup. Remarkably, for these solutions, close to the AdS-boundary the $a$-function  \cite{Freedman:1999gp,Myers:2010xs,Myers:2010tj} grows towards the bulk, suggesting an increase in the number of degrees of freedom along the renormalization group flow.

The paper is structured as follows. In section \ref{sect:Holographic model} we present the holographic model. We pay special attention to the proper definition of $\mathcal{PT}$ and to the construction of the Dyson map in both the CFT and the gravitational dual.
In section \ref{sec:NonHermitianLattice} we define the non-Hermitian lattice, discuss the numerical setup and present our results. We distinguish two types of solutions: with imaginary current and without imaginary current. We identify the Dyson map for the latter and explicitly check that they can be described in terms of a Hermitian theory. In subsection \ref{subsect:NHLattice_Stability} we compute the quasinormal frequencies for the solutions with imaginary current and discuss their stability. The NEC and the $a$-function are studied in subsection \ref{subsect:NonHermitianLattice_NEC}. To elucidate the ground state of the model, the low-temperature behavior is studied in subsection \ref{subsect:LowTbehaviour}.
In section \ref{sec:NonHermitianJJ} we define the non-Hermitian junction and present some solutions. We find that the imaginary current found in the non-Hermitian lattice localizes at the junction.
In section \ref{sec:Conclusions} we summarize our findings, present our conclusions and give an outlook on further studies.
The holographic dictionary is constructed in appendix \ref{appendix:Holographic Dictionary} and a detailed discussion of the calculation of the quasinormal frequencies can be found in appendix \ref{appendix:DetailsQNFs}.


\section{Holographic Model}\label{sect:Holographic model}
In this section we discuss the relevant aspects of our model. We begin by describing our  setup from the CFT side, with special emphasis on the proper definition of $\mathcal{PT}$. We follow closely the discussion of \cite{Xian:2023zgu}. Subsequently, we examine the action of the Dyson map from the CFT perspective and its relation to external gauge transformations. Next, we discuss the gravitational dual, presenting the relevant action, the equations of motion, the boundary conditions, and the holographic dictionary. The derivation of the latter can be found in appendix \ref{appendix:Holographic Dictionary}. Lastly, we end this section by discussing the symmetries on the gravitational side of the duality and comparing them to those found in the CFT.


\subsection{Non-Hermitian CFT}\label{subsect:Non-Hermitian CFT}

As stated in the introduction, we want to add non-Hermiticity through the sources $\bar{s}$ and $s$ of a scalar operator $\mathcal{O}$ charged under a $U(1)$ symmetry and its Hermitian conjugate $\bar{\mathcal{O}}$. Thus, the relevant action to study should be 
\begin{equation}\label{eq:PTAction_NoCurrent}
    \mathcal{S}=\mathcal{S}_{\text{CFT}}+\int d^\text{d}x\,\,\left(\bar{s}\mathcal{O}+s\bar{\mathcal{O}}\right)=\int d^\text{d}x\,\,\mathcal{L}(x)\,,
\end{equation}
where $\mathcal{S}_{\text{CFT}}$ is the action in absence of non-Hermitian sources and $\mathcal{L}$ is the Lagrangian of the full theory.

Let us see that the action \eqref{eq:PTAction_NoCurrent} is $\mathcal{PT}$-symmetric. We define the effect of $\mathcal{PT}$ on the coordinates to be
\begin{equation}
    x=(t,x^1,x^2,...,x^{d-1})\xrightarrow{\mathcal{PT}}\mathcal{PT}x=(-t,-x^1,x^2,...,x^{d-1})\,,
\end{equation}
 where we take $x^1$ as the direction along which the inhomogeneities are present.

We choose $\mathcal{O}$ to be a scalar under $\mathcal{PT}$, i.e. 
\begin{equation}
    \mathcal{O}(x)\xrightarrow{\mathcal{PT}}\mathcal{O}(\mathcal{PT}x)\,,\qquad \bar{\mathcal{O}}(x)\xrightarrow{\mathcal{PT}}\bar{\mathcal{O}}(\mathcal{PT}x)\,,
\end{equation}
meanwhile, the sources transform as conventional functions \footnote{We are taking a view of the symmetry where the sources are treated as classical spurion fields transforming under the symmetry at the level of the Lagrangian. However, one could instead take an alternative view where sources do not transform and only get conjugated due to the antiunitary nature of $\mathcal{PT}$ and similar results would follow. \label{ftnote:PassivevsActive}}
\begin{equation}
    s(x)\xrightarrow{\mathcal{PT}}s^*(\mathcal{PT}x)\,,\qquad \bar{s}(x)\xrightarrow{\mathcal{PT}}\bar{s}^*(\mathcal{PT}x)\,.
\end{equation}
Then, assuming that $\mathcal{PT}$ acts trivially on $\mathcal{S}_{\text{CFT}}$ and that $s$ and $\bar{s}$ are real, we conclude that the Lagrangian in the action \eqref{eq:PTAction_NoCurrent} is a scalar under $\mathcal{PT}$
\begin{equation}\label{eq:Lagrangian_PT}
    \mathcal{L}(x)\xrightarrow{\mathcal{PT}}\mathcal{L}(\mathcal{PT}x)\,,
\end{equation}
and thus that $\mathcal{PT}$ is a symmetry of the quantum theory. We implicitly assume that the path integral measure is also invariant. Remarkably the assumption of reality of the sources differs from demanding the theory to be Hermitian, as we can have real sources that do not satisfy $s^*=\bar{s}$. Hence, this setup does indeed allow for a straightforward implementation of non-Hermiticity.

Let us now take a moment to make explicit what would constitute spontaneous symmetry breaking (SSB) of $\mathcal{PT}$ in this setup. We distinguish two possibilities for the vacuum to break $\mathcal{PT}$
\begin{itemize}
    \item \underline{Sources are not invariant under $\mathcal{PT}$}: In this case, $s^*(\mathcal{PT}x)\neq s(x)$ and/or $\bar{s}^*(\mathcal{PT}x)\neq \bar{s}(x)$. Therefore, although the Lagrangian is $\mathcal{PT}$-symmetric, we generically expect the state to break $\mathcal{PT}$ as we are imposing it explicitly through the sources.
    \item \underline{Sources are invariant under $\mathcal{PT}$}. Here $s^*(\mathcal{PT}x)= s(x)$ and $\bar{s}^*(\mathcal{PT}x)= \bar{s}(x)$. Thus, we do not expect the state to always break $\mathcal{PT}$ as no external source breaks it. Hence, the SSB would be related to a dynamical process.
\end{itemize}
Clearly the former situation is not particularly interesting while the second possibility is much more physically relevant. For that reason, we will refer as SSB only to the second scenario.\footnote{This choice matches the one that would follow from the alternate view of the transformation discussed at the end of footnote \ref{ftnote:PassivevsActive}.}

For later convenience, we also consider the possibility of adding to the action \eqref{eq:PTAction_NoCurrent} a term coupling the $U(1)$ current to an external gauge field $a_\mu$
\begin{equation}\label{eq:PTAction_YesCurrent}
    \mathcal{S}'=\mathcal{S}_{\text{CFT}}+\int d^\text{d}x\,\,\left(\bar{s}\mathcal{O}+s\bar{\mathcal{O}}+a_\mu J^\mu\right)\,.
\end{equation}
The current and the gauge field transform under $\mathcal{PT}$ as\footnote{Note that these are the usual transformation rules for a current operator and an external 1-form gauge field. The only remarkable feature is that we allow for the external gauge field to be a complex function. As we will see, this complexification plays an important role in the construction of the Dyson map.} 
\begin{equation}\label{eq:PT_transformation_Gauge}
    a_\mu(x)\xrightarrow{\mathcal{PT}}(-1)^\mu a_\mu^*(\mathcal{PT}x)\,,\qquad J^\mu(x)\xrightarrow{\mathcal{PT}}-(-1)^\mu J^\mu(\mathcal{PT}x)\,,
\end{equation}
where $(-1)^\mu=-1$ for $\mu=0,1$ and $+1$ otherwise. Hence the theory \eqref{eq:PTAction_YesCurrent} is invariant under $\mathcal{PT}$ provided that the external gauge field satisfies 
\begin{equation}
    - a_\mu^*(\mathcal{PT}x)= a_\mu(\mathcal{PT}x)\,.
\end{equation}

We stress that although the equation above indicates that $\mathcal{PT}$ is broken for real external gauge fields, this is not in conflict with the use of real gauge fields in $\mathcal{PT}$-symmetric QFTs. The key point here is that $a_\mu$ is not a quantum field that fluctuates in the path integral, instead it is a classical 1-form that transforms under $\mathcal{PT}$ as such.\footnote{Under $\mathcal{PT}$ an external 1-form $a=a_\mu dx^\mu$ transforms as a function, i.e., 
\begin{equation*}
    a(x)\xrightarrow{\mathcal{PT}} a^*(\mathcal{PT}x)=a_\mu^*(\mathcal{PT}x)d(\mathcal{PT}x)^\mu
\end{equation*}
Hence the component $a_\mu$ satisfies $a_\mu(x)\xrightarrow{\mathcal{PT}}(-1)^\mu a_\mu^*(\mathcal{PT}x)$.} Hence the restrictions to $a_\mu$ are completely different from those imposed on a conventional $U(1)$ quantum gauge field.


\subsection{Dyson map as an external gauge transformation}\label{subsect:Dyson map as an external gauge transformation}

As stated in the introduction, in the $\mathcal{PT}$-symmetric phase, a non-Hermitian theory can be mapped to a Hermitian one through a Dyson map. This immediately ensures the existence of unitary evolution for some cleverly chosen inner product.

In general constructing the Dyson map in a field theory is a highly non-trivial task. However, as shown in \cite{Morales-Tejera:2022hyq}, for our model we can easily build  
a family of non-Hermitian theories with fine-tuned gauge field such that we can identify the inverse Dyson map with an external gauge transformation.
Let us consider a Hermitian theory with action $\eqref{eq:PTAction_NoCurrent}$ and $s=\bar{s}=M$. To obtain a non-Hermitian theory we could consider the similarity transformation such that
\begin{equation}\label{eq:Similarity_operators}
    \mathcal{O}\rightarrow S\mathcal{O}\,,\qquad \bar{\mathcal{O}}\rightarrow S^{-1}\bar{\mathcal{O}}\,,
\end{equation}
where $S=e^{\beta(x)}$ is a complexified $U(1)$ transformation. At leading order, this transformation yields the following non-Hermitian action 
\begin{equation}
    \mathcal{S}=\mathcal{S}_{\text{CFT}}+\int d^\text{d}x\,\,\left(MS\mathcal{O}+MS^{-1}\bar{\mathcal{O}}+\frac{i}{q}(S^{-1} \partial_\mu S)  J^\mu +O(S^{-1} \partial_\mu S)^2 \right)\,,
\end{equation}
where $q$ is the $U(1)$ charge of $\mathcal{O}$ and $J^\mu$ is the $U(1)$ current. This action 
has the form of \eqref{eq:PTAction_YesCurrent} as 
$iq^{-1}S^{-1} \partial_\mu S$ behaves as an external gauge field. The term proportional to the current follows from the existence of the $U(1)$ symmetry in $\mathcal{S}_\text{CFT}$. We know that if $S$ were to be a conventional infinitesimal $U(1)$ matrix, the variation of $\mathcal{S}_\text{CFT}$ under \eqref{eq:Similarity_operators} would give a term $iq^{-1}(S^{-1} \partial_\mu S)  J^\mu$ plus higher order corrections; all of which could be eliminated by coupling the theory to a $U(1)$ gauge field. However, now under the complexified $U(1)$, the gauge field needed to eliminate the extra terms cannot consistently be made into a dynamical quantum gauge field, instead it will be an external classical gauge field \cite{Chernodub:2024lkr}.

In terms of the external sources $s, \bar{s}$ and the external gauge field $a_\mu$ we can construct a Dyson map as the following external gauge transformation
\begin{equation}\label{eq:Dyson_Map}
   s\rightarrow sS^{-1}\,,\qquad \bar{s}\rightarrow \bar{s}S\,,\qquad a_\mu\rightarrow a_\mu+\frac{i}{q}S^{-1}\partial_\mu S\,.
\end{equation}
It is important to note that, in general, this Dyson map does not bring any action \eqref{eq:PTAction_YesCurrent} to a Hermitian one. We can use it to set $s=\bar{s}$, but it will typically induce an $a_\mu\neq0$ unless the initial $a_\mu$ is properly fine-tuned. Nonetheless, we will abuse notation and always call this complexified external gauge transformation Dyson map.


\subsection{Gravitational dual}\label{subsect:Gravitational dual}
For our holographic model, we consider the 4-dimensional minimal model introduced in \cite{Arean:2019pom}, whose bulk action is given by
\begin{equation}\label{eq:Bulk_Grav_Action}
    \mathcal{S}=\int d^4y \sqrt{-g}\left(R-2\Lambda-\frac{1}{4}F_{MN}F^{MN}-\mathcal{D}_M\phi \mathcal{D}^M\bar{\phi}-m^2\phi\bar{\phi}-\frac{v}{2}\phi^2\bar{\phi}^2\right)\,,
\end{equation}
where $\phi$ and $\bar{\phi}$ are scalars charged under a $U(1)$ symmetry, $A$ is the gauge field of the $U(1)$ symmetry, $F=dA$ is the field strength tensor, $\Lambda=-3/l^2$ is the cosmological constant and $l$ is the AdS radius, which we set to unity. The scalar mass is taken to be $m^2=-2$ and the coupling $v=3$. The action of the $U(1)$ transformation is 
\begin{equation}\label{eq:Bulk_GT}
    \phi\rightarrow e^{-iq\alpha}\phi\,,\qquad \bar{\phi}\rightarrow e^{iq\alpha}\bar{\phi}\,,\qquad A_M\rightarrow A_M-\partial_M\alpha\,,
\end{equation}
and the covariant derivatives are defined as 
\begin{equation}
    \mathcal{D}_M\phi=\partial_M\phi-iqA_M\phi\,,\qquad \mathcal{D}_M\bar{\phi}=\partial_M\bar{\phi}+iqA_M\bar{\phi}\,,
\end{equation}
where $q$ is the scalar charge.

The equations of motion derived from the action \eqref{eq:Bulk_Grav_Action} are 
\begin{subequations}\label{eq:EoMs_NoDeTurck}
\begin{align}
    &\mathcal{M}_{N}=\nabla_M \tensor{F}{^M_N}+iq\left(\phi \mathcal{D}_N\bar{\phi}-\bar{\phi} \mathcal{D}_N\phi\right)=0\,,\label{eq:EoMs_NoDeTurck_Maxwell}\\
    &\mathcal{F}=\left(\nabla_M-iqA_M\right) \mathcal{D}^M\phi-m^2\phi-v\phi^2\bar{\phi}=0\,,\\
    &\mathcal{\bar{F}}=\left(\nabla_M+iqA_M\right) \mathcal{D}^M\bar{\phi}-m^2\bar{\phi}-v\bar{\phi}^2\phi=0\,,\\
    &\mathcal{E}_{MN}=R_{MN}-\Lambda g_{MN}-\frac{1}{2}\left(T_{MN}-\frac{1}{2}g_{MN}\tensor{T}{^{P}_P} \right)=0\,,\label{eq:EoMs_NoDeTurck_Einstein}
\end{align}
\end{subequations}
where the bulk energy-momentum tensor is given by
\begin{equation}
    T_{MN}=\tensor{F}{_{PM}}\tensor{F}{^{P}_N}+2 \mathcal{D}_{(M}\phi\mathcal{D}_{N)}\bar{\phi}-g_{MN}\left(\frac{1}{4}F_{PQ}F^{PQ}+\mathcal{D}_P\phi \mathcal{D}^P\bar{\phi}+m^2\phi\bar{\phi}+\frac{v}{2}\phi^2\bar{\phi}^2\right)\,.
\end{equation}

We are interested in static solutions at nonzero temperature and vanishing chemical potential with inhomogeneities along the direction $x^1$ only. Hence, we choose the following ansatz compatible with those assumptions~\cite{Hartnoll:2015faa}
\begin{align}\label{eq:Ansatz_Poincare}
    ds^2=\frac{1}{z^2}&\left[- \left(1-z^3\right)h_1 dt^2 + h_3 \left(dx^1 + h_5 dz\right)^2 +h_4 (dx^2)^2  + \frac{h_2}{1-z^3}  dz^2 \right]\,,\nonumber\\
    &h_1=h_1(x^1,z)\,,\qquad  h_2=h_2(x^1,z)\,,\qquad  h_3=h_3(x^1,z)\,,\nonumber\\
    &\qquad\qquad h_4=h_4(x^1,z)\,,\qquad  h_5=h_5(x^1,z)\,,\nonumber\\
    &\phi=\phi(x^1,z)\,,\qquad \bar{\phi}=\bar{\phi}(x^1,z)\,,\qquad A=A_1(x^1,z)dx^1\,.
\end{align}
where $z$ is the holographic radial coordinate. The AdS boundary is located at $z=0$ and we have a black brane horizon at $z=1$.   

Regarding the boundary conditions, at the AdS boundary we impose
\begin{align}\label{eq:UV_Boundary_Conditions}
    ds^2&(z\rightarrow0)=\frac{1}{z^2}\left[- dt^2 +(dx^1)^2 + (dx^2)^2  + dz^2 \right]\,,\nonumber\\
    \partial_z\phi(x^1,0)=&s(x^1)\,,\qquad \partial_z\bar{\phi}(x^1,0)=\bar{s}(x^1)\,,\qquad A_1(x^1,0)=a_1(x^1)\,,
\end{align}
where $s$, $\bar{s}$ and $a_1$ are the external sources and gauge field in the action \eqref{eq:PTAction_YesCurrent}.

Throughout this paper we focus only on two distinct cases characterized by the values of $s$, $\bar{s}$ and $a_1$
\begin{enumerate}

    \item[A.] Here we take the following parameters
    \begin{equation}\label{eq:Def_Sources_PTbroken}
        s=(1-\eta)M\,,\qquad  \bar{s}=(1+\eta)M\,,\qquad a_1=0\,,
    \end{equation}
    where $M$ is a constant and $\eta=\eta(x^1)$ is a function representing the non-Hermiticity. 
    \item[B.] In this case we instead take the family of sources
    \begin{equation}\label{eq:Def_Sources_PTunbroken}
        s=(1-\eta)M\,,\qquad  \bar{s}=(1+\eta)M\,,\qquad a_1=\frac{i}{q}\frac{\eta'}{1-\eta^2}\,,
    \end{equation}
    which, following the discussion of section \ref{subsect:Dyson map as an external gauge transformation} can be mapped through a Dyson map to the Hermitian case 
    \begin{equation}\label{eq:Def_Sources_Hermitian}
        s=\sqrt{1-\eta^2}M\,,\qquad  \bar{s}=\sqrt{1-\eta^2}M\,,\qquad a_1=0\,.
    \end{equation}
\end{enumerate}

Note that case B is ill-defined if $|\eta(x^1)|=1$ as $a_1$ in \eqref{eq:Def_Sources_PTunbroken} diverges. Moreover, the alternative picture \eqref{eq:Def_Sources_Hermitian} obtained through the Dyson map, also ceases to be $\mathcal{PT}$-symmetric and Hermitian as the sources $s$ and $\bar{s}$ become complex. Hence, as we would like the non-Hermiticity parametrized by $\eta$ to be a feature that we can continuously turn on starting from the Hermitian case with $\eta=0$; we only consider $|\eta(x^1)|<1$ in case B for consistency. On the other hand, in case A we can, a priori, take $|\eta(x^1)|\geq 1$ without any apparent issues. However, as we will see, the regime $|\eta(x^1)|\geq1$ does not admit solutions with real metrics for all values of $\eta$.

To conclude this section we summarize here the relevant holographic dictionary needed to analyze the dual CFT. An explicit derivation can be found in appendix \ref{appendix:Holographic Dictionary}. Using the following asymptotic expansion for the fields
\begin{subequations}\label{eq:Asymptotic_Expansion}
\begin{align}
    h_1(x^1,z)&=1-\frac{1}{4}s(x^1)\bar{s}(x^1)\,z^2+h_1^{(3)}(x^1)\,z^3+...\,,\\
    h_2(x^1,z)&=1-\frac{1}{3}\left[h_1^{(3)}(x^1)+h_3^{(3)}(x^1)+h_4^{(3)}(x^1)\right]\,z^3+...\,,\\
    h_3(x^1,z)&=1-\frac{1}{4}s(x^1)\bar{s}(x^1)\,z^2+h_3^{(3)}(x^1)\,z^3+...\,,\\
    h_4(x^1,z)&=1-\frac{1}{4}s(x^1)\bar{s}(x^1)\,z^2+h_4^{(3)}(x^1)\,z^3+...\,,\\
    h_5(x^1,z)&=\frac{1}{8}\partial_1\left[\bar{s}(x^1)s(x^1)\right]\,z^3 +...\,,\\ 
    \phi(x^1,z)&=s(x^1)\,z+\phi^{(2)}(x^1)\,z^2+...\,,\\
    \bar{\phi}(x^1,z)&=\bar{s}(x^1)\,z+\bar{\phi}^{(2)}(x^1)\,z^2+...\,,\\
    A_1(x^1,z)&=a_1(x^1)+A_1^{(1)}(x^1)\,z+...\,,
\end{align}
\end{subequations}
obtained from solving the equations of motion for $z\rightarrow0$; we conclude that the vacuum expectation values (VEVs) of the operators $\mathcal{O}$ and $\bar{\mathcal{O}}$, the current $J_1$, and the boundary energy-momentum tensor $T_{\mu\nu}$; are given by
\begin{align}\label{eq:VEVs}
    \expval{J_1}&=A_1^{(1)}\,,\qquad \expval{\mathcal{O}}=\phi^{(2)}\,,\qquad \expval{\bar{\mathcal{O}}}=\bar{\phi}^{(2)}\,, \nonumber\\
    \expval{T_{tt}}&=2-\frac{2}{3}h_1^{(3)}+\frac{7}{3}(h_3^{(3)}+h_4^{(3)})+s\bar{\phi}^{(2)}+\bar{s}\phi^{(2)}\,,\nonumber\\
    \expval{T_{11}}&=1+\frac{2}{3}\,h_3^{(3)}-\frac{7}{3}(h_1^{(3)}+h_4^{(3)})-s\bar{\phi}^{(2)}-\bar{s}\phi^{(2)}\,,\nonumber\\
    \expval{T_{22}}&=1+\frac{2}{3}\,h_4^{(3)}-\frac{7}{3}(h_1^{(3)}+h_3^{(3)})-s\bar{\phi}^{(2)}-\bar{s}\phi^{(2)}\,.
    \end{align}
Moreover, we also find the following Ward identity associated with the $U(1)$
symmetry of the boundary CFT
\begin{equation}\label{eq:U(1)_WardIdentity}
    \partial_{1} \expval{J_1}=iq\left(s\expval{\bar{\mathcal{O}}} -\bar{s}\expval{\mathcal{O}}\right)\,.
\end{equation}


\subsection{Matching symmetries}
Lastly, to conclude our description of  the holographic model, we describe the matching of symmetries between the CFT and gravitational dual in greater detail.

First, let us note that a $z$-independent gauge transformation \eqref{eq:Bulk_GT} leaves the action \eqref{eq:Bulk_Grav_Action} invariant but changes the boundary conditions as in \eqref{eq:Dyson_Map}.
Hence, we conclude that, as expected from the usual holographic prescription, the external gauge transformations in the Dyson map correspond to complexified local gauge transformations in the gravitational dual.

Regarding $\mathcal{PT}$, in the gravitational dual we have the following transformation rules inherited from the corresponding dual operators
 \begin{subequations}
\begin{align}
    y=(x,z)=(t,x^1,x^2,z)\xrightarrow{\mathcal{PT}}\mathcal{PT}y=(\mathcal{PT}x,z)=(-t,-x^1,x^2,z)\,,\\
    \phi(y)\xrightarrow{\mathcal{PT}}\phi(\mathcal{PT}y)\,,\qquad \bar{\phi}(y)\xrightarrow{\mathcal{PT}}\bar{\phi}(\mathcal{PT}y)\,,\qquad A_1(y)\xrightarrow{\mathcal{PT}}A_1(\mathcal{PT}y)\,.
\end{align}
\end{subequations}
Thus, the action \eqref{eq:Bulk_Grav_Action} is trivially invariant under $\mathcal{PT}$. However, we also need to account for the boundary conditions \eqref{eq:UV_Boundary_Conditions} which might break $\mathcal{PT}$. Under $\mathcal{PT}$ they transform as
 \begin{subequations}
\begin{align}
    \partial_z\phi(x,0)=s(x) &\xrightarrow{\mathcal{PT}} \partial_z\phi(\mathcal{PT}x,0)=s^*(\mathcal{PT}x)\,, \\
    \partial_z\bar{\phi}(x,0)=\bar{s}(x) &\xrightarrow{\mathcal{PT}} \partial_z\bar{\phi}(\mathcal{PT}x,0)=\bar{s}^*(\mathcal{PT}x)\,, \\
    A_1(x,0)=a_1(x)&\xrightarrow{\mathcal{PT}}  A_1(\mathcal{PT}x,0)=-a_1^*(\mathcal{PT}x)\,,
\end{align}
\end{subequations}
where $\bar{s}$, $s$ have been treated as functions under $\mathcal{PT}$ and $a_1$ as an external gauge field. Then, in order for the theory to be invariant under $\mathcal{PT}$ as in \eqref{eq:Lagrangian_PT}, we conclude that we have to restrict ourselves to
\begin{align}
    s(x)=s^*(x)\,,\qquad \bar{s}(x)=\bar{s}^*(x)\,,\qquad -a_1^*(x)=a_1(x)\,,
\end{align}
which remarkably matches the criteria derived in section \ref{subsect:Non-Hermitian CFT}. In the following, to ensure that the sources do not explicitly break $\mathcal{PT}$ we further restrict  $s$, $\bar{s}$ to be even functions of $x^1$.

\section{Non-Hermitian lattice}\label{sec:NonHermitianLattice}
In this section we consider $\eta$ to be a smooth, even, periodic function with period $L$. Physically, we consider that this setup describes a lattice with lattice constant $L$, where there is inflow/outflow that smoothly extends throughout each site. For simplicity we take $\eta$ to be a cosine function
\begin{equation}\label{eq:eta_Lattice}
    \eta(x^1)=a\cos\left(\frac{2\pi}{L} x^1\right)\,,
\end{equation}
and we study solutions for different values of $a$.

It is worth noting that, due to the conformal symmetry of the UV, our results depend only on dimensionless quantities. Thus, without loss of generality, we choose to parametrize them in terms of $\{a,T/M,LM,q\}$. Moving forward, we set $LM=3$ and $q=1$ unless stated otherwise. This allows us to characterize the solutions by their temperature $T/M$ and the amplitude $a$.


\subsection{Numerical method}\label{subsect:NonHermitianLattice_NumericalMethod}
In order to solve the equations of motion \eqref{eq:EoMs_NoDeTurck}, we employ the DeTurck trick to make the Einstein equations \eqref{eq:EoMs_NoDeTurck_Einstein} elliptic \cite{Headrick:2009pv,Horowitz:2012ky,Dias:2015nua}. In summary, this trick consists of replacing the Einstein equations by the Einstein-DeTurck equations
\begin{equation}
    R_{MN}-\Lambda g_{MN}-\frac{1}{2}\left(T_{MN}-\frac{1}{2}\tensor{T}{^P_P}g_{MN}\right)-\nabla_{(M}\xi_{N)}=0\,,
    \label{eq:EoMs_DeTurck_Einstein}
\end{equation}
where $\xi^P= g^{MN}\left(\tensor{\Gamma}{^P_M_N}-\tensor{\bar{\Gamma}}{^P_M_N}\right)$ with $\Gamma$ and $\bar{\Gamma}$ the Christoffel symbols of the metric we are solving for (target metric) and of a reference metric, respectively. The Einstein-DeTurck equations are hyperbolic by construction and thus they are numerically better behaved. Hence we choose to solve them and check that the DeTurck term vanishes for the solution; thus ensuring that it is also a solution to the Einstein equations. Numerically we guarantee this by checking that the DeTurck constraint $\xi_M\xi^M$ is smaller than a given tolerance. We refer the reader to \cite{Dias:2015nua} for further details on this method.

Importantly, the reference metric should have the same asymptotic behavior as the target metric. Thus we choose the reference metric to be given by
\begin{equation}\label{eq:RefMetric_Poincare}
    ds^2=\frac{1}{z^2}\left[- \left(1-z^3\right) dt^2 + \left(dx^1\right)^2 + (dx^2)^2  + \frac{dz^2}{1-z^3}   \right]\,.
\end{equation}

Besides employing DeTurck's trick, we also note that when solving Maxwell's equations \eqref{eq:EoMs_NoDeTurck_Maxwell} we get a constraint and two second-order equations. We choose to only solve the second-order equations and check that the constraint is satisfied for the solutions up to a given tolerance.

With all this in mind, we discretize the system using a Chebyshev grid in the $z$ direction and a Fourier grid with periodicity $L$ on the $x^1$ direction. We solve the final equations using the Newton-Raphson method. All numerical calculations have been carried out using \textit{Mathematica}.


\subsection{Numerical solutions}
\label{subsect:NonHermitianLattice_NumericalSolutions}
As indicated in section \ref{subsect:Gravitational dual}, we distinguish between two scenarios A and B, defined by equations \eqref{eq:Def_Sources_PTbroken} and \eqref{eq:Def_Sources_PTunbroken}, respectively. 
We label scenario A as the phase with current and B as the phase without current, for reasons that will become apparent below.

Regarding the numerics, in this section we employ a Chebyshev grid with 20 points for the $z$ direction and a Fourier grid with 40 points for the $x^1$ direction; we demand a tolerance of $10^{-8}$ in both the Newton-Raphson method as well as for the DeTurck constraint and the constraint arising from the Maxwell equations.


\subsubsection{Phase with current}
\label{subsubsect:NonHermitianLattice_PTbroken}

\begin{figure}
\centering
\begin{subfigure}{\textwidth}
    \includegraphics[width=\textwidth]{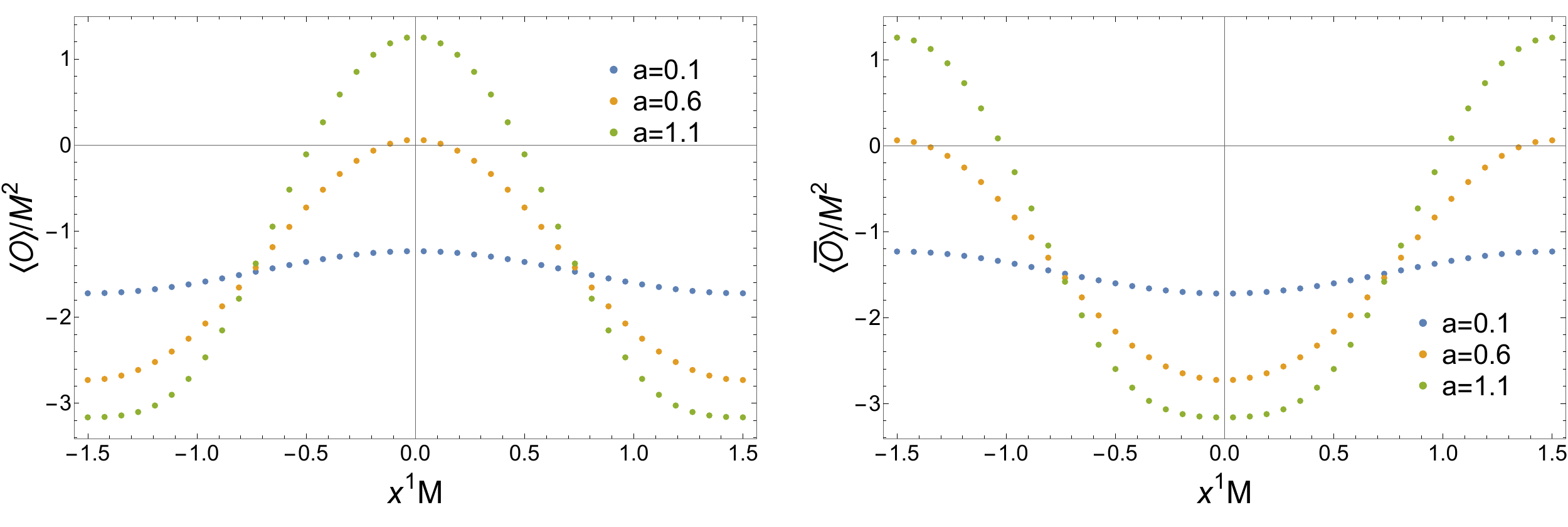}
    \caption{$T/M=0.5$}
    \label{fig:Os_PTbroken_T0d5}
\end{subfigure}
\hfill
\hfill
\begin{subfigure}{\textwidth}
    \includegraphics[width=\textwidth]{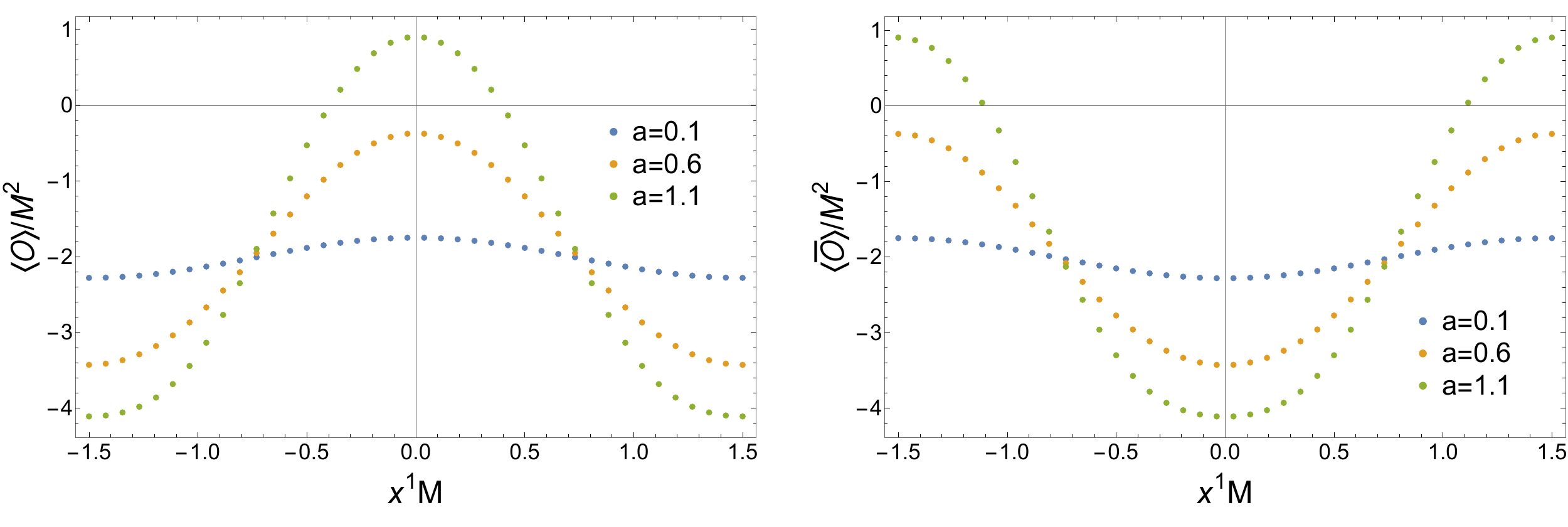}
    \caption{$T/M=1$}
    \label{fig:Os_PTbroken_T1d0}
\end{subfigure}        
\caption{Expectation values of the condensates $\expval{{\mathcal{O}}}$ and $\expval{\bar{\mathcal{O}}}$ in the phase with current.}
\label{fig:Os_PTbroken}
\end{figure}

\begin{figure}
\centering
\begin{subfigure}{.49\textwidth}
    \includegraphics[width=\textwidth]{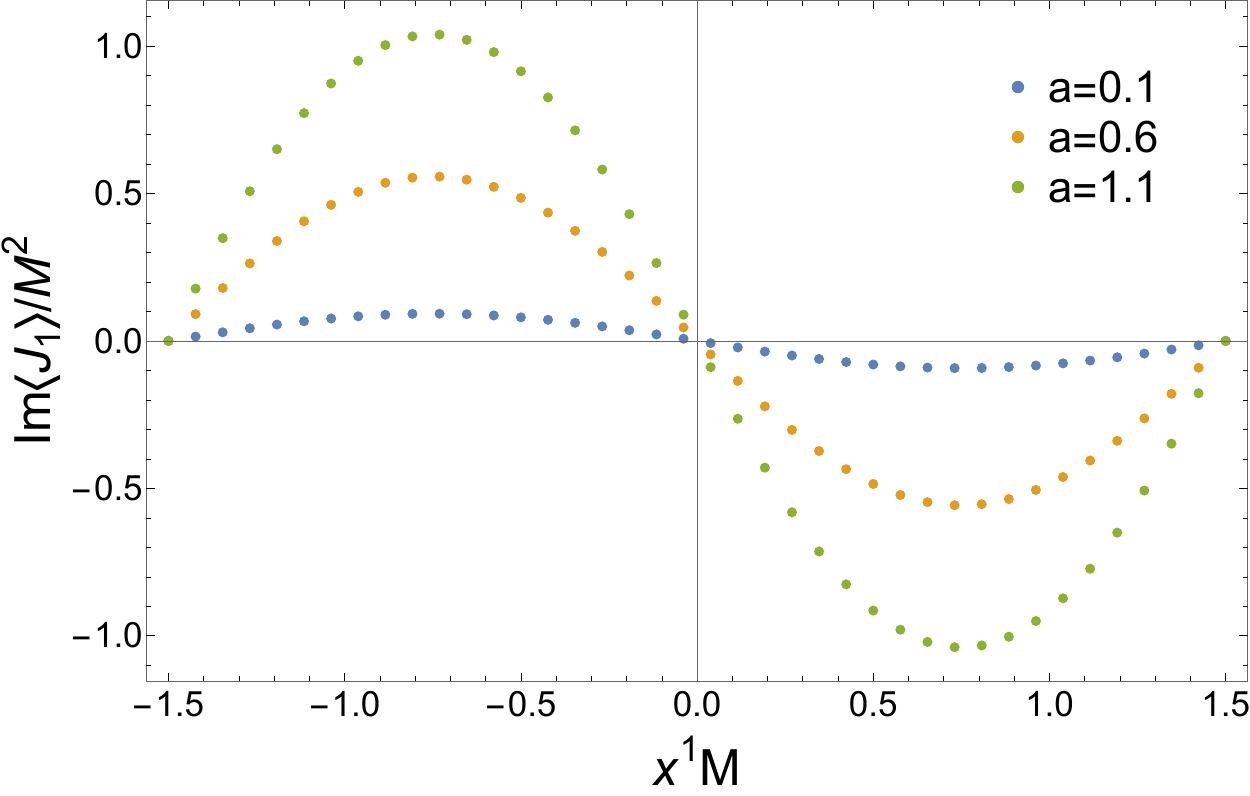}
    \caption{$T/M=0.5$}
    \label{fig:Current_PTbroken_T0d5}
\end{subfigure}
\hfill
\begin{subfigure}{.49\textwidth}
    \includegraphics[width=\textwidth]{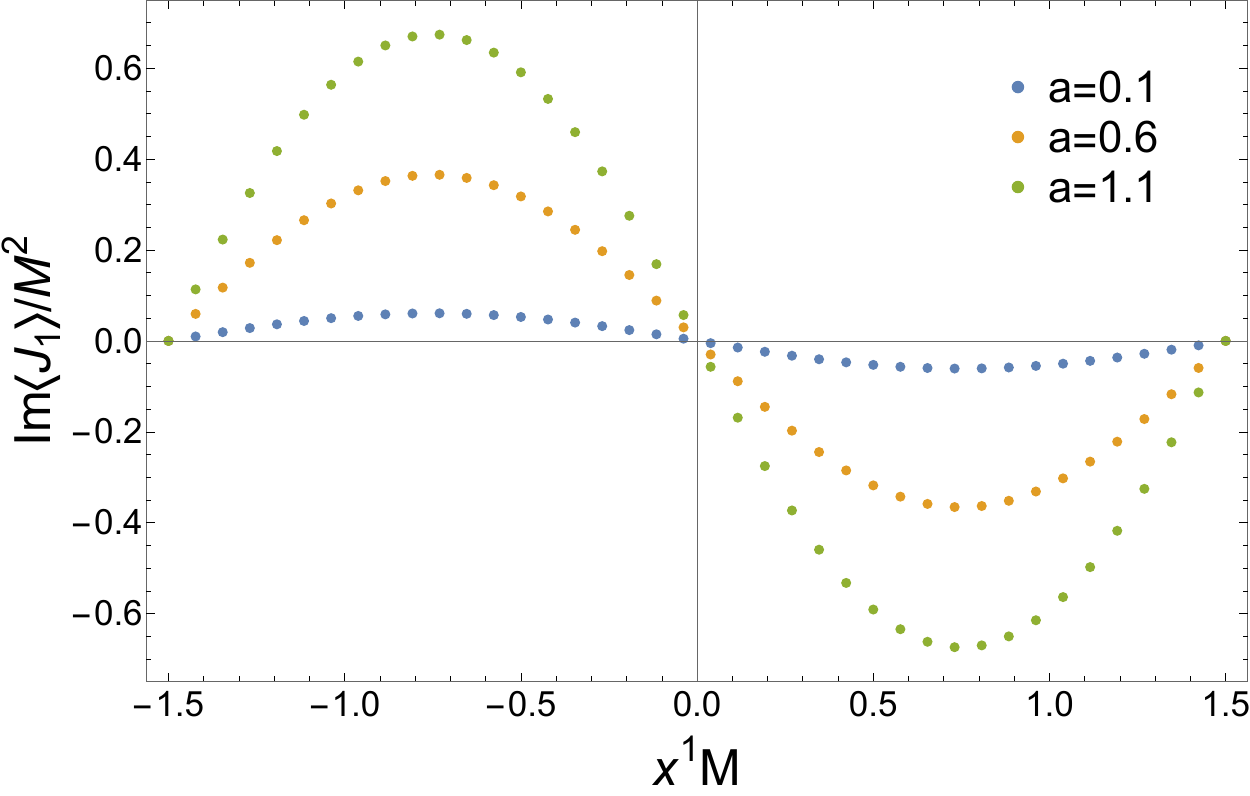}
    \caption{$T/M=1$}
    \label{fig:Current_PTbroken_T1d0}
\end{subfigure}
\caption{Imaginary part of the expectation value of the current $\expval{J_1}$ in the phase with current. The real part is zero for any $\{a,T/M\}$. }
\label{fig:Current_PTbroken}
\end{figure}

\begin{figure}
\centering
\begin{subfigure}{.49\textwidth}
    \includegraphics[width=\textwidth]{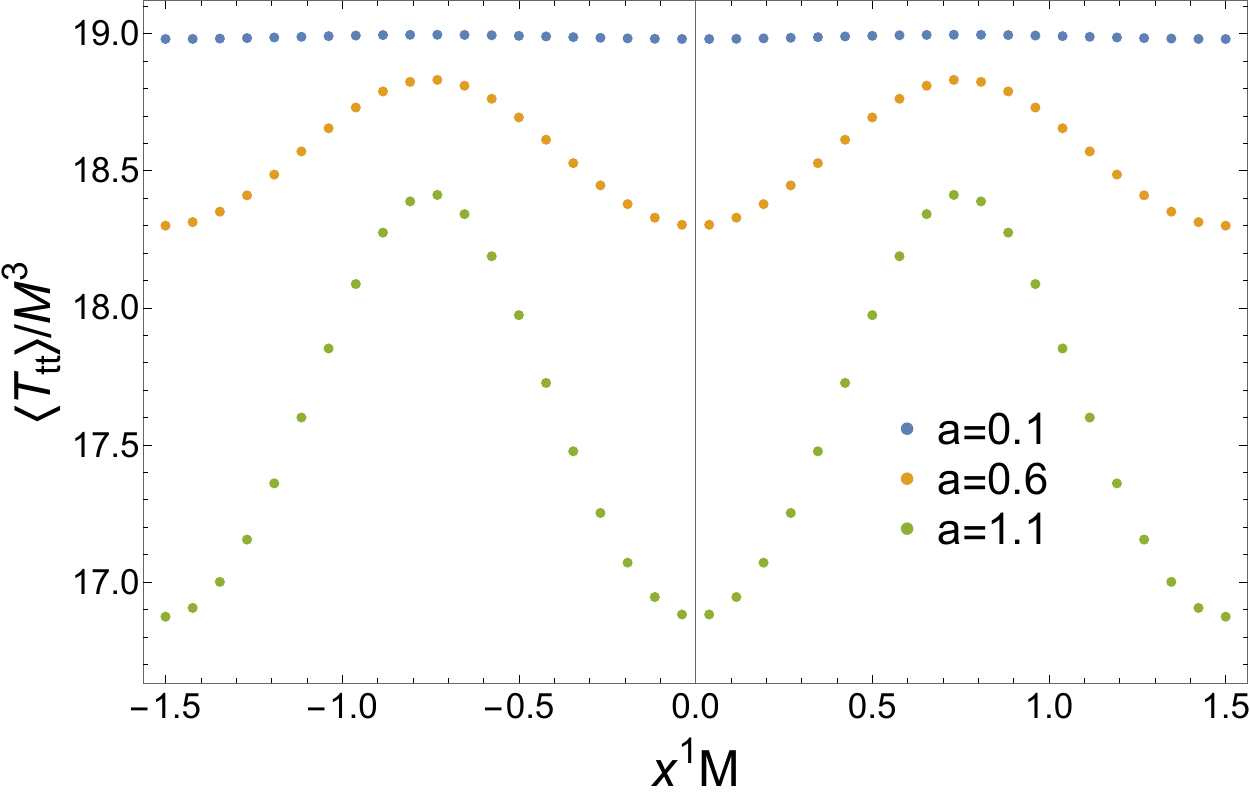}
    \caption{$T/M=0.5$}
    \label{fig:Ttt_PTbroken_T0d5}
\end{subfigure}
\hfill
\begin{subfigure}{.49\textwidth}
    \includegraphics[width=\textwidth]{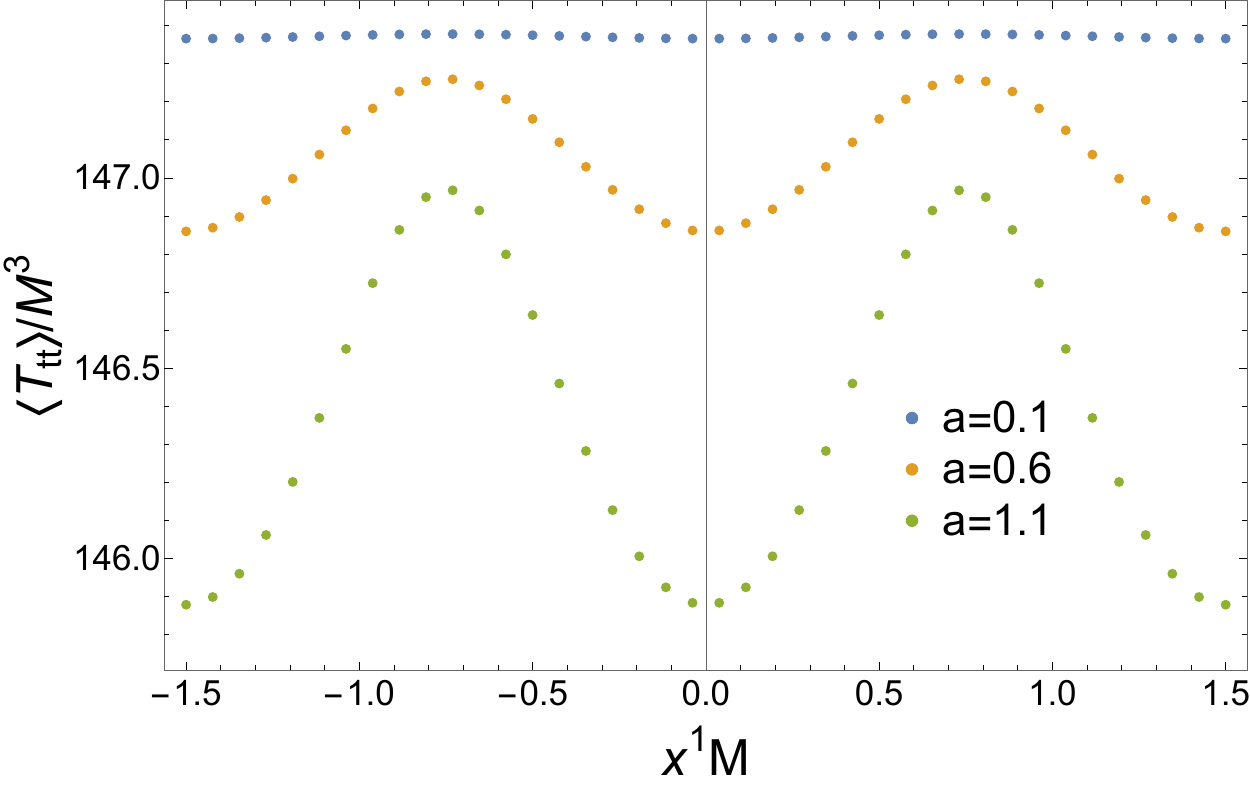}
    \caption{$T/M=1$}
    \label{fig:Ttt_PTbroken_T1d0}
\end{subfigure}
\caption{Expectation value of the energy density $\expval{T_{tt}}$ in the phase with current. }
\label{fig:Ttt_PTbroken}
\end{figure}

In figures \ref{fig:Os_PTbroken}-\ref{fig:Ttt_PTbroken} we plot the expectation values of the condensates $\expval{\mathcal{O}}$, $\expval{\bar{\mathcal{O}}}$, the current $\expval{J_1}$ and the energy density $\expval{T_{tt}}$ for amplitudes $a=\{0.1,0.6,1.1\}$ and temperatures $T/M=\{0.5,1\}$. 
We observe that the inhomogeneous non-Hermiticity yields a purely imaginary current which was previously absent in the homogeneous setup.
This current is an odd function of $x^1$ and thus it
does not spontaneously break $\mathcal{PT}$ despite its imaginary character:\footnote{Note that due to the antiunitary nature of $\mathcal{PT}$, when acting on a vacuum expectation value we get the $\mathcal{PT}$-transformed operator and an overall conjugation. }
\begin{equation}\label{eq:J_underPT}
    \expval{J_1(x)}\xrightarrow{\mathcal{PT}}\expval{J_1(\mathcal{PT}x)}^*=-\expval{J_1(\mathcal{PT}x)}=\expval{J_1(x)}\,,
\end{equation}

In view of the $\mathcal{PT}$-preserving nature of all expectation values,
this model allows for unitary time-evolution despite not being mapped to
a Hermitian theory via the Dyson map \eqref{eq:Dyson_Map}.
We could use the map \eqref{eq:Dyson_Map} with $S=\sqrt{\frac{1-\eta}{1+\eta}}$ to have $s=\bar{s}=\sqrt{1-\eta^2}M$ and thus eliminate the non-Hermiticity from the sources. However, by doing so we introduce a purely imaginary external gauge field $a_1$, proportional to $\partial_1\eta$, which couples to the current $J_1$, which is a Hermitian operator. Hence, even if we remove the non-Hermiticity from the sources of $\mathcal{O}$ and $\bar{\mathcal{O}}$, the Lagrangian always remains non-Hermitian as long as $\partial_1\eta\neq0$.
It would be interesting to find a Dyson map between this $\mathcal{PT}$-symmetric phase with an imaginary current and its equivalent Hermitian theory.
However, we have not managed to find it as we expect it to correspond to some non-trivial field redefinition.
It is worth noting that the current we observe is the same up to multiplication by $i$ as the current arising in the Hermitian setup resulting from changing $\eta\rightarrow i\eta$. Nonetheless, the scalar expectation values and the dual geometry are different in both scenarios.

\begin{figure}
\centering
\begin{subfigure}{.49\textwidth}
    \includegraphics[width=\textwidth]{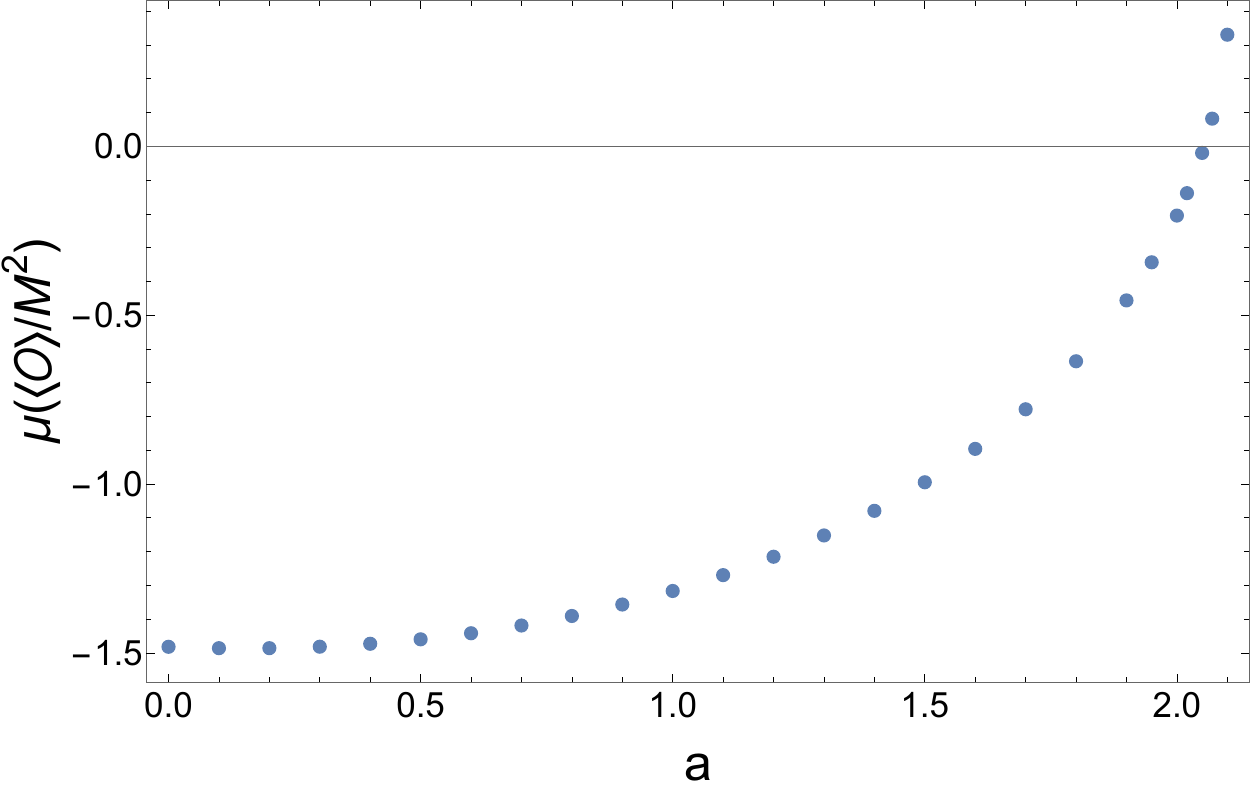}
    \caption{$T/M=0.5$}
    \label{fig:Ttt_PTbroken_T0d5}
\end{subfigure}
\hfill
\begin{subfigure}{.49\textwidth}
    \includegraphics[width=\textwidth]{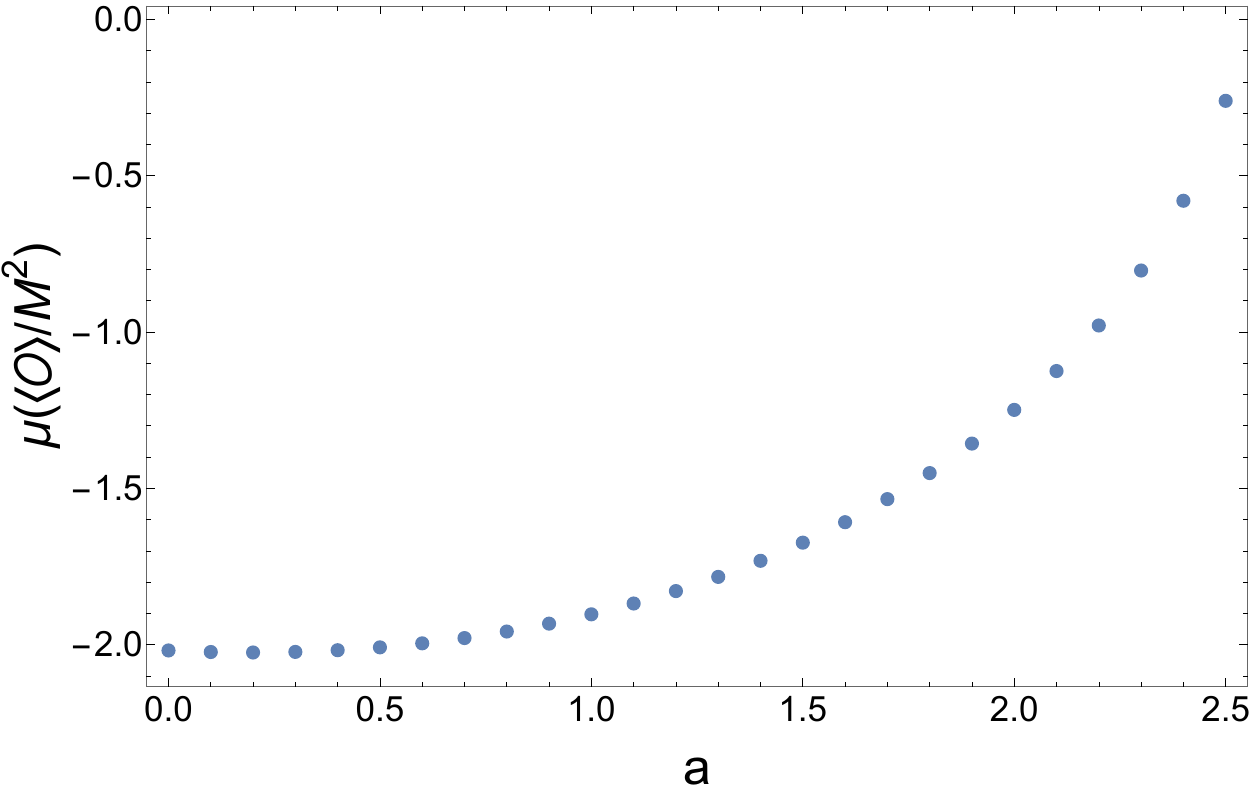}
    \caption{$T/M=1$}
    \label{fig:Ttt_PTbroken_T1d0}
\end{subfigure}      
\caption{Mean of the condensate $\mu(\expval{{\mathcal{O}}})$ in the phase with current as a function of the amplitude $a$. We do not find solutions with real metric for $a>a_\text{max}\approx 2.1$ in (a) and $a>a_\text{max}\approx 2.5$ in (b). We note that  $\mu(\expval{{\mathcal{O}}})=\mu(\expval{\bar{\mathcal{O}}})$.}.
\label{fig:OsCurve_PTbroken}
\end{figure}

Interestingly, we cannot find solutions with real metric for all values of $\eta$. Indeed, as illustrated in figure \ref{fig:OsCurve_PTbroken} where we plot the mean of $\expval{\mathcal{O}}$ 
as a function of $a$, we observe that there is a maximum $a_\text{max}$ beyond which solutions are not found. This behavior is very reminiscent of the one observed in the homogeneous setup in \cite{Arean:2019pom}. However, as we will see in the next section, we observe that in our setup, some of the solutions with $\max|\eta|>1$ are linearly stable.


\subsubsection{Phase without current}
\begin{figure}
\centering
\begin{subfigure}{\textwidth}
    \includegraphics[width=\textwidth]{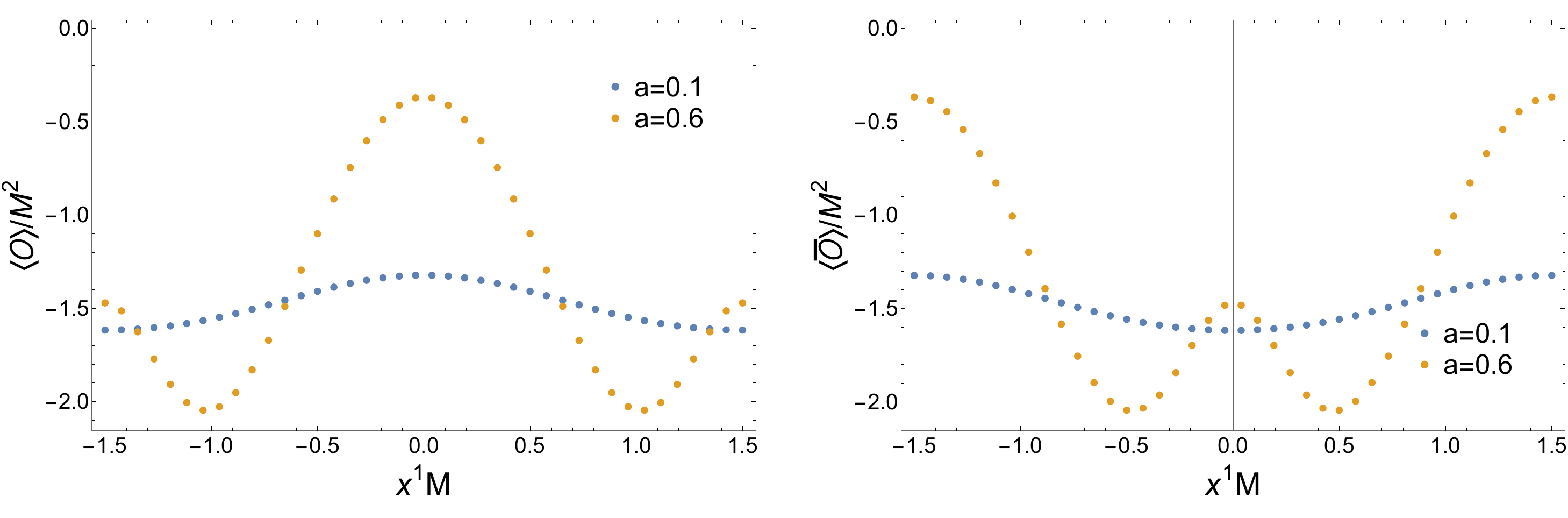}
    \caption{$T/M=0.5$}
    \label{fig:Os_PTunbroken_T0d5}
\end{subfigure}
\hfill
\hfill
\begin{subfigure}{\textwidth}
    \includegraphics[width=\textwidth]{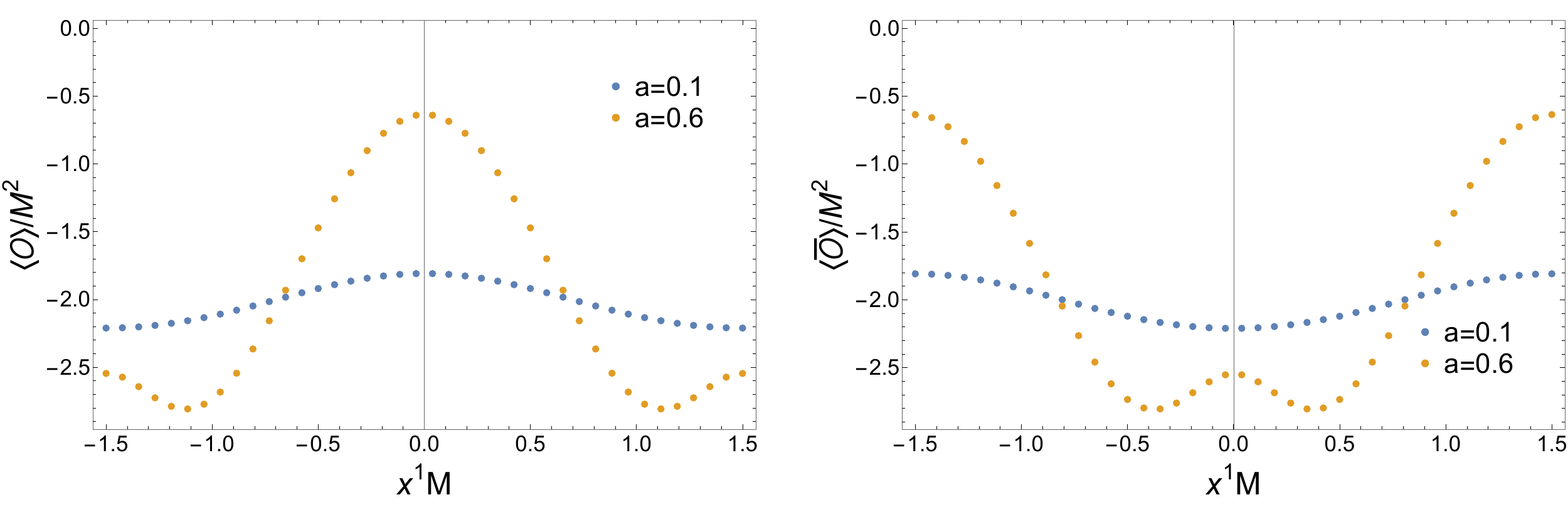}
    \caption{$T/M=1$}
    \label{fig:Os_PTunbroken_T1d0}
\end{subfigure}        
\caption{Expectation values of the condensates $\expval{{\mathcal{O}}}$ and $\expval{\bar{\mathcal{O}}}$ in the phase without current.}
\label{fig:Os_PTunbroken}
\end{figure}

\begin{figure}
\centering
\begin{subfigure}{.49\textwidth}
    \includegraphics[width=\textwidth]{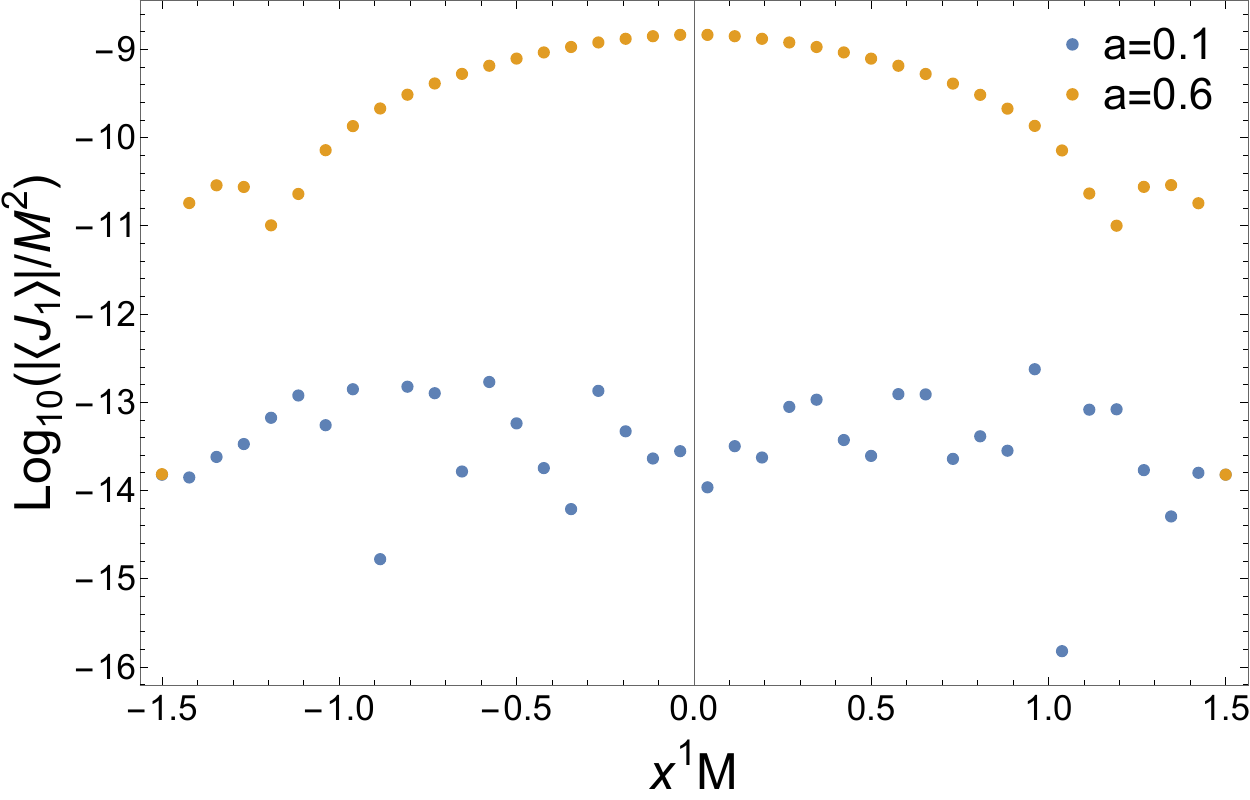}
    \caption{$T/M=0.5$}
    \label{fig:Current_PTunbroken_T0d5}
\end{subfigure}
\hfill
\begin{subfigure}{.49\textwidth}
    \includegraphics[width=\textwidth]{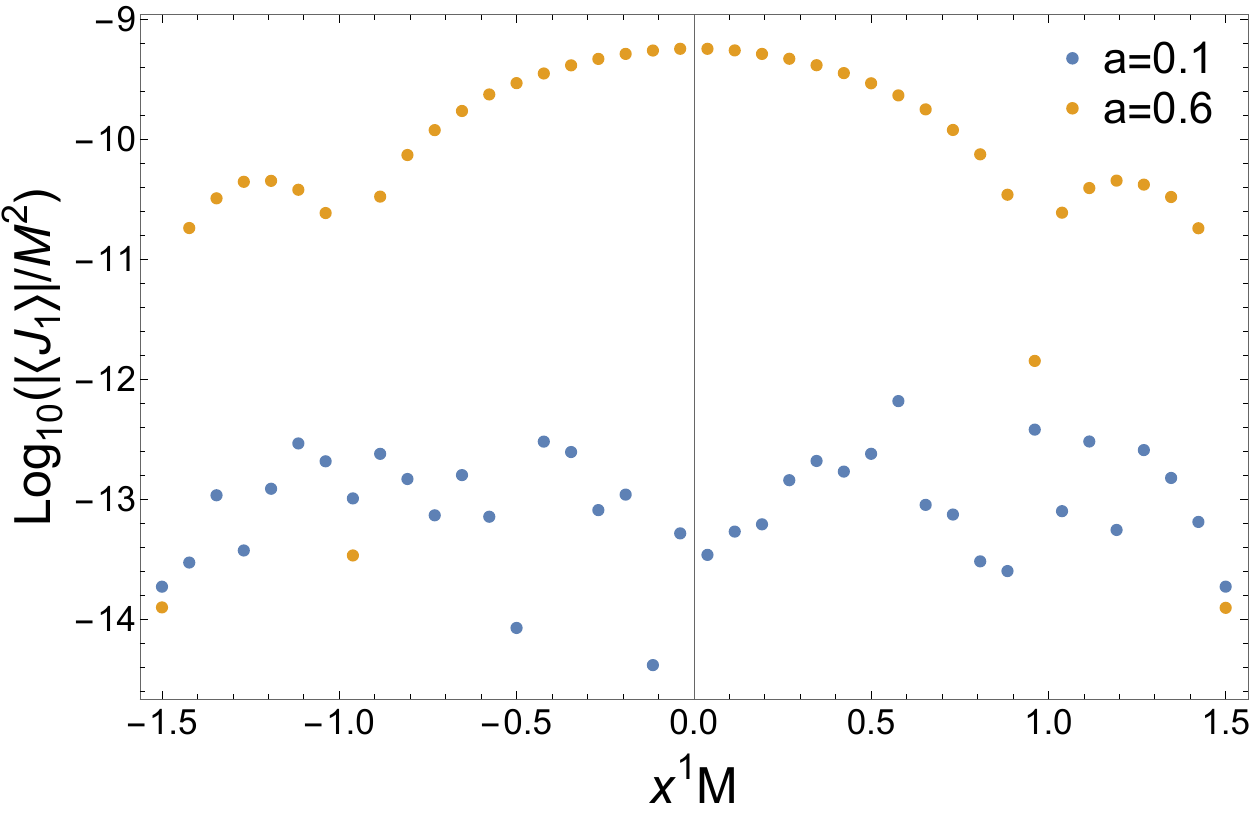}
    \caption{$T/M=1$}
    \label{fig:Current_PTunbroken_T1d0}
\end{subfigure}
\caption{Absolute value of the expectation value of the current $\expval{J_1}$ in the phase without current. Note that in both plots the current vanishes up to the accuracy of the numerics. }
\label{fig:Current_PTunbroken}
\end{figure}

\begin{figure}
\centering
\begin{subfigure}{.485\textwidth}
    \includegraphics[width=\textwidth]{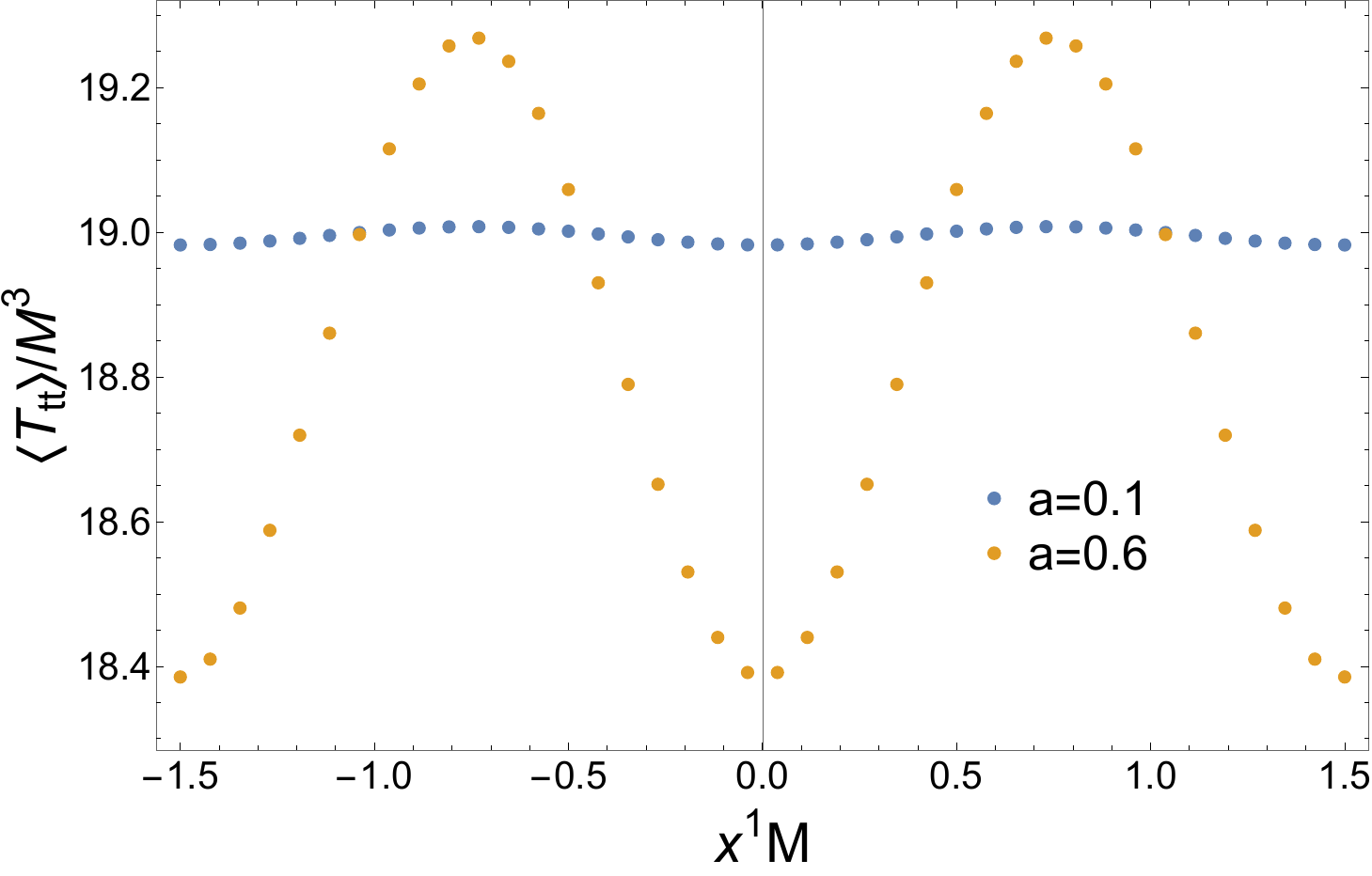}
    \caption{$T/M=0.5$}
    \label{fig:Ttt_PTunbroken_T0d5}
\end{subfigure}
\hfill
\begin{subfigure}{.495\textwidth}
    \includegraphics[width=\textwidth]{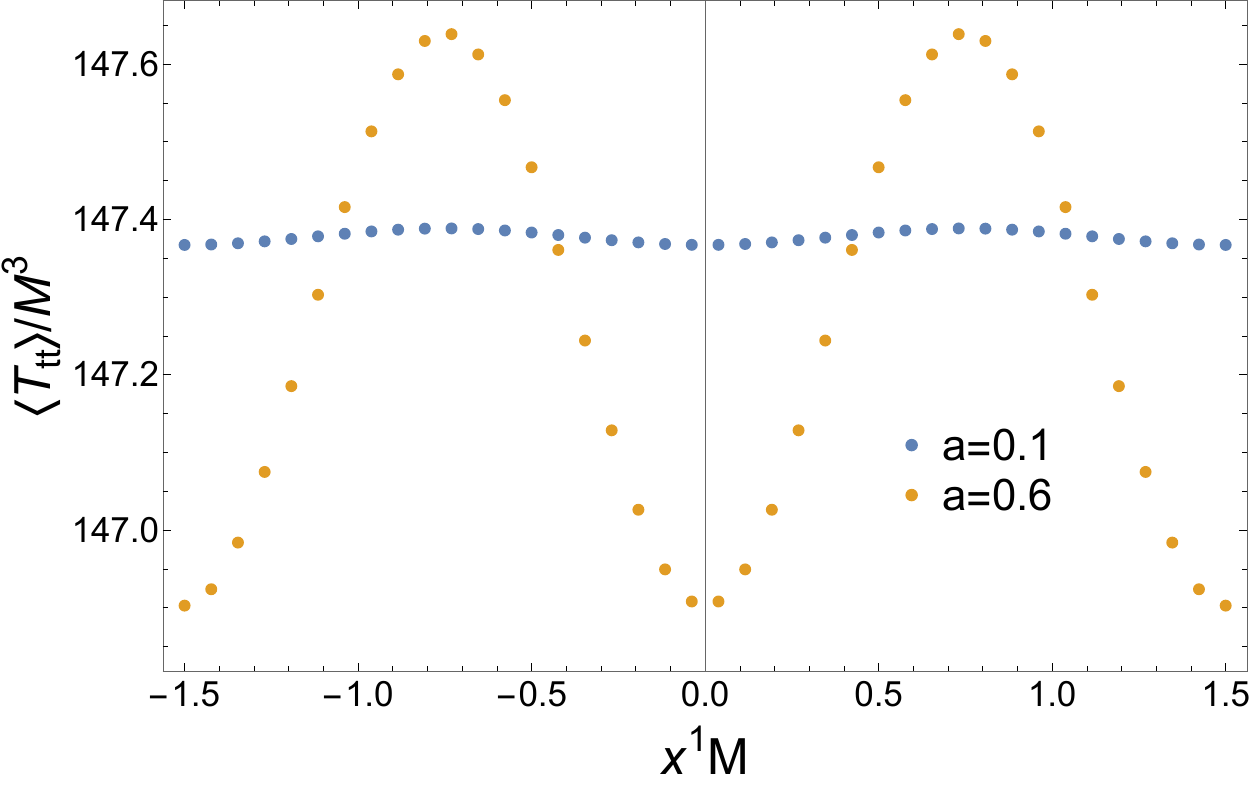}
    \caption{$T/M=1$}
    \label{fig:Ttt_PTunbroken_T1d0}
\end{subfigure}
\caption{Expectation value of the energy density $\expval{T_{tt}}$ in the phase without current. }
\label{fig:Ttt_PTunbroken}
\end{figure}

In figures \ref{fig:Os_PTunbroken}-\ref{fig:Ttt_PTunbroken} we plot the expectation values of the condensates $\expval{\mathcal{O}}$, $\expval{\mathcal{\bar{O}}}$, the current $\expval{J_1}$ and the energy density $\expval{T_{tt}}$ for amplitudes $a=\{0.1,0.6\}$ and temperatures $T/M=\{0.5,1\}$\footnote{Recall that in this setup, corresponding to scenario B in section \ref{subsect:Gravitational dual}, we restrict $|\eta(x^1)|<1$ as discussed in the aforementioned section. For $\eta$ given by \eqref{eq:eta_Lattice} this implies that we must take $|a|<1$.}.
In this case, we observe that the current $\expval{J_1}$ vanishes and 
thus,
as indicated in section \eqref{subsect:Gravitational dual}, we can perform a Dyson map \eqref{eq:Dyson_Map} with $S=\sqrt{\frac{1-\eta}{1+\eta}}$ to obtain a Hermitian $\mathcal{PT}$-symmetric theory with sources $\eqref{eq:Def_Sources_Hermitian}$. In order to explicitly check the equivalence between both descriptions, we would like to compare the VEVs $\expval{\mathcal{O}}_\text{H}$ and $\expval{\bar{\mathcal{O}}}_{\text{H}}$ computed in the Hermitian theory \eqref{eq:Def_Sources_Hermitian} to the VEVs $\expval{\mathcal{O}}_{\text{NH}}$ and $\expval{\bar{\mathcal{O}}}_{\text{NH}}$ computed in the non-Hermitian one \eqref{eq:Def_Sources_PTunbroken}. We can do so by recalling that the Dyson map \eqref{eq:Dyson_Map}, which we identified with an external gauge transformation, can also be regarded as a similarity transformation \eqref{eq:Similarity_operators} generated by an operator $\mathcal{Q}$ such that
\begin{equation}
    \mathcal{Q}\mathcal{O}\mathcal{Q}^{-1}=S\mathcal{O}\,,\qquad  \mathcal{Q}\mathcal{\bar{O}}\mathcal{Q}^{-1}=S^{-1}\bar{\mathcal{O}}\,,
\end{equation}
\begin{equation}
    \mathcal{Q}\mathcal{S}[s,\bar{s},a_\mu]\mathcal{Q}^{-1}=\mathcal{S}\left[S^{-1}s,S\bar{s},a_\mu+\frac{i}{q}S^{-1}\partial_\mu S\right]\,.
\end{equation}
Then, as the trace is invariant under similarity transformations we have that
\begin{align}
    \expval{\mathcal{\bar{O}}}_{\text{NH}}&=\Tr[\mathcal{\bar{O}}e^{-\beta H_{\text{NH}}}]=\Tr[\mathcal{Q}\mathcal{\bar{O}}e^{-\beta H_{\text{NH}}}\mathcal{Q}^{-1}]=\Tr[S^{-1}\mathcal{\bar{O}}e^{-\beta H_\text{H}}]=S^{-1}\expval{\bar{O}}_\text{H}\,,\nonumber\\
    \expval{\mathcal{O}}_{\text{NH}}&=\Tr[\mathcal{O}e^{-\beta H_{\text{NH}}}]=\Tr[\mathcal{Q}\mathcal{O}e^{-\beta H_{\text{NH}}}\mathcal{Q}^{-1}]=\Tr[S\mathcal{O}e^{-\beta H_\text{H}}]=S\expval{O}_\text{H}\,.
\end{align}
where $H_{\text{H}}$ and $H_{\text{NH}}$ are the generators of time evolution, i.e. the Hamiltonians, in the Hermitian and non-Hermitian theories, respectively. 

\begin{figure}
    \centering
    \includegraphics[width=\textwidth]{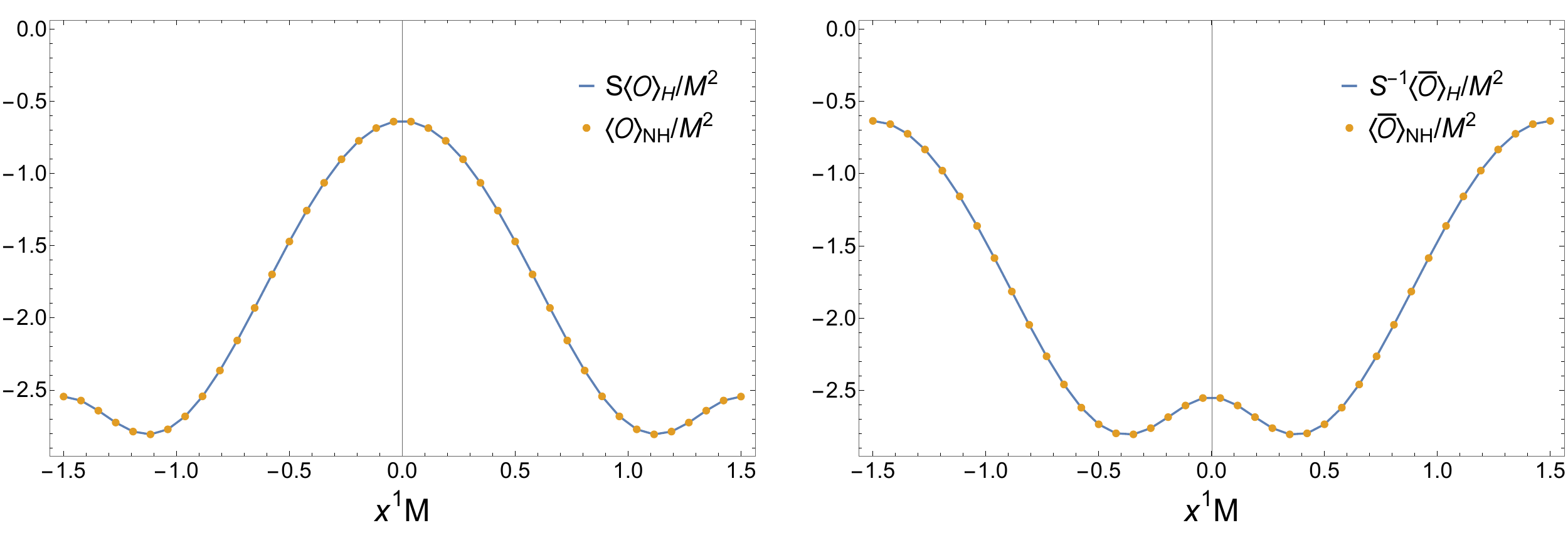}
    \caption{Comparison between the Hermitian and non-Hermitian descriptions for the phase without current phase at $\{a,T/M\}=\{0.6,1\}$.}
    \label{fig:Os_PTunbroken_Comparison}
\end{figure}

Then, provided we can equivalently use the non-Hermitian and Hermitian descriptions, we should find that $\expval{\mathcal{{O}}}_{\text{NH}}=S\expval{\mathcal{{O}}}_{\text{H}}$ and $\expval{\mathcal{\bar{O}}}_{\text{NH}}=S^{-1}\expval{\mathcal{\bar{O}}}_{\text{H}}$ for $S=\sqrt{\frac{1-\eta}{1+\eta}}$. This is confirmed by figure \ref{fig:Os_PTunbroken_Comparison} where we clearly see that for $\{a,T/M\}=\{0.6,1\}$ the equivalence is indeed satisfied.


\subsection{Quasinormal frequencies and stability}\label{subsect:NHLattice_Stability}
In order to assess the stability of the solutions we have found in the previous section, we study the quasinormal frequencies (QNFs) of the system, looking for potential instabilities. 
For the phase without current we know the Dyson map and the corresponding Hermitian theory, which we do not expect to be unstable. Hence, here we focus only on the phase with imaginary current whose Hermitian counterpart is unknown. 

QNFs are the eigenfrequencies of the linearized fluctuations over our background solution which satisfy infalling boundary conditions at the horizon and sourceless boundary conditions at the AdS boundary. Consequently, we choose to write the corresponding quasinormal modes (QNMs) as 
\begin{equation}
    \delta \varphi=\delta\varphi(x^1,z)e^{-i\omega t +i k x^1} \,,
\end{equation}
where $\varphi$ denotes any field of the gravity side satisfying the aforementioned boundary conditions, $\omega$ is the QNF and $k$ is the momentum.\footnote{Our definition of the QNMs is motivated by the usual Bloch decomposition frequently used in solid state physics. As the lattice is periodic, we can write any eigenfunction as a plane wave with momentum $k$ in the first Brillouin zone times a periodic function.} Due to the periodic structure of the lattice, the momentum $k$ satisfies $k\sim k+\frac{2\pi}{L}$.

As we defined QNMs without any momentum in the $x^2$ direction, we are not breaking the reflection symmetry along the $x^2$ direction. Hence, we can decouple fluctuations into two sectors characterized by their transformation rules under reflections along the $x^2$ axis
\begin{enumerate}
    \item[i.] Odd: Changes sign under reflection along $x^2$.
    \item[ii.] Even: Does not change sign under reflection along $x^2$.
\end{enumerate}
A detailed discussion on the numerical method and the subtleties involved in the computation of the QNFs can be found in appendix \ref{appendix:DetailsQNFs}. Henceforth we focus only on the even sector as we have found it to be the only one playing a relevant role in the stability analysis. 

\begin{figure}
\centering
\begin{subfigure}{.485\textwidth}
    \includegraphics[width=\textwidth]{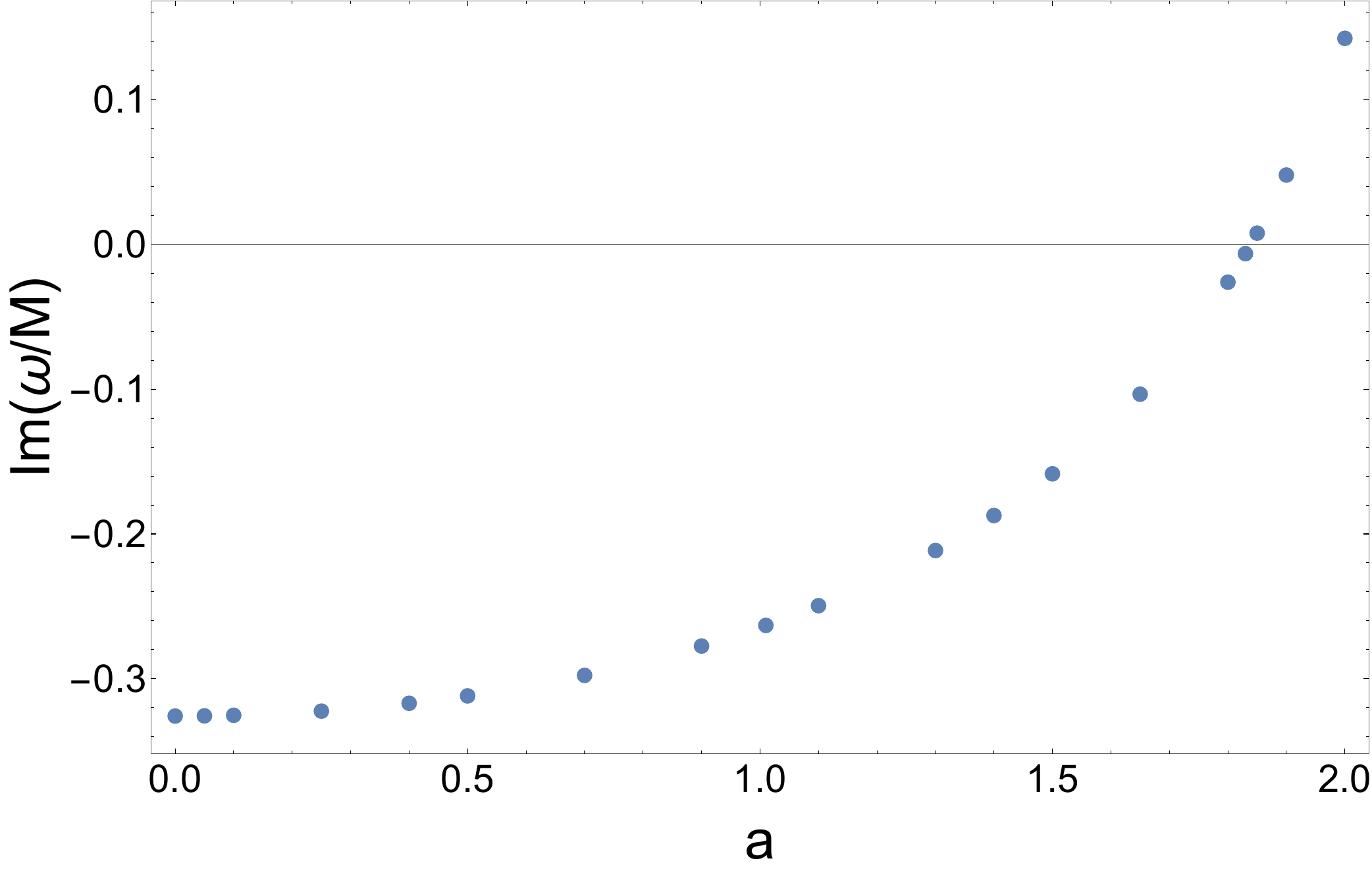}
    \caption{$T/M=0.5$}
    \label{fig:QNMsT0d5}
\end{subfigure}
\hfill
\begin{subfigure}{.495\textwidth}
    \includegraphics[width=\textwidth]{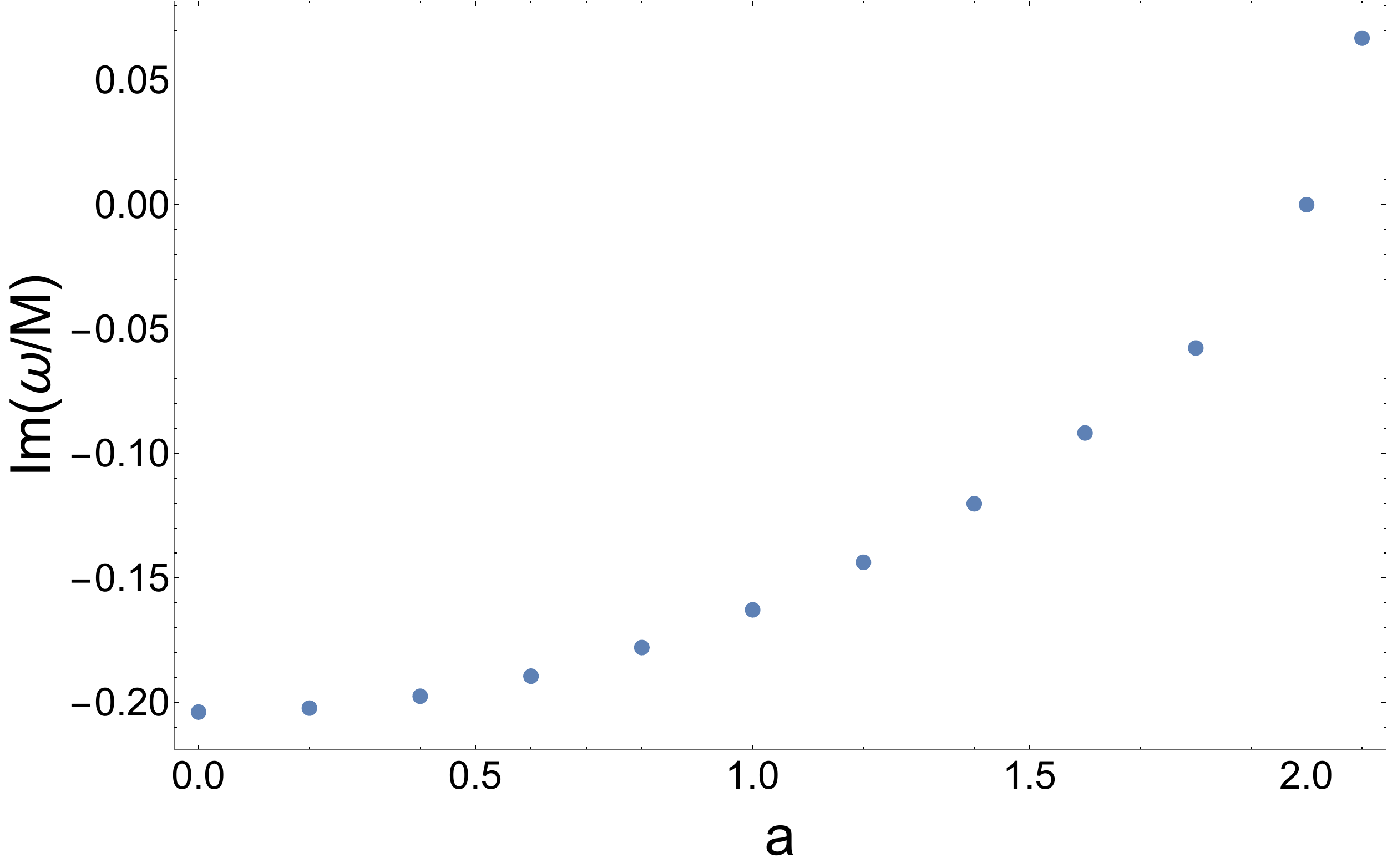}
    \caption{$T/M=1$}
    \label{fig:QNMsT1d0}
\end{subfigure}
\caption{Imaginary part of the lowest lying QNF for the solutions with current at small but nonzero momentum $k/M$. Concretely, in (a) the momentum is $k/M=\frac{2\pi}{3}\cdot 10^{-4}$ and in (b) it is $k/M=\frac{4\pi}{3}\cdot 10^{-4}$. Stable solutions are found for $a\lesssim1.8$ in (a) and $a\lesssim2$ in (b).}
\label{fig:QNMsLattice}
\end{figure}

In figure \ref{fig:QNMsLattice} we present the imaginary part of the lowest lying QNFs in the even sector as a function of $a$ for $T/M=0.5$ and $T/M=1$. Remarkably, we find that these
solutions are always stable for $a<1$. This indicates that the existence of an imaginary current does not generically drive the system unstable. Eventually, for some $a>1$ but smaller than the $a_{max}$ defined in the previous section, instabilities kick in. Remarkably, this indicates that there are stable solutions that locally have $|\eta(x^1)|>1$. This contrasts with the homogeneous case, where solutions were found to be unstable the moment $|\eta|$ crossed 1. 
It would be interesting to study how the existence and stability of solutions with $|\eta(x^1)|>1$ depends on temperature $T/M$ and system size $L\,M$. This would be numerically intensive and we leave it for future work.

\begin{figure}
\centering
\includegraphics[width=\linewidth]{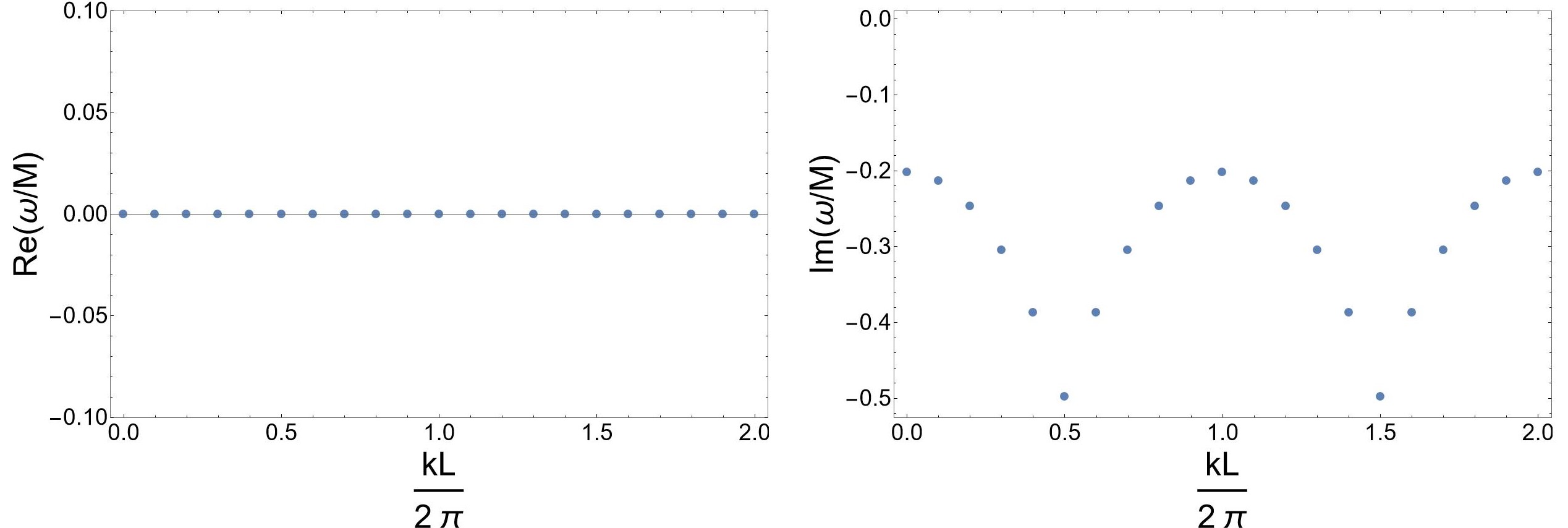}
\caption{Dependence of the lowest lying QNF on the dimensionless momentum $kL$.}
\label{fig:1stQNF_Kdependence}   
\end{figure}

In figure \ref{fig:1stQNF_Kdependence} we plot real and imaginary parts of the lowest lying QNF at $\{a,T/M\}=\{0.2,1\}$ as a function of the dimensionless momentum $kL=3k/M$. We note that the equivalence relation $k\sim k+\frac{2\pi}{L}$, arising from the lattice structure, manifests itself in the periodicity of the QNF. This confirms that we only need to consider momenta in the first Brillouin zone to describe the system. Within this first Brillouin zone we observe that the real part of the QNF is always zero and that the imaginary part decreases to a minimum $\Im(\omega/M)\approx -0.5$ and then rises back again to a maximum $\Im(\omega/M)\approx -0.2$.

\subsection{Null energy condition}\label{subsect:NonHermitianLattice_NEC}

We have found that in the
phase with imaginary current we can have stable solutions with $|\eta|>1$ in some region of space. Now, in this subsection we further explore the nature of such solutions by studying whether they violate the bulk null energy condition (NEC) 
\begin{equation}
    T_{MN}\,u^M u^N\geq0\,,
\end{equation}
where $u$ is any vector tangent to a null geodesic.

Let us commence by performing a near-boundary analysis. Then the fields can be approximated by their asymptotic expansion \eqref{eq:Asymptotic_Expansion}. In particular, in this approximation the metric reads
\begin{equation}
    ds^2= \frac{1}{z^2}\left(dt^2+(dx^1)^2+(dx^2)^2+dz^2\right)+...\,,
\end{equation}
and admits a null radial geodesic with tangent vector $u=\partial_t+\partial_z$. For this particular geodesic we then have 
\begin{equation}
    T_{MN}\,u^M u^N=2s\bar{s}+...= 2(1-\eta^2)M^2+...\,,
\end{equation}
where in the last equality we have plugged in the definition of the sources for the phase with imaginary current \eqref{eq:Def_Sources_PTbroken}. Hence, the above analysis tells us that, at least near the AdS boundary, the NEC is violated in the region where $|\eta|>1$.

To determine whether this violation persists deep in the bulk we need to find an expression for the tangent vector of the radial null geodesics corresponding to our metric ansatz \eqref{eq:Ansatz_Poincare}. However, as the metric components depend on the coordinate $x^1$, we cannot find an analytic expression and we instead need to resort to a numerical approach to compute the geodesics.

Heuristically, what we do is numerically solve the trajectory of a massless particle in our background and then compute the tangent vector $u$. 
We find that it is helpful to use $z$ to parameterize the trajectory and consider the reduced interval $\{\epsilon,1-\epsilon\}$, where $\epsilon$ is a small cutoff.
The geodesic equations in terms of $z$ then read
\begin{equation}\label{eq:Geodesic_equation}
    \ddot{y}^\mu+\tensor{\Gamma}{^\mu_M_N}\dot{y}^M\dot{y}^N-\dot{y}^\mu\tensor{\Gamma}{^z_M_N}\dot{y}^M\dot{y}^N=0\,,
\end{equation}
where $\mu$ denotes a boundary coordinate ($y^\mu=\{t,x^1,x^2\}$), $M$, $N$ represent bulk coordinates ($y^M=\{t,x^1,x^2,z\}$) and the dot denotes derivatives with respect to $z$.

To obtain null radial geodesics such that near the AdS boundary $u\approx\partial_t+\partial_z$, we choose the following initial conditions
\begin{equation}
    y^\mu(\epsilon)=\{0,x^1_0,0\}\,,\qquad \dot{y}^\mu(\epsilon)=\{\dot{t}_0,0,0\}\,,
\end{equation}
where $\dot{t}_0$ is the positive root of the equation
$g_{MN}(x^1_0,\epsilon)\dot{y}^M(\epsilon)\dot{y}^N(\epsilon)=0$. 
It is worth noting that the equations of motion of $x^2$ decouple and thus we get $x^2(z)=0$ for all $z$.

With this, we can solve numerically the geodesic equation \eqref{eq:Geodesic_equation} to obtain the tangent vector $u$ given by 
\begin{equation}
    u=\dot{y}^\mu\partial_\mu+\partial_z\,.
\end{equation}
This then allows us to compute the contraction $T_{MN}u^Mu^N$ and thus to test the NEC for a given radial geodesic. We repeat this process for different values of $x^1_0$ to obtain a complete foliation of the spacetime.

\begin{figure}
\centering
\begin{subfigure}{.49\textwidth}
    \includegraphics[width=\textwidth]{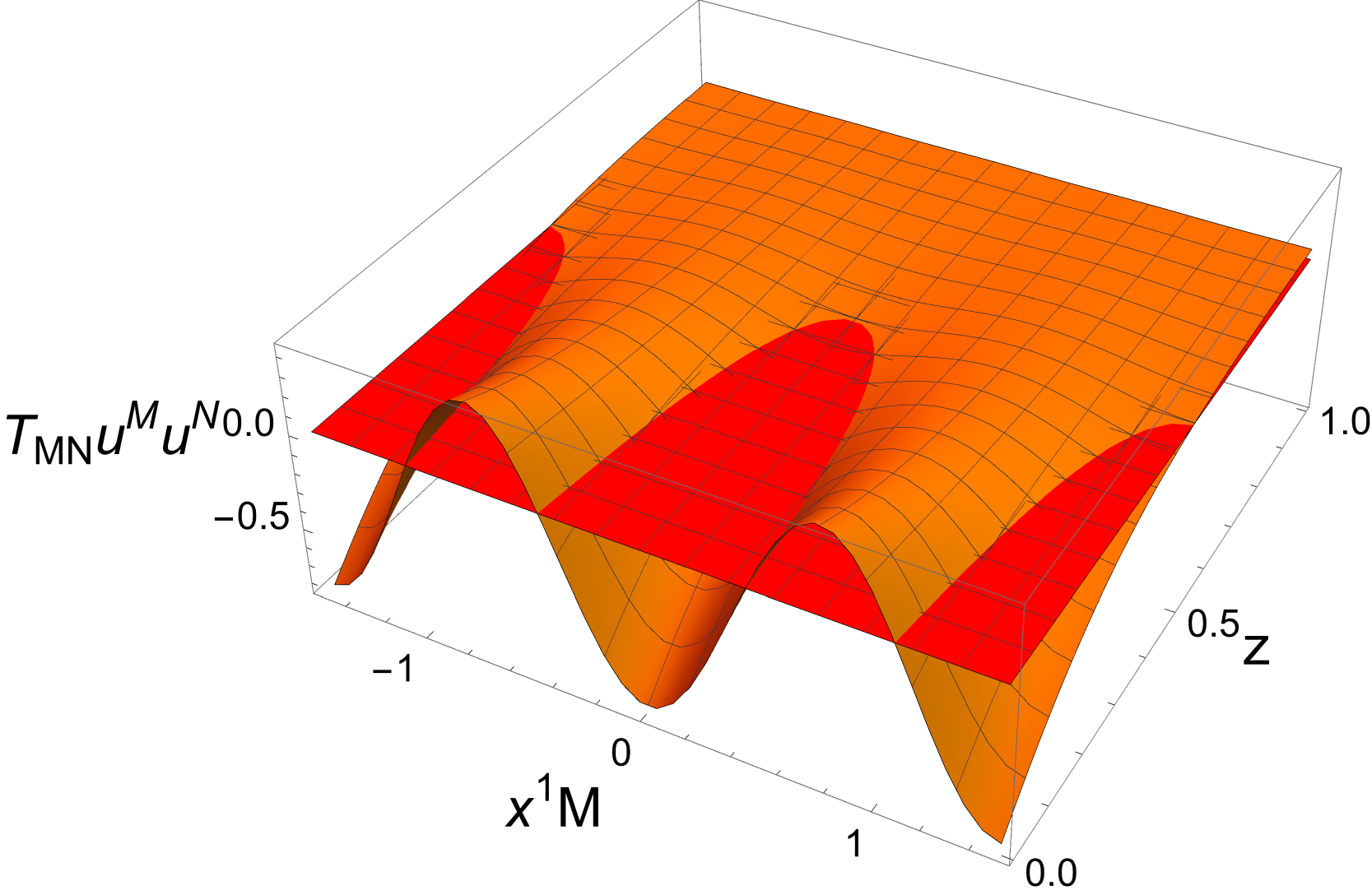}
    \caption{Stable solution with $a=1.7$.}
    \label{fig:NEC1d7}
\end{subfigure}
\hfill
\begin{subfigure}{.49\textwidth}
    \includegraphics[width=\textwidth]{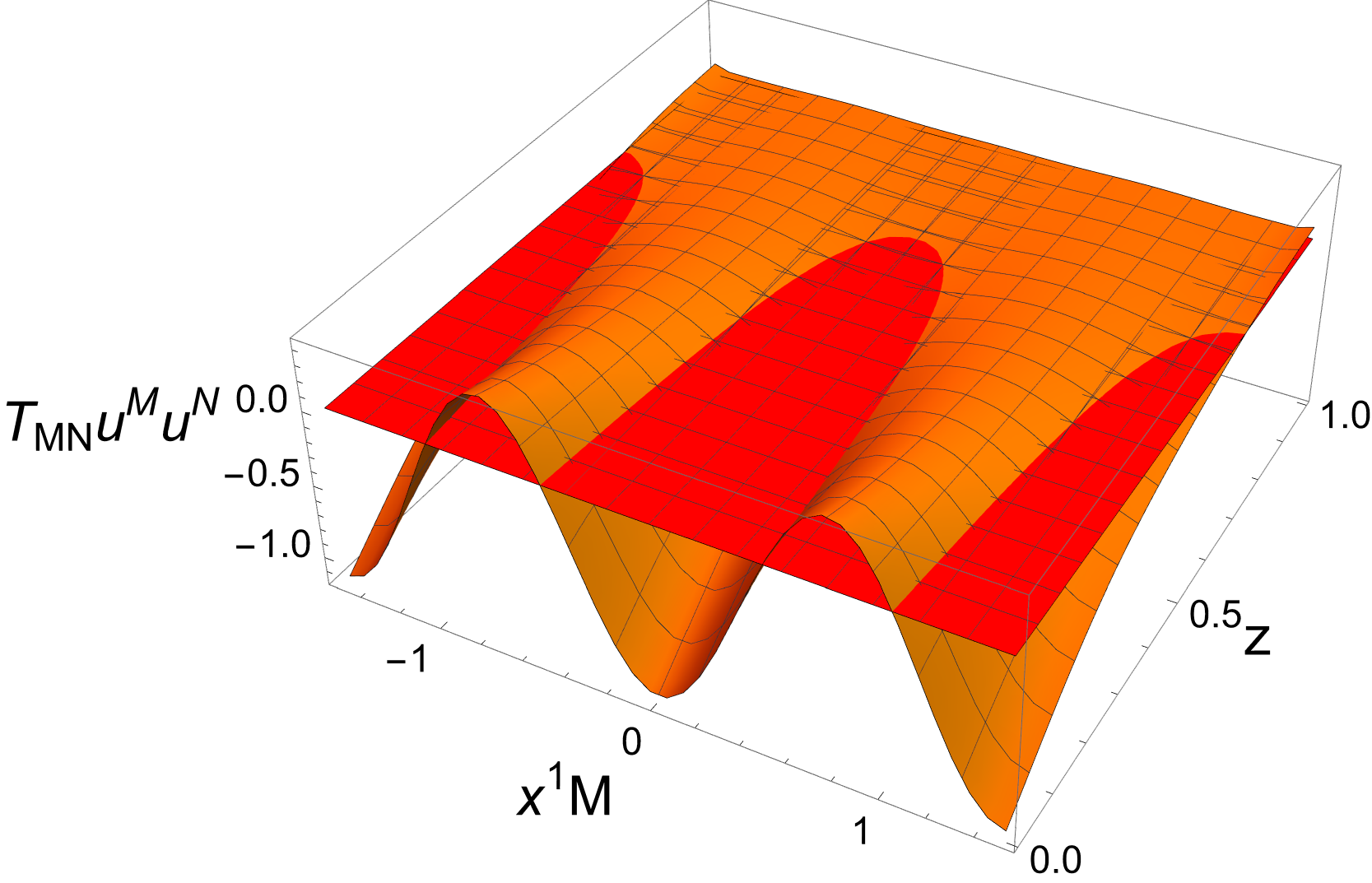}
    \caption{Unstable solution with $a=1.9$.}
    \label{fig:NEC1d9}
\end{subfigure}
\caption{Violation of the NEC for the solutions with current at $T/M=0.5$. The orange surface denotes $T_{MN}u^Mu^N$ and the red one is the surface $T_{MN}u^Mu^N=0$ for reference.}
\label{fig:NEC}
\end{figure}

In figure \ref{fig:NEC} we plot the final result for two 
solutions with imaginary current, a stable one with $\{a,T/M\}=\{1.7,0.5\}$ and an unstable one with $\{a,T/M\}=\{1.9,0.5\}$. We observe that the ``degree of violation'' of the NEC decreases towards the horizon. This is in stark contrast with \cite{Morales-Tejera:2022hyq}, where for time-dependent sources the NEC was violated only deep in the bulk.
As we shall see in the next section, our observed behavior seems to be closely related to the fact that 
the geometry becomes homogeneous deep in the IR.
In \cite{Xian:2023zgu} the authors found that the NEC is generically violated everywhere in the bulk only for homogeneous solutions with $|\eta|>1$; hence, as our homogeneous limit corresponds to $\eta=0$, we expect the NEC to be preserved towards the IR of the theory. Thus to interpolate continuously between the UV violation and the IR preservation, one expects the ``degree of violation'' of the NEC to decrease as we indeed observe. 

Perhaps most remarkably, we find that both stable and unstable solutions have a very similar profile for $T_{MN}u^M u^N$. This seems to suggest that a priori one cannot determine the stability of the solution with knowledge of the violation of the NEC. This contrasts with the homogeneous setup where solutions that violated the NEC were also found to always be unstable \cite{Arean:2019pom,Xian:2023zgu}.

An interesting conclusion of the violation of the NEC is that the standard $a$-function \cite{Freedman:1999gp,Myers:2010xs,Myers:2010tj}, defined to be monotonically decreasing towards the IR, increases near the boundary. To prove this we proceed perturbatively in $z$. 

Using the near-boundary expansion \eqref{eq:Asymptotic_Expansion}, we write the metric as
\begin{equation}\label{eq:Metric_UV_afunction}
    ds^2=\frac{(1-\delta(x^1)z^2)}{z^2}\left[-dt^2+(dx^1)^2+(dx^2)^2\right]+\frac{dz^2}{z^2}+...\,,
\end{equation}
where the ellipsis denotes terms of order $z$ and we have defined $\delta=s\bar{s}/4$. In order to define the $a$-function we introduce the domain wall coordinate 
\begin{equation}
    \rho=-\log(z)+...\,,
\end{equation}
where now the ellipsis denotes terms of order $z^3$.
When written in terms of this radial coordinate $\rho$, the metric \eqref{eq:Metric_UV_afunction} takes the standard domain wall form
\begin{equation}
    ds^2=e^{2A}\left[-dt^2+(dx^1)^2+(dx^2)^2\right]+d\rho^2+...\,, \qquad A=\frac{1}{2}\log\left(\frac{1-\delta \, z^2}{z^2}\right)\,.
\end{equation}
Then, using the definition of the $a$-function of \cite{Freedman:1999gp,Myers:2010xs,Myers:2010tj} we get
\begin{equation}
    a=\frac{\pi^{3/2}}{\Gamma(3/2)l_p^2}(\partial_\rho A)^{-2}=\frac{\pi^{3/2}}{\Gamma(3/2)l_p^2}(1-\delta \, z^2)^2=\frac{\pi^{3/2}}{\Gamma(3/2)l_p^2}\left(1-\frac{s\bar{s}}{4} z^2\right)^2\,,
\end{equation}
where $l_p$ is the Planck length. Remarkably, $a$ increases towards $z>0$ for $s\bar{s}<0$, i.e., for $|\eta|>1$.
Therefore, we find that for lattices with a non-Hermitian function $\eta$ that locally satisfies $|\eta|>1$, near the boundary, the $a$-function increases towards the IR.
Hence, given that the $a$-function counts the degrees of freedom along the renormalization group (RG) flow, it seems that such lattices are problematic as after integrating out the UV, we (locally) gain degrees of freedom. Furthermore, defining the averaged $a$-function $a_{\text{av}}$ as \footnote{We thank K. Landsteiner for suggesting the inclusion of this computation.}
\begin{equation}
    a_{\text{av}}=\frac{\int_{-L/2}^{L/2} dx^1\sqrt{g_{11}}\, a }{\int_{-L/2}^{L/2} dx^1\sqrt{g_{11}}}=1-\frac{z^2}{2L}\int_{-L/2}^{L/2} dx^1 s\bar{s}\,,
\end{equation}
we conclude that lattices with amplitudes greater than 2, globally gain degrees of freedom.\footnote{Although we have not presented stable lattices with amplitudes $a>2$ in this paper, we expect them to exist for high enough temperatures.}
However, one should note that the failure of the above $a$-functions to be monotonically decreasing is not enough to prove the previous assessments. In particular, to ensure that indeed we gain degrees of freedom along the RG flow, we would need to prove the non-existence of a monotonically decreasing function behaving as the central charge for $z\rightarrow0$. 

\begin{figure}
\centering
\begin{subfigure}{.49\textwidth}
    \includegraphics[width=\textwidth]{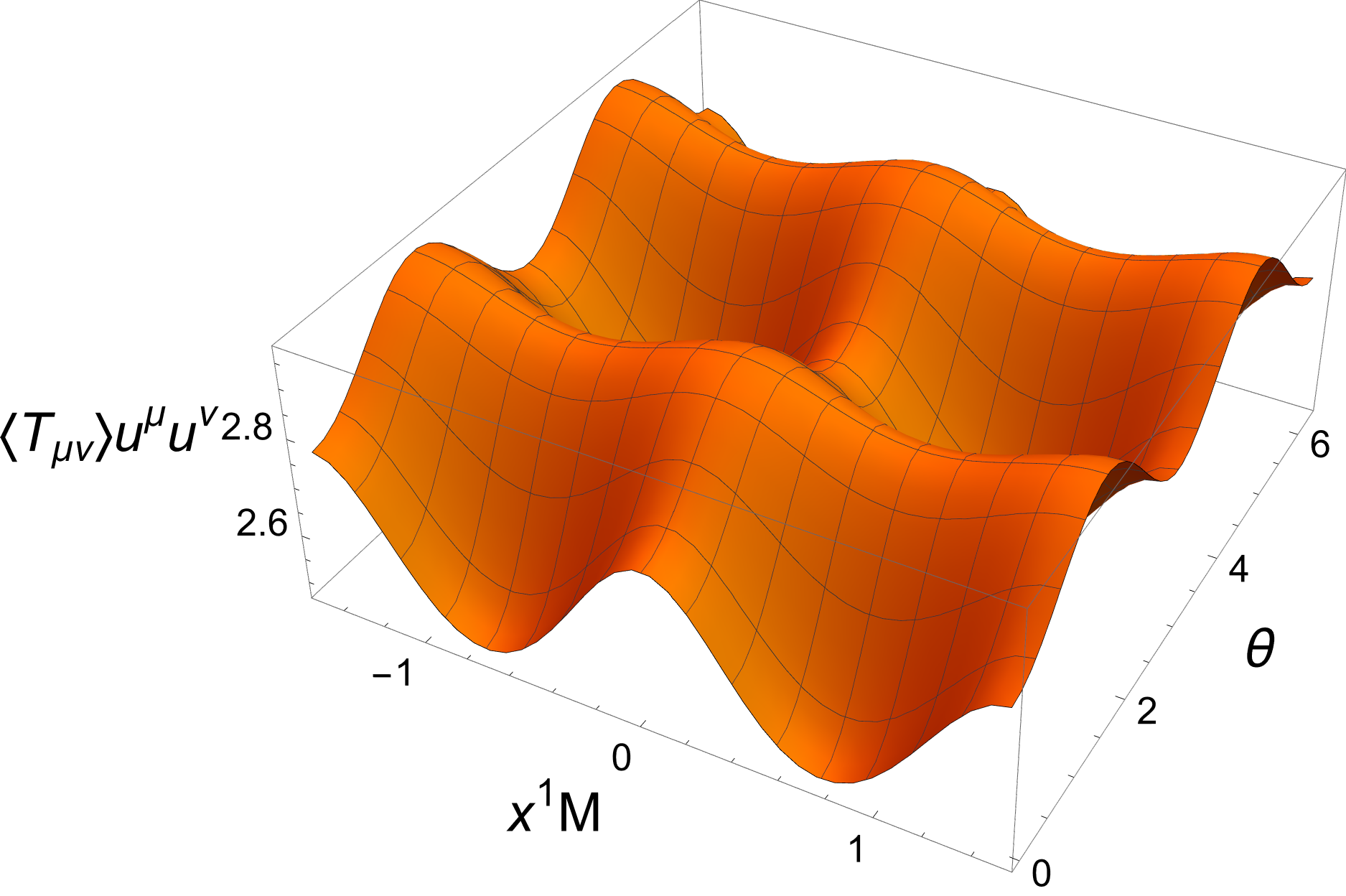}
    \caption{Stable solution with $a=1.7$.}
    \label{fig:bdryNEC1d7}
\end{subfigure}
\hfill
\begin{subfigure}{.49\textwidth}
    \includegraphics[width=\textwidth]{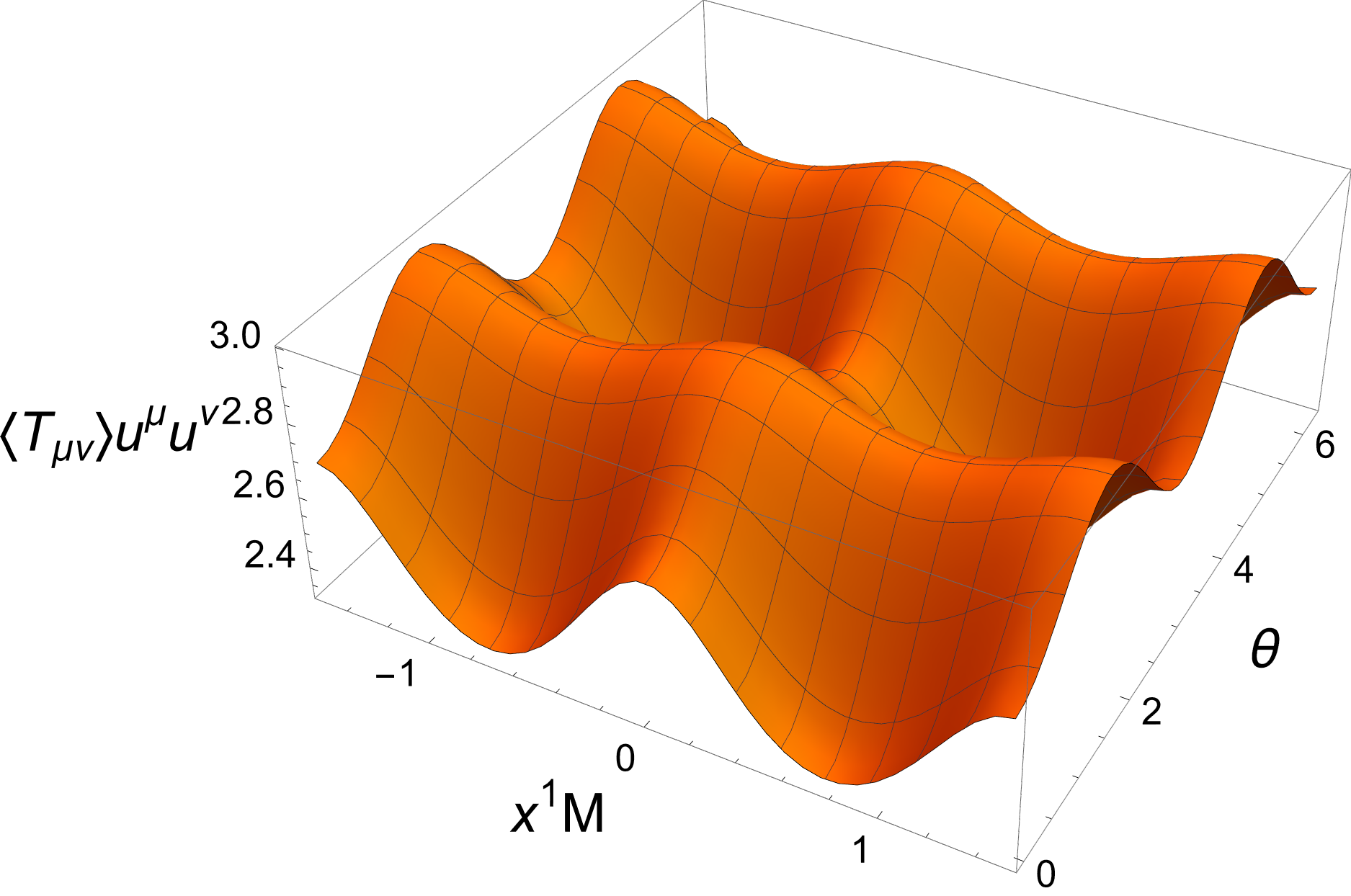}
    \caption{Unstable solution with $a=1.9$.}
    \label{fig:bdryNEC1d9}
\end{subfigure}
\caption{Boundary NEC for the solutions with current at $T/M=0.5$. The orange surface denotes $\expval{T_{\mu\nu}}u^\mu u^\nu$ where here the (boundary) null vector is $u=\partial_t+\cos(\theta)\partial_1+\sin(\theta)\partial_2$. Remarkably for both stable and unstable solutions the boundary NEC is not violated.}
\label{fig:bdryNEC}
\end{figure}

Lastly, we want to stress that despite the UV violation of the NEC, we have managed to find solutions that are seemingly stable at the linear level and which do not 
violate the boundary NEC (see figure \ref{fig:bdryNEC}).
However, one should note that our stability analysis is not fully conclusive. Concretely, it can only find unstable QNMs crossing to the upper half of the complex plane near $\Re\omega=0$. Hence, if there were unstable modes crossing with $|\Re\omega|>>1$, we would be insensitive to them. 
Moreover, due to our ansatz we are also incapable of detecting modes crossing with non-zero momentum in the $x^2$ direction.\footnote{Given the magnitude of the gap observed in figure~\ref{fig:QNMsLattice} 
for values of $a$ sufficiently below $a_\text{max}$ (but above 1) one can
reasonably expect that, if they occurred at all, these instabilities would happen only for sizable values of the momentum in the homogeneous spatial direction.}


\subsection{Low-temperature behavior}\label{subsect:LowTbehaviour}

In this section we look into the low-temperature behavior of the non-Hermitian lattice.
As we describe next we observe that our geometry becomes homogeneous in the $T\to0$ limit. 
This will allow us to construct the dual of the ground state of the non-Hermitian lattice. It is given by an inhomogeneous domain wall between two homogeneous AdS geometries
where the IR of the $T=0$ lattice corresponds to a 
state that can be mapped via
a complexified $U(1)$ transformation~\eqref{eq:Bulk_GT}
to the ground state of the homogeneous Hermitian system.

\begin{figure}
\centering
\begin{subfigure}[]{.49\textwidth}
    \includegraphics[width=\textwidth]{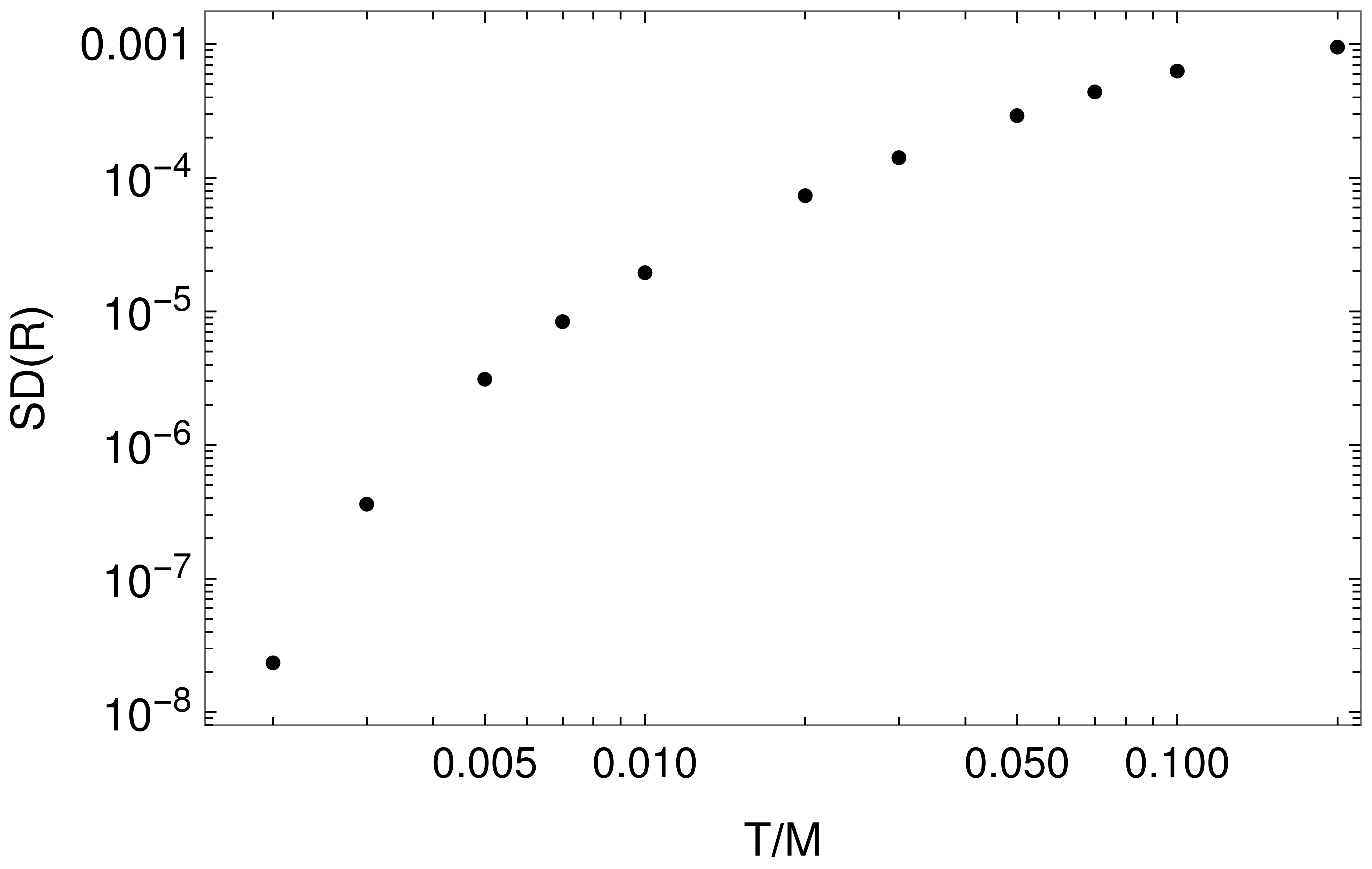}
\end{subfigure}
\hfill
\begin{subfigure}[]{.49\textwidth}
    \includegraphics[width=\textwidth]{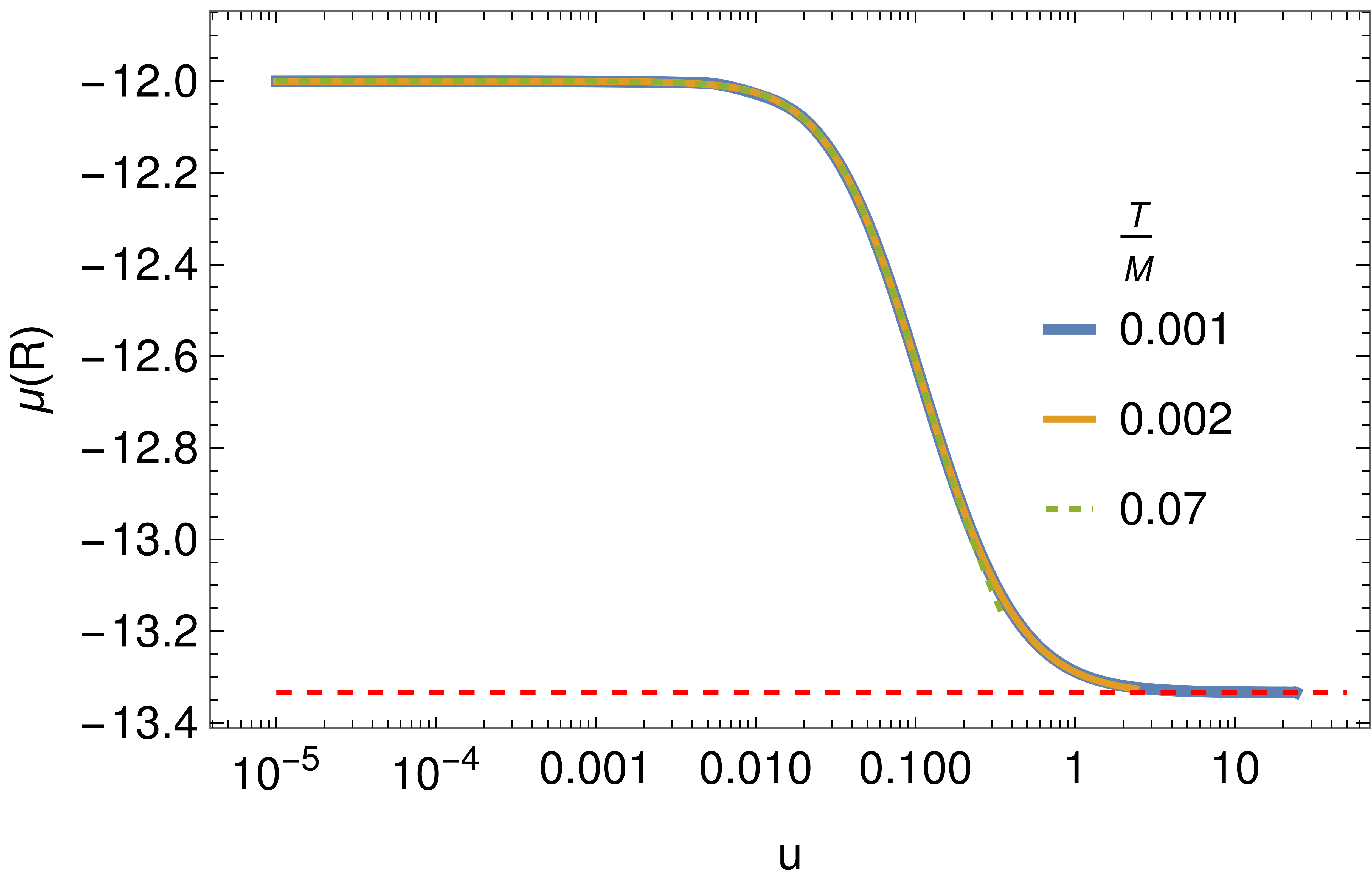}
\end{subfigure}
\caption{Left: standard deviation of the Ricci scalar at the horizon as a function of $T/M$.
Right: spatial average of the Ricci scalar as a function of the radial coordinate $u=zM/M_0$, where $M_0=10$ is the value of the source for the $T=0$ solution of section~\ref{ssubsect:zeroT}.
The red dashed line indicates the value $R=-12-4/v$ corresponding to the IR fixed point~\eqref{eq:HermitianIR}.}
\label{fig:lowTRicci}
\end{figure}

We study the low-temperature limit of the non-Hermitian lattice by computing numerical solutions at fixed $LM=251.3$ and $a=0.1$. We further set $q=2$.\footnote{It was noted in \cite{Gubser:2009cg} that in order to have a conformal IR fixed point
$q>1/(L_\text{IR}\,\phi_\text{IR})$, where $L_\text{IR}$ is the radius of the AdS IR geometry and $\phi_\text{IR}$ the IR value of the scalar field. For our setup this means $q>\sqrt{(1+3v)/6}\approx1.29$ since $v=3$.} We find that these choices are better suited for the numerics of this section.
At the lowest temperatures we employ a Chebyshev grid with 220 points along the $z$ direction and a Fourier grid with 30 points along the $x^1$ periodic direction.
As illustrated by Fig.~\ref{fig:lowTRicci} we find that as $T$ is lowered the near-horizon geometry becomes less inhomogeneous.
In order to further characterize the IR geometry it is useful to recall that the zero temperature solution of the Hermitian theory features an IR fixed point given by
\begin{equation}
h_1=h_2=h_3=h_4={3v\over 1+3v}\,,\quad h_5=0\,,\quad
\phi=\bar\phi=\sqrt{2\over v}\,,\quad
A_1=0\,,
\label{eq:HermitianIR}
\end{equation}
which corresponds to an AdS$_{3+1}$ metric with $R=-12 -4/v$.
In the left panel of Fig.~\ref{fig:lowTRicci} we plot the standard deviation of the Ricci scalar at the horizon as a function of temperature. 
It is clear that the spatial dependence is washed away as temperature is lowered. Below $T/M\approx 0.002$, the standard deviation of the Ricci scalar at the horizon is $O(10^{-8})$ and thus the Ricci scalar at the horizon is homogeneous within our numerical precision.
Moreover, one can actually check that all metric functions tend to the homogeneous solution above. We next plot the spatial average of the Ricci scalar along the bulk in the right panel of Fig.~\ref{fig:lowTRicci}.
One can see that at low enough temperature the near-horizon geometry corresponds to the fixed point~\eqref{eq:HermitianIR}. The dual geometry is then a domain wall interpolating between two homogeneous AdS spaces.

\begin{figure}
\centering
\begin{subfigure}[t]{.48\textwidth}
    \includegraphics[width=\textwidth]{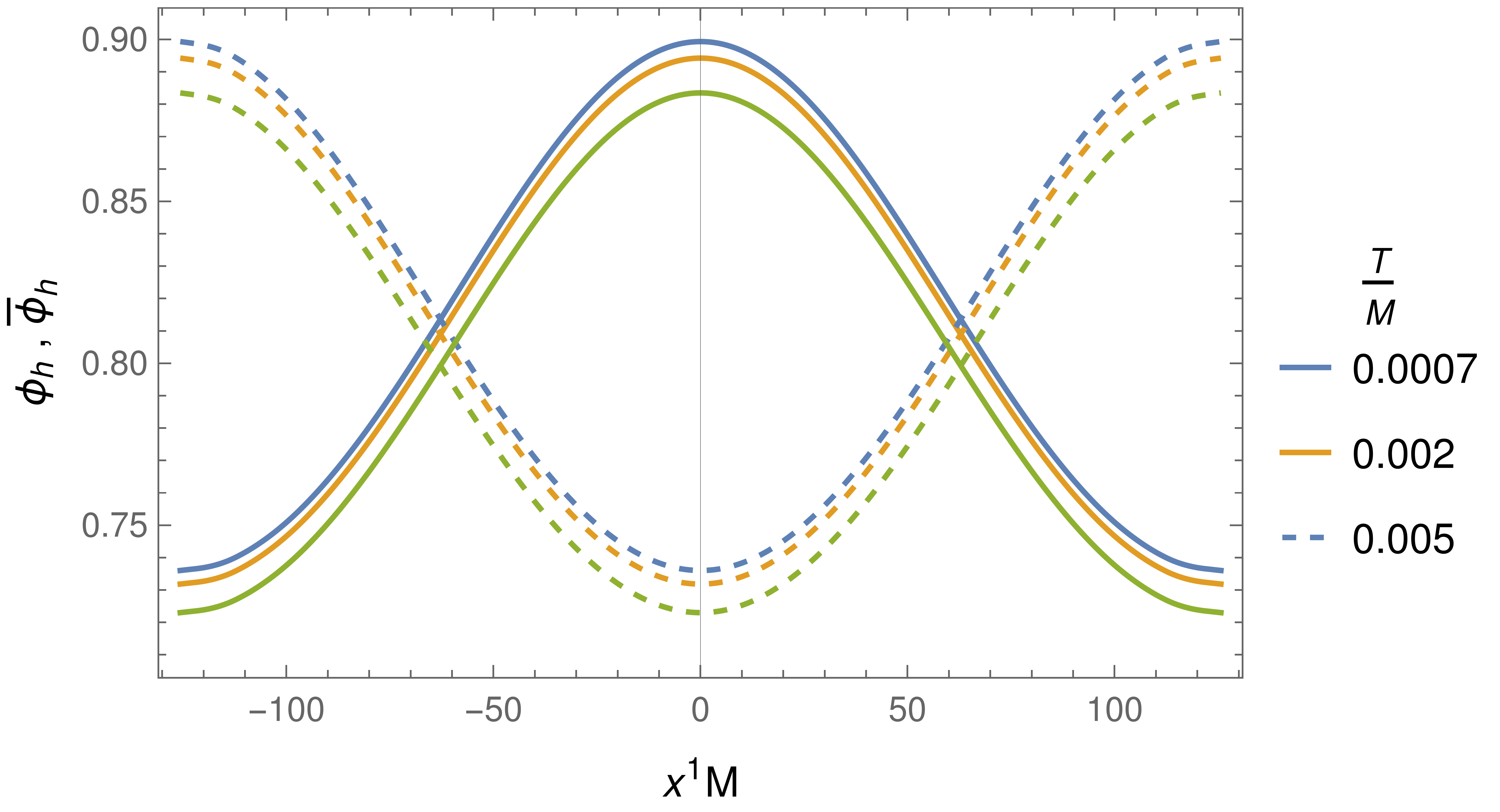}
    \label{fig:}
\end{subfigure}
\hfill
\begin{subfigure}[t]{.50\textwidth}
    \includegraphics[width=\textwidth]{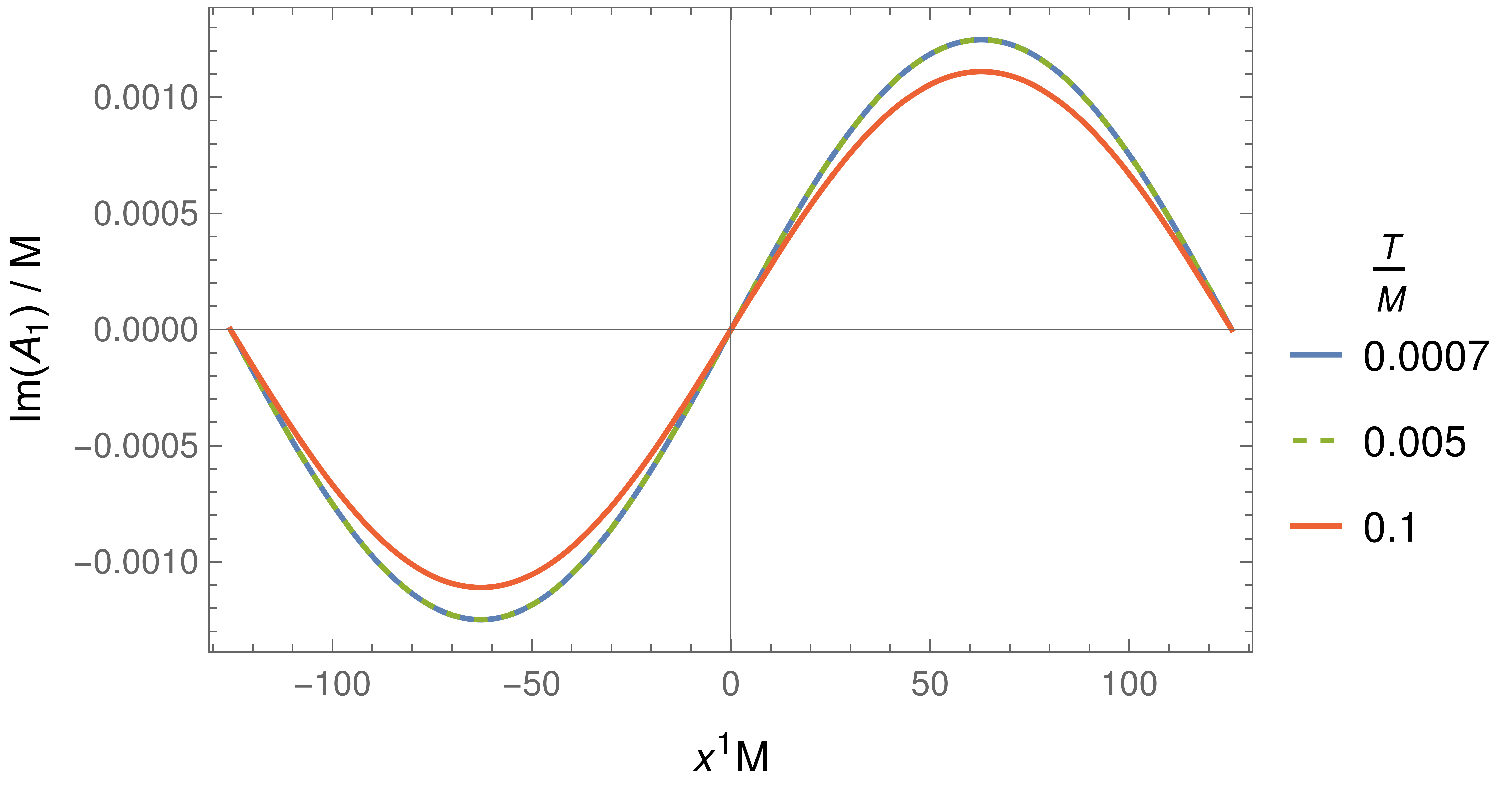}
    \label{fig:}
\end{subfigure}
\caption{Left: $\phi$ (solid lines) and $\bar\phi$ (dashed lines) at the horizon for three different temperatures.
Right: imaginary part of $A_1$ at the horizon. For temperatures below $T/M\approx0.02$ plots of $A_1$ are indistinguishable from each other hence we also included the line for $T/M=0.1$.}
\label{fig:lowTmatter}
\end{figure}

By examining the matter sector of the system one finds that the IR of the low-temperature solution of the non-Hermitian lattice is not exactly~\eqref{eq:HermitianIR}. As shown in Fig.~\ref{fig:lowTmatter}, when $T$ is lowered the matter fields $\phi$,  $\bar\phi$, and $A_1$ tend to inhomogeneous functions.
We will see that in the $T\to0$ limit these are a complexified $U(1)$ rotation~\eqref{eq:Bulk_GT} of the homogeneous solution~\eqref{eq:HermitianIR}. Therefore, our IR is
the low energy counterpart of the UV configuration of
case B in section~\ref{subsect:Gravitational dual}.
The geometry becomes that of AdS$_{3+1}$ and the matter fields asymptote to
a configuration as in~\eqref{eq:Def_Sources_PTunbroken}, namely
\begin{equation}
\phi(1,x^1)=S \sqrt{2\over v}
\,,\quad
\bar\phi(1,x^1)=S^{-1} \sqrt{2\over v}
\,,\quad
A_1(1,x^1)=-\frac{i}{q}S^{-1}\partial_{1}S\,,
\label{eq:lowTmatterIR_S}
\end{equation}
with
\begin{equation}
    S=\sqrt{1-\tilde \eta\over 1+\tilde\eta}\,,
    \label{eq:sdeflowT}
\end{equation}
where $\tilde\eta=\tilde\eta(x^1)$ is a function we can read from our numerical solutions.

\begin{figure}
\centering
\includegraphics[width=0.49\textwidth]{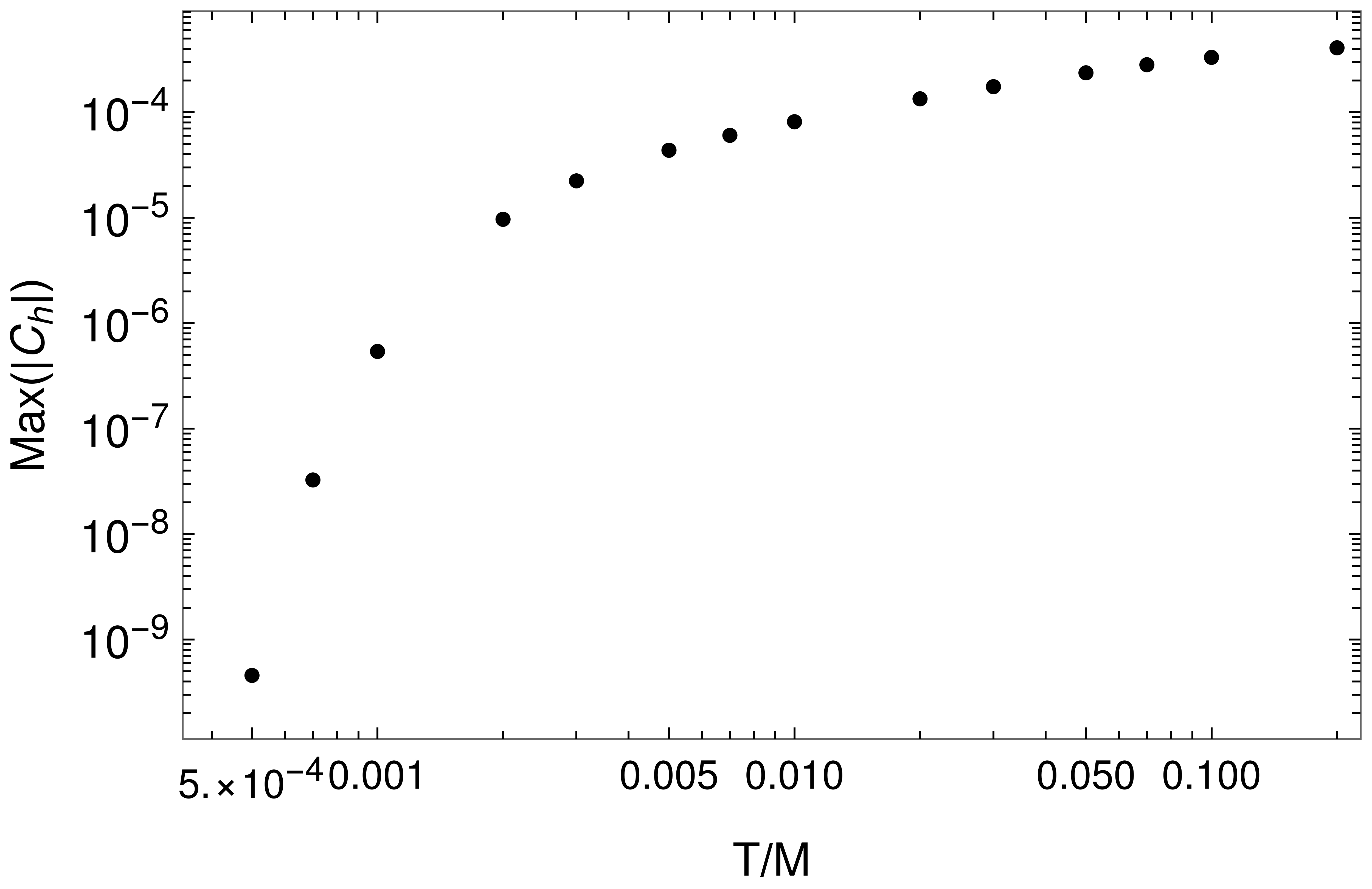}
\caption{Plot of the IR constraint equation~\eqref{eq:IRconst}. We have defined $C= A_x - i\,(\phi\,\partial_{1}\bar\phi
 -\bar\phi\,\partial_{1}\phi)/(2q\phi\bar\phi)$ and here we plot the maximum of its absolute value at the horizon. }
\label{fig:IRconst}
\end{figure}

In order to verify that the low-temperature IR of the system indeed obeys~\eqref{eq:lowTmatterIR_S} one can check that the constraint
\begin{equation}
 A_1\big|_{z=1} = i\,{\phi\,\partial_{1}\bar\phi
 -\bar\phi\,\partial_{1}\phi
 \over2q\,\phi\bar\phi}\bigg|_{z=1}\,,
 \label{eq:IRconst}
\end{equation}
which follows readily from~\eqref{eq:lowTmatterIR_S},\footnote{Notice that $\log(S^2)=\log(\phi/\bar\phi)$, and thus one has
that $A_1=-i/(2q)\,\partial_{1}\log(\phi/\bar\phi)
$.} 
is satisfied in the IR of our solution.
We plot this constraint in Fig~\ref{fig:IRconst} where one can check that indeed it holds in the $T\to0$ limit.
It would then be interesting to determine the function $\tilde\eta$ characterizing the IR of our solutions and compare it to the function $\eta(x^1)$ that sets the UV non-Hermitian deformation. We will do that in the next subsection after constructing solutions at zero temperature.


\subsubsection{The zero temperature solution}
\label{ssubsect:zeroT}
The features of the low-temperature solution hint at the existence of a zero temperature geometry given by a domain wall between two AdS fixed points.
To look for zero temperature non-Hermitian lattices we follow~\cite{Santos:T0} and adopt the
metric ansatz
\begin{align}\label{eq:Ansatz_T0metric}
    ds^2=\frac{\left(z-1\right)^2}{z^2}&\left[-  h_1\,dt^2 + h_3 
        \left(dx^1 + h_5\,dz\right)^2 +h_4 (dx^2)^2  + \frac{h_2}{\left(z-1\right)^4}\, dz^2 \right]\,,
\end{align}
where $h_1$, $h_2$, $h_3$, $h_4$, $h_5$ are functions of $z$ and $x^1$.
As in the finite temperature ansatz we also switch on the matter fields
$\phi$, $\bar\phi$ and $A_1$ as functions of $z$ and $x^1$.

We look for numerical solutions of the equations of motion~\eqref{eq:EoMs_NoDeTurck} with the same UV boundary 
conditions~\eqref{eq:UV_Boundary_Conditions}. 
As in the low-temperature solutions above, we set $LM=251.3$ and the amplitude of the non-Hermitian lattice to $a=0.1$.
Again we apply the DeTurck trick and modify the Einstein equations
as discussed in section \ref{subsect:NonHermitianLattice_NumericalMethod}.
In the IR we require the metric~\eqref{eq:Ansatz_T0metric} to describe a zero temperature horizon and thus fulfill
$h_5(1,x)=0$, while the remaining metric functions are nonvanishing and, moreover,
$h_1(1,x)=h_2(1,x)$~\cite{Santos:T0}.

In our numerical simulations we set an IR cutoff at $z_{IR}=1-10^{-3}$ and check that our solution at that cutoff approaches that of a zero temperature horizon.\footnote{We checked that, as in~\cite{Santos:T0}, towards the IR the metric functions approach constant values but their derivatives are divergent. These divergences are ill-suited for our numerical methods and thus the need for an IR cutoff.} We find that $|h_5(z_{IR},x)|\leq10^{-12}$ while all the remaining metric functions have converged to their values at the fixed point~\eqref{eq:HermitianIR} within 5 significant digits.
Our solutions satisfy the Maxwell constraint equation to $O(10^{-8})$. 
We have run these zero temperature numerical simulations employing
a Chebyshev grid with 80 points for the $z$ direction and 
a Fourier grid with 20 points for the $x^1$ direction.
An extra subtlety is that due to the non-analytic behavior expected towards the IR we employ a fourth order finite-difference discretization along the $z$ direction.

\begin{figure}
\centering
\begin{subfigure}[t]{.49\textwidth}
    \includegraphics[width=\textwidth]{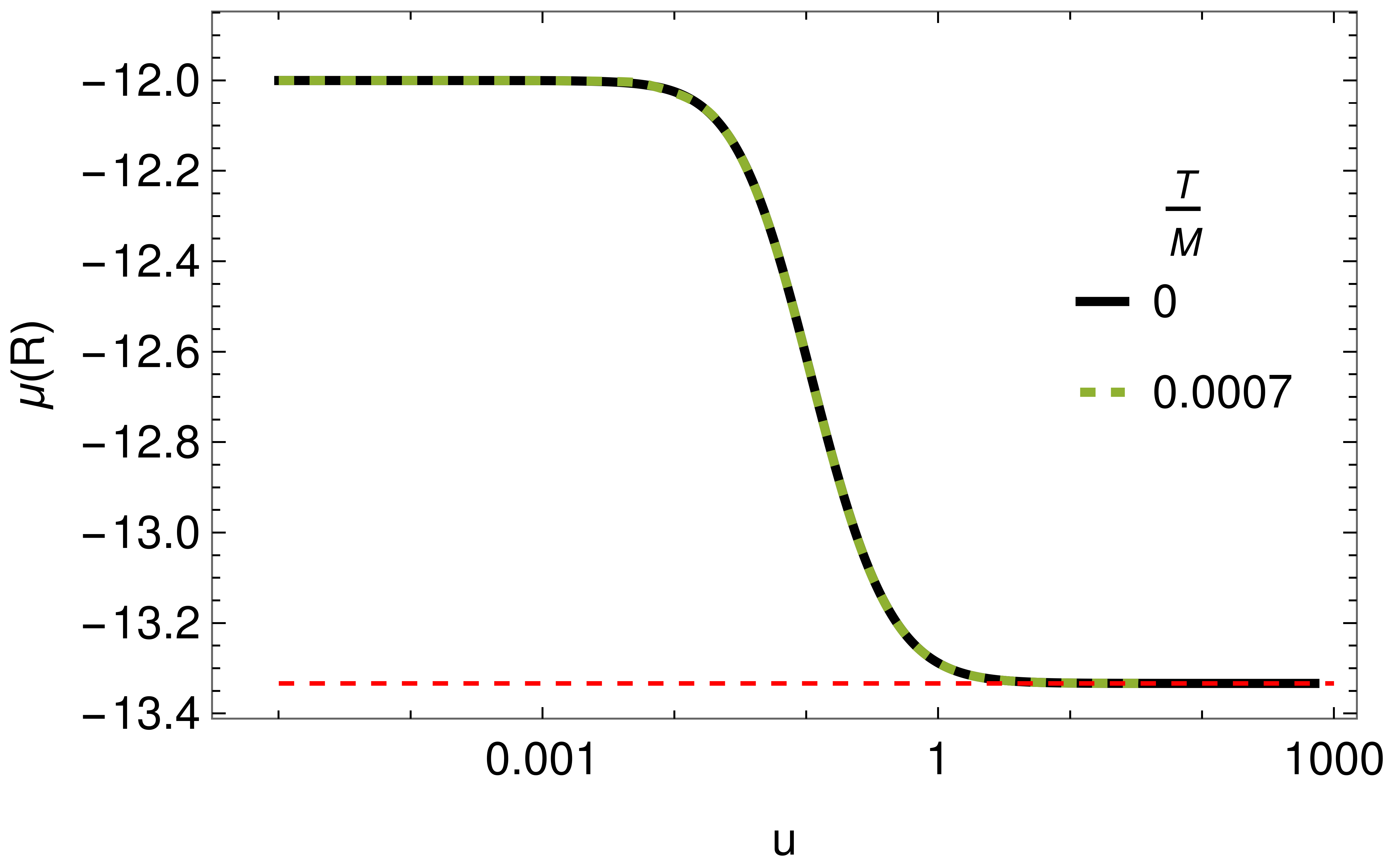}
    \label{fig:}
\end{subfigure}
\hfill
\begin{subfigure}[t]{.49\textwidth}
    \includegraphics[width=\textwidth]{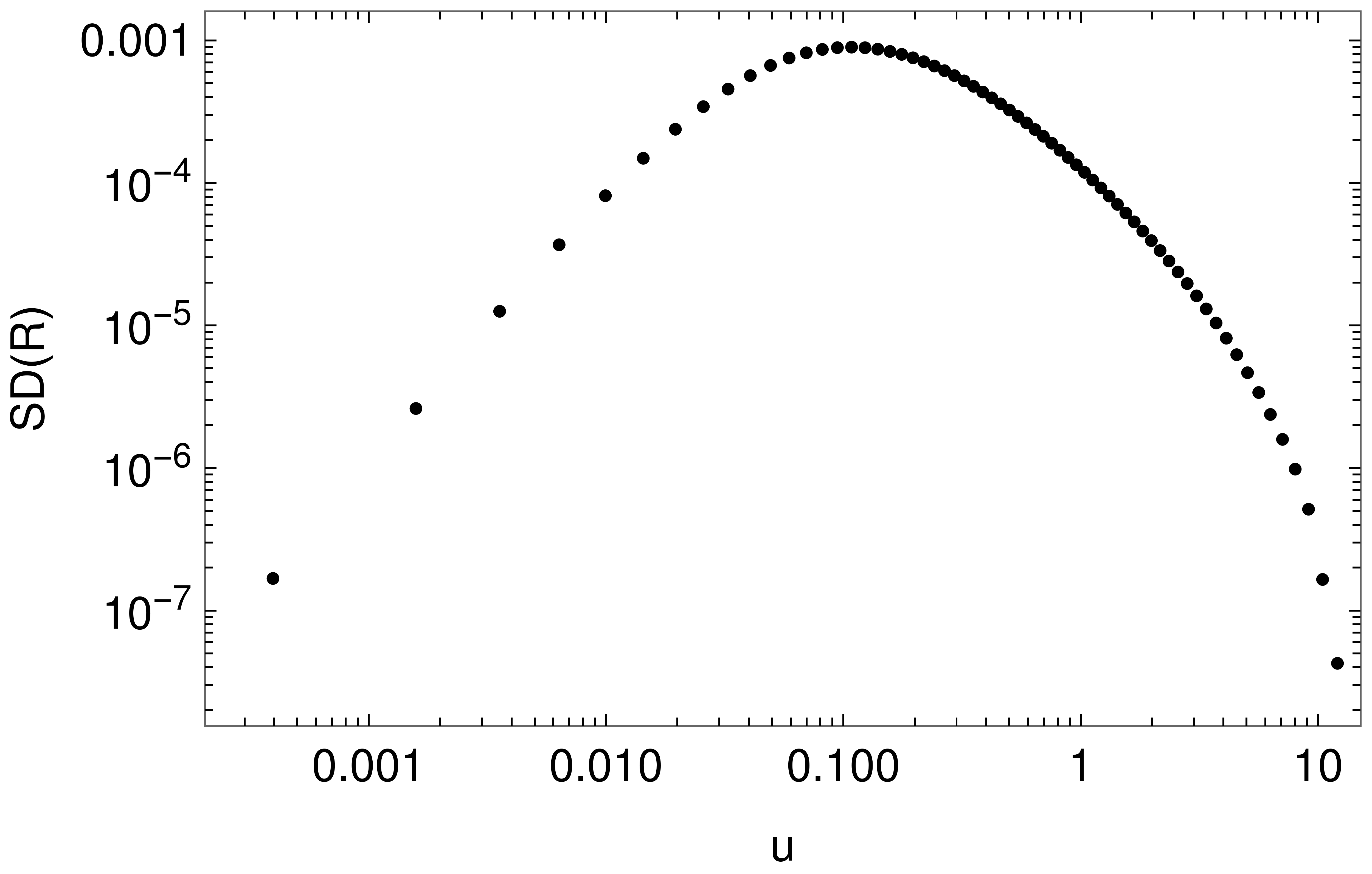}
    \label{fig:}
\end{subfigure}
\caption{Left: spatial average of the Ricci scalar as a function of the radial coordinate for
zero and very low temperature.
The red dashed line indicates the value $R=-12-4v$ corresponding to the IR fixed point~\eqref{eq:HermitianIR}.
Right: standard deviation of the Ricci scalar as a function of the radial coordinate for $T=0$. Beyond $u\approx 10$, SD$(R)$ vanishes within our numerical precision.
In both plots the radial coordinate is defined as $u=zM/M_0$ at finite $T$ and $u=z/(1-z)$ for $T=0$.}
\label{fig:T0}
\end{figure}

In the left panel of Fig.~\ref{fig:T0} we plot the spatial average of the Ricci scalar, $\mu(R)$,  as a function of the radial coordinate for our zero temperature geometry. One can see that the scalar curvature indeed interpolates between the values corresponding to the UV and IR AdS fixed points. We have superimposed the result for $\mu(R)$ at a very low finite temperature to show that indeed the finite temperature geometry converges to the zero temperature domain wall in the $T\to0$ limit.
Our geometry is of course inhomogeneous inside the bulk, but the spatial dependence vanishes in the IR. To illustrate this, in the right panel of Fig.~\ref{fig:T0} we plot the standard deviation of $R$: it peaks at small values of the radial 
coordinate\footnote{This definition of the radial coordinate brings the metric ansatz~\eqref{eq:Ansatz_T0metric} to a form that allows direct comparison with
the finite temperature ansatz, in terms of $u=zM/M_0$, in the low $T/M$ limit reached as $M\to\infty$.}
$u=z/(1-z)$ and then quickly vanishes towards $u\to\infty$.

\begin{figure}
\centering
\begin{subfigure}[t]{.49\textwidth}
    \includegraphics[width=\textwidth]{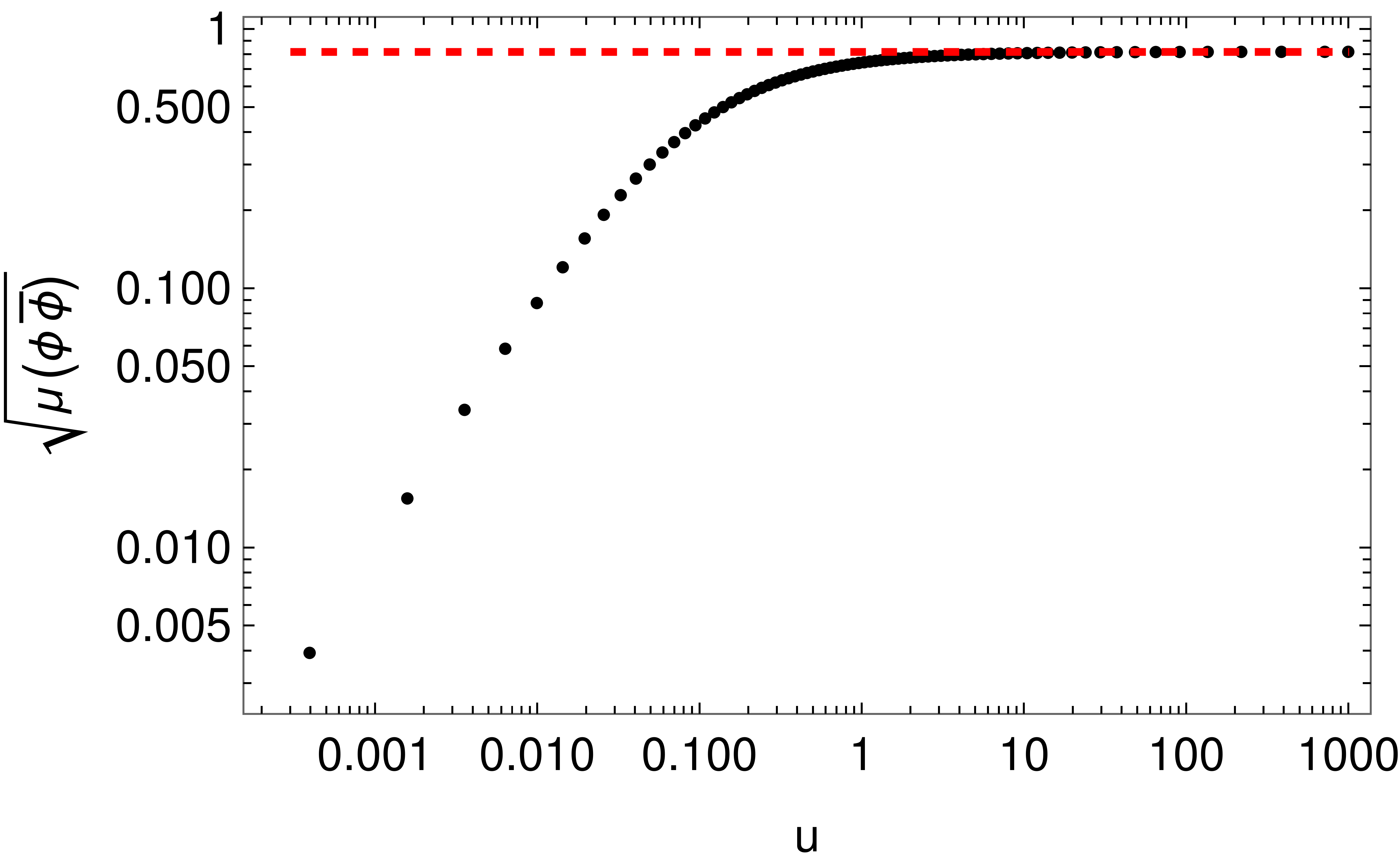}
    \label{fig:}
\end{subfigure}
\hfill
\begin{subfigure}[t]{.49\textwidth}
    \includegraphics[width=\textwidth]{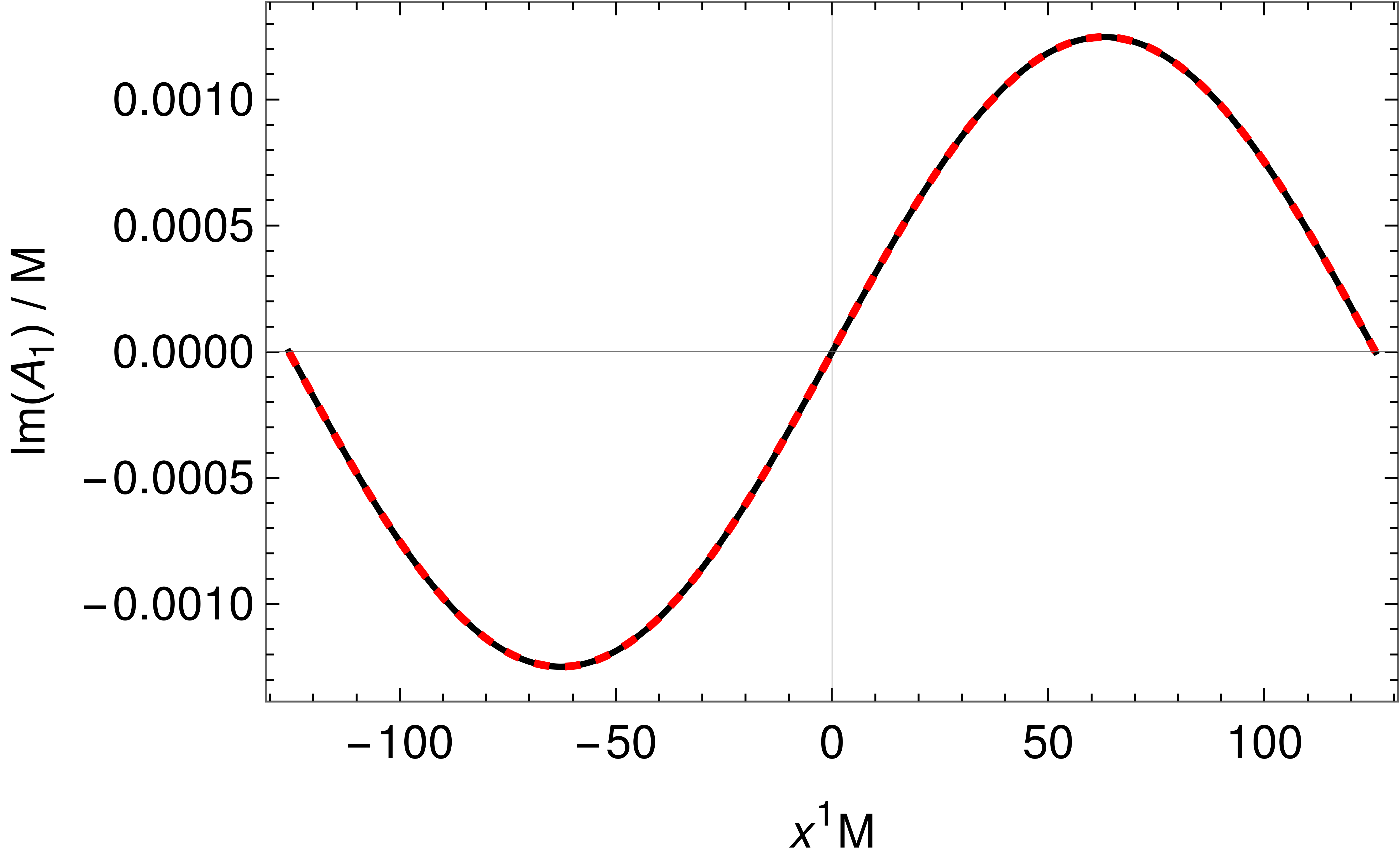}
    \label{fig:}
\end{subfigure}
\caption{Left: square root of the spatial average of $\phi\,\bar\phi$ versus the radial coordinate. The red dashed line indicates the value $\sqrt{2/v}$
corresponding to the IR fixed point~\eqref{eq:HermitianIR}.
Right: imaginary part of $A_1$ at the horizon. The solid black line shows the numerical solution for $A_1$ while the dashed red line results from solving for $A_1$ from the IR constraint~\eqref{eq:IRconst}.}
\label{fig:T0matter}
\end{figure}

\begin{figure}
\centering
\includegraphics[width=0.51\textwidth]{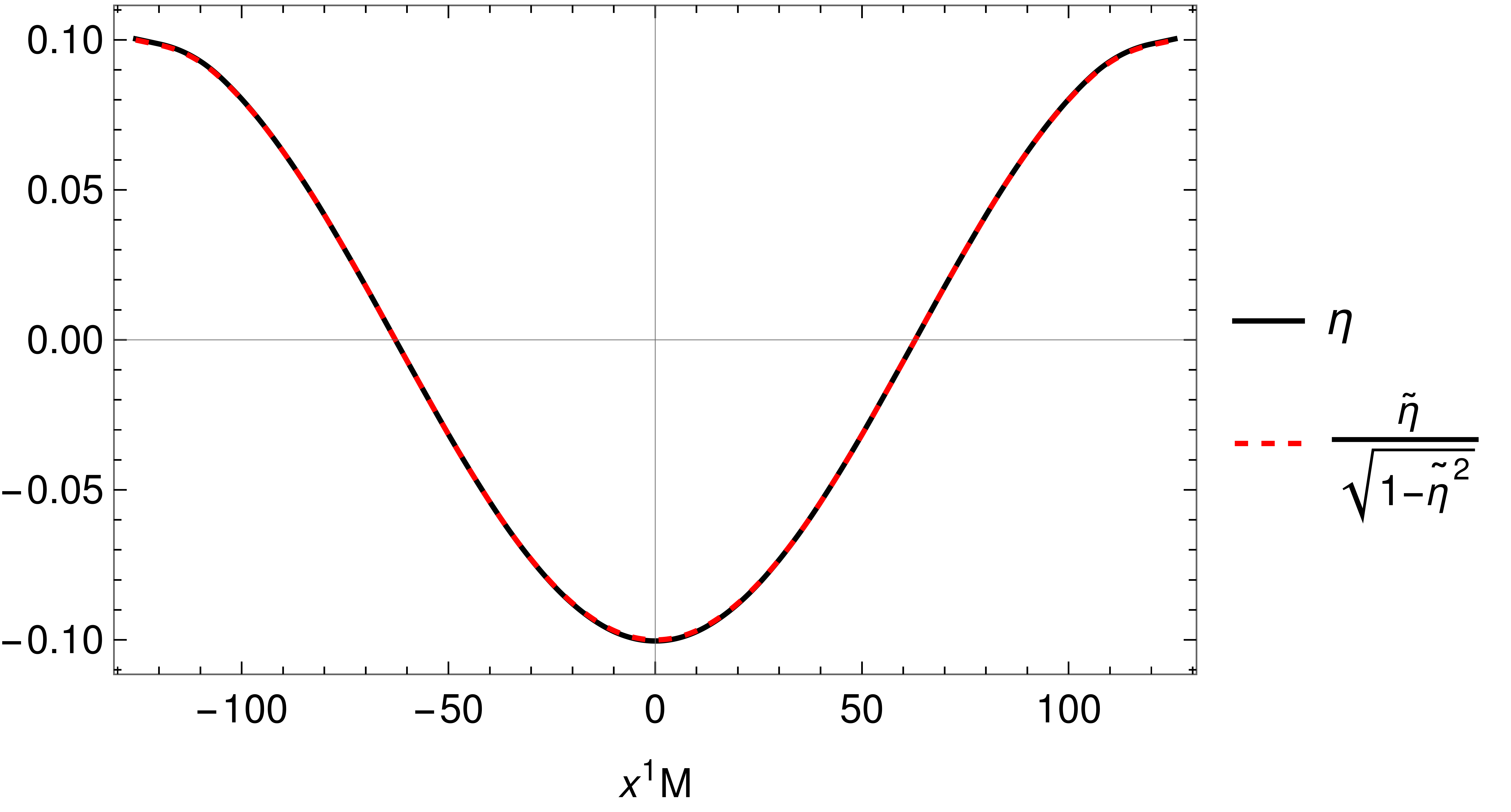}
\caption{$\eta$ (black solid line) and $\tilde\eta/\sqrt{1-\tilde\eta^2}$ (red dashed line).}
\label{fig:tildeEta}
\end{figure}

Next we check that the matter sector indeed realizes the IR fixed point~\eqref{eq:lowTmatterIR_S}. In the left panel of Fig.~\ref{fig:T0matter} we look at the spatial average of the product $\phi\bar\phi$. Notice that if~\eqref{eq:lowTmatterIR_S} holds, $\sqrt{\phi\bar\phi}=\sqrt{2/v}$ in the IR, and, of course, the spatial averaging becomes superfluous there. Our numerics confirm that this is the case in the IR of the geometry. In the right panel of Fig~\ref{fig:T0matter} we show the gauge field $A_1$ at the horizon; it obeys the IR constraint~\eqref{eq:IRconst}. The matter sector is therefore realizing the 
fixed point~\eqref{eq:lowTmatterIR_S}, and we can now determine $\tilde\eta(x^1)$ from the IR of the zero temperature solution. Indeed, from~\eqref{eq:lowTmatterIR_S} it follows that
\begin{equation}
{\tilde \eta\over \sqrt{1-\tilde\eta^2}}=\sqrt{v\over8}\left(\bar\phi-\phi\right)\big|_{z=1}\,.
\end{equation}
In Fig.~\ref{fig:tildeEta} we plot $\tilde \eta/ \sqrt{1-\tilde\eta^2}$
read from our numerical solution and compare it to the function $\eta(x^1)$ that defines the non-Hermitian UV sources. 
To numerical accuracy we find the following relation tying $\tilde\eta(x^1)$ to $\eta(x^1)$:
\begin{equation}
{\tilde \eta\over \sqrt{1-\tilde\eta^2}}=\eta\,.   
\end{equation}

In summary, the zero temperature solutions confirm that the IR limit of our inhomogeneous non-Hermitian holographic theories is a
complexified $U(1)$ transformation of the usual homogeneous conformal fixed point. Trivially this is the same type of IR as for the solutions without current for which the Dyson map is explicitly known. This is not surprising as it has been observed in similar holographic settings that the spatial modulation introduced via inhomogeneous sources can be washed away along the RG flow~\cite{Donos:2012js,Donos:2014uba,Donos:2014yya}. This in turn implies that both models flow to the a homogeneous IR geometry where the non-Hermiticity is encoded in a complex $U(1)$ rotation of the matter sector.

Remarkably despite not knowing the Dyson map taking the solutions with imaginary current to their Hermitian counterparts, we can still construct it for the IR of the theory. This suggests that perhaps one could attempt to formulate the Dyson map by perturbatively correcting the one found for the IR. However due to the non-analytic behavior of the matter and metric fields such computation seems highly non-trivial.


\section{Non-Hermitian junction}\label{sec:NonHermitianJJ}

In section \ref{sec:NonHermitianLattice}, we considered $\eta(x^1)$ to be a smooth periodic function. This was interpreted as constructing theories where the inflow/outflow is not highly localized, but rather smoothly extends throughout each site in a lattice. 

\begin{figure}
    \centering
    \includegraphics[width=0.7\linewidth]{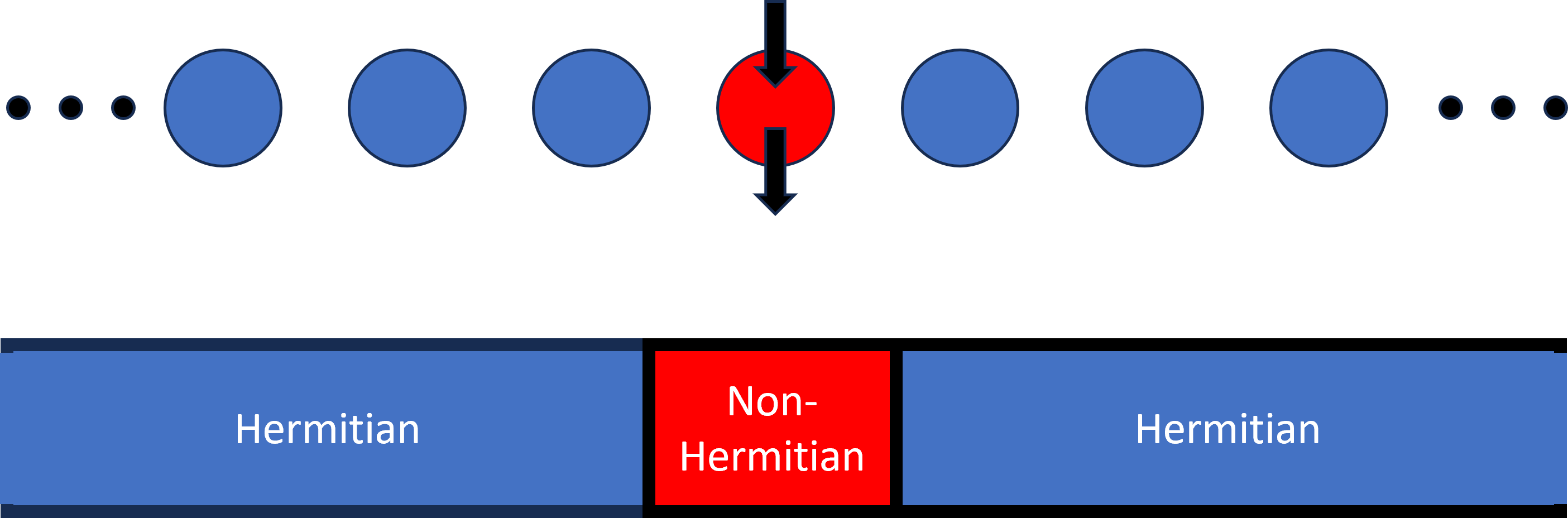}
    \caption{Schematic plot illustrating the construction of the non-Hermitian junction. (\textit{Upper}) Lattice where in one of the sites (red circle) there is flow of matter from/to the exterior while in the remaining sites (blue circles) there is none. (\textit{Lower}) As the lattice structure is broken by the existence of a special site, the lattice picture can be replaced by a junction consisting of a semi-infinite Hermitian system, a non-Hermitian system and another semi-infinite Hermitian system.}
    \label{fig:JJScheme}
\end{figure}

\begin{figure}
    \centering
    \includegraphics[width=0.49\textwidth]{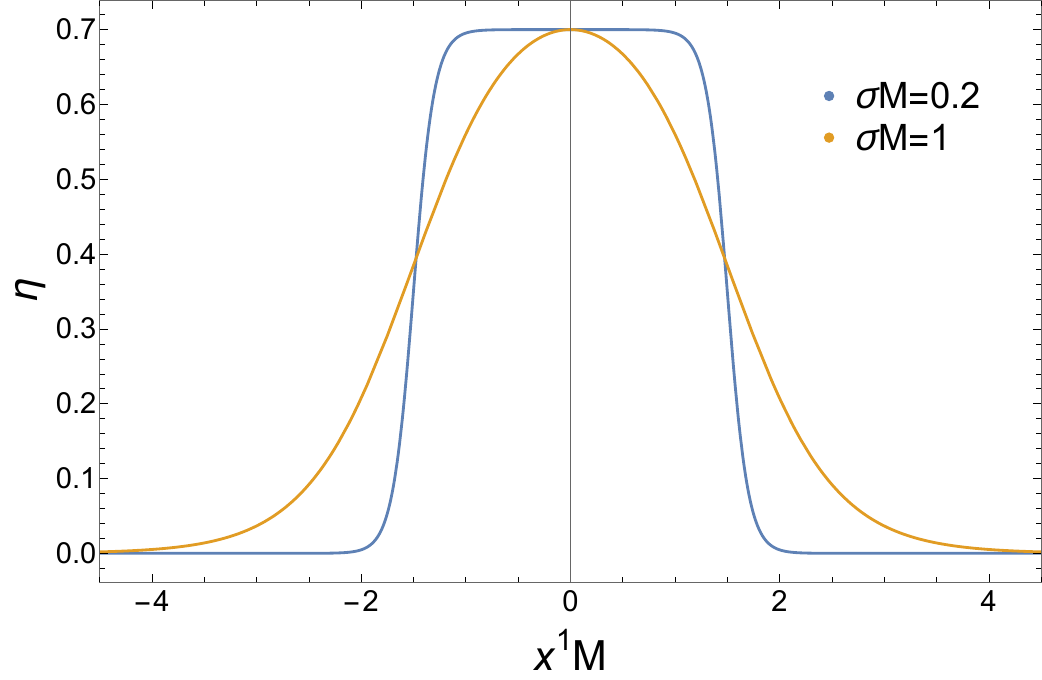}
    \caption{Non-Hermitian function $\eta(x^1)$ for the non-Hermitian junction.}
    \label{fig:etas_Junction}
\end{figure}

It would also be interesting to explore a model where, rather than having inflow/outflow distributed across all sites in the lattice, we have a single cell where non-Hermiticities are present while the rest of the lattice is Hermitian (see figure \ref{fig:JJScheme} for a schematic). In this section we explore such a setup in the 
phase featuring an imaginary current
characterized by the sources \eqref{eq:Def_Sources_PTbroken} with the following expression for $\eta(x^1)$  
\begin{equation}
    \eta(x^1)=\frac{a}{2\tanh(\frac{L}{2\sigma})}\left[\tanh(\frac{x^1+L/2}{\sigma})-\tanh(\frac{x^1-L/2}{\sigma})\right]\,,
\end{equation}
where $L$ is the width of the region where $\eta\not\approx0$, $a$ is the maximum and $\sigma$ controls the steepness. We label this model \textit{non-Hermitian Junction} due to its resemblance to a Josephson junction where instead of having  superconductor/insulator/superconductor we now have Hermitian/non-Hermitian/Hermitian. Note that as $\sigma$ and $L$ are dimensionful quantities, we can exploit the conformal symmetry of the UV to work in terms of their dimensionless equivalents $\sigma M$ and $LM$. 
Thus, solutions are parametrized by $\{a,T/M,LM,\sigma M,q\}$. For conciseness, we fix $q=1$, $a=0.7$, $T/M=1$ and $LM=3$; and we consider two junctions: a smooth one with $\sigma M=1$ and a steep one with $\sigma M=0.2$. In figure \ref{fig:etas_Junction} we plot $\eta(x^1)$ for these two junctions.


Regarding the numerical approach, we proceed exactly as in the previous sections, with only a few minor modifications. Instead of imposing periodic boundary conditions, we now consider $x^1\in[-x^1_{m},x^1_{m}]$, with $x^1_{m}$ such that $\partial_1\eta(\pm x^1_{m})\approx 0$; and impose that the $x^1$-derivatives of all functions vanish on the boundaries $x^1=\pm x^1_{m}$. Additionally, we use 4th order finite difference for the $x^1$-derivatives instead of the Fourier derivative matrices that we have used previously.
More explicitly, in the $x^1$-direction we work with a 4th order finite difference grid with 250 points and in the $z$-direction we employ a Chebyshev-Gauss-Lobatto grid with 20 points.\footnote{With this grid size the study of QNFs becomes unfeasible with our methods.} For the junctions with $\sigma M=0.2$ and $\sigma M=1$, we take $x^1_m=1.5L$ and $x^1_m=4L$, respectively.

\begin{figure}
    \centering
    \includegraphics[width=\textwidth]{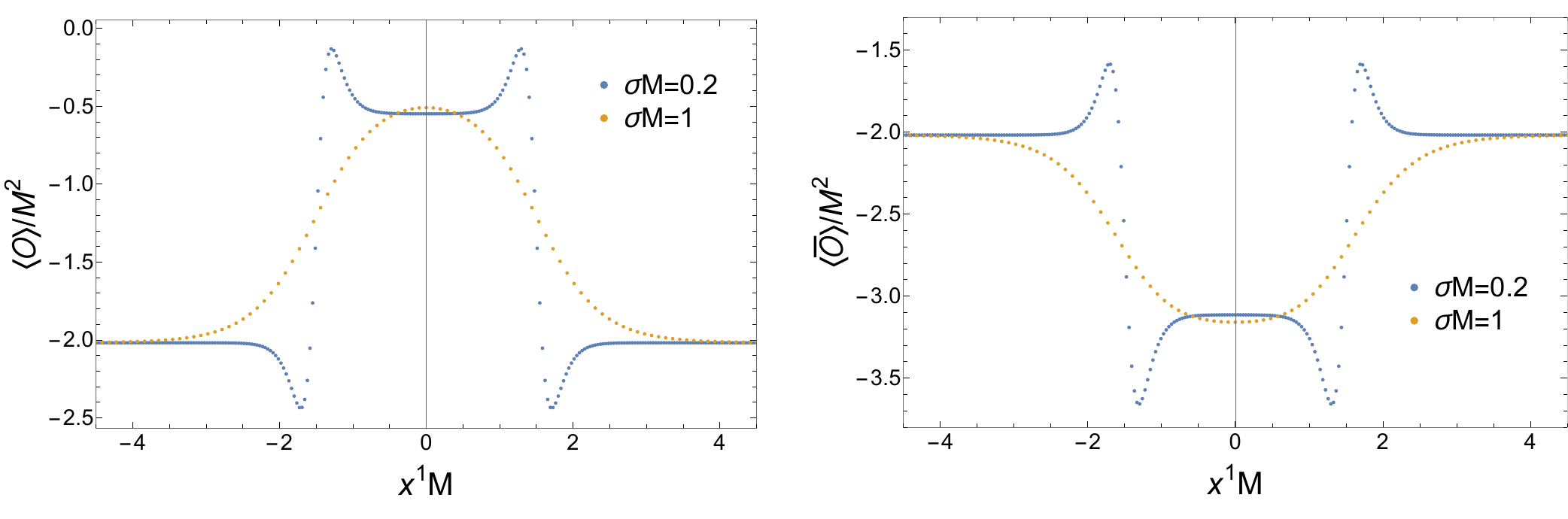}
    \caption{Expectation values of the condensates $\expval{\mathcal{O}}$ and $\expval{\bar{\mathcal{O}}}$ in the non-Hermitian junction.}
    \label{fig:Os_JJ}
\end{figure}

\begin{figure}
\centering
\begin{subfigure}{.487\textwidth}
    \includegraphics[width=\textwidth]{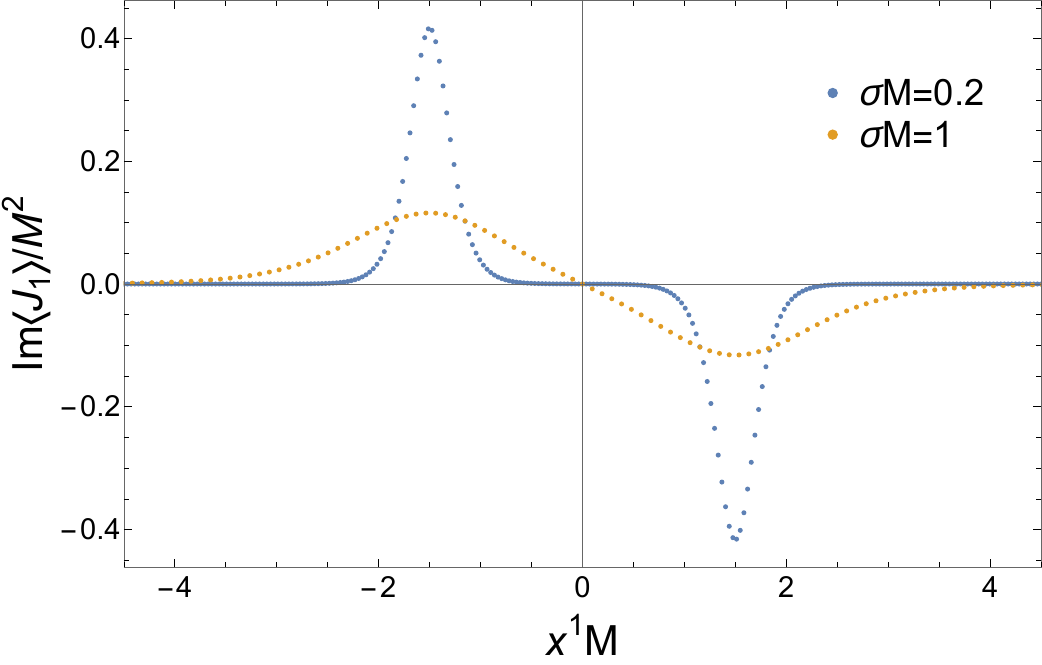}
\end{subfigure}
\hfill
\begin{subfigure}{.492\textwidth}
    \includegraphics[width=\textwidth]{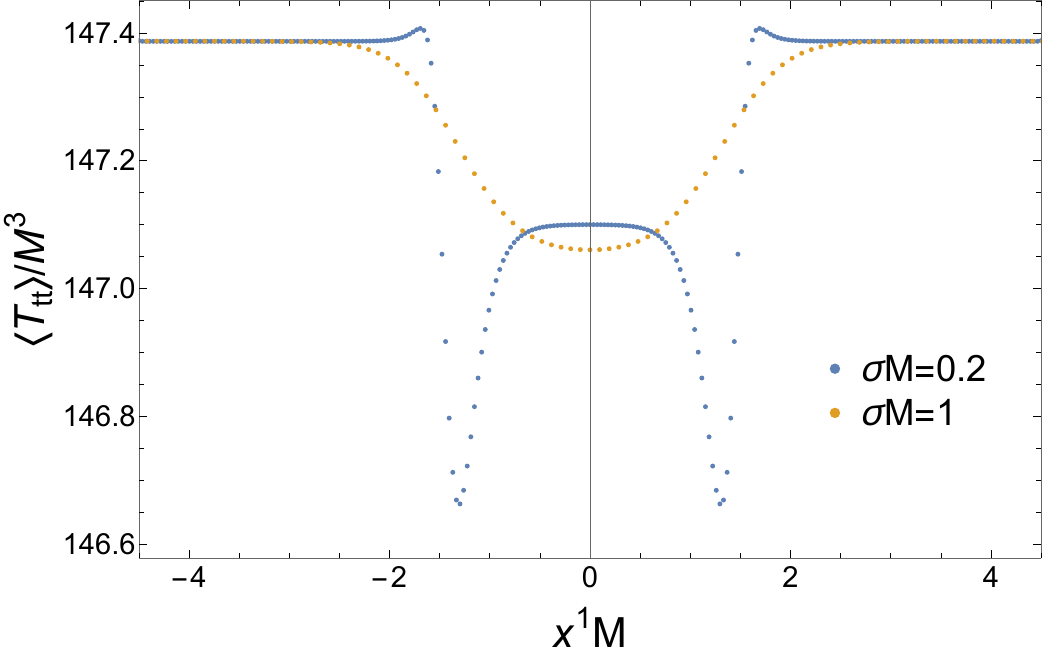}
\end{subfigure}
\caption{Expectation values of the current $\expval{J_1}$ (left) and the energy density $\expval{T_{tt}}$ (right) for the non-Hermitian junction. We note that $\expval{J_1}$ is purely imaginary.}
\label{fig:TttandJ_JJ}
\end{figure}

\begin{figure}
\centering
\begin{subfigure}{.49\textwidth}
    \includegraphics[width=\textwidth]{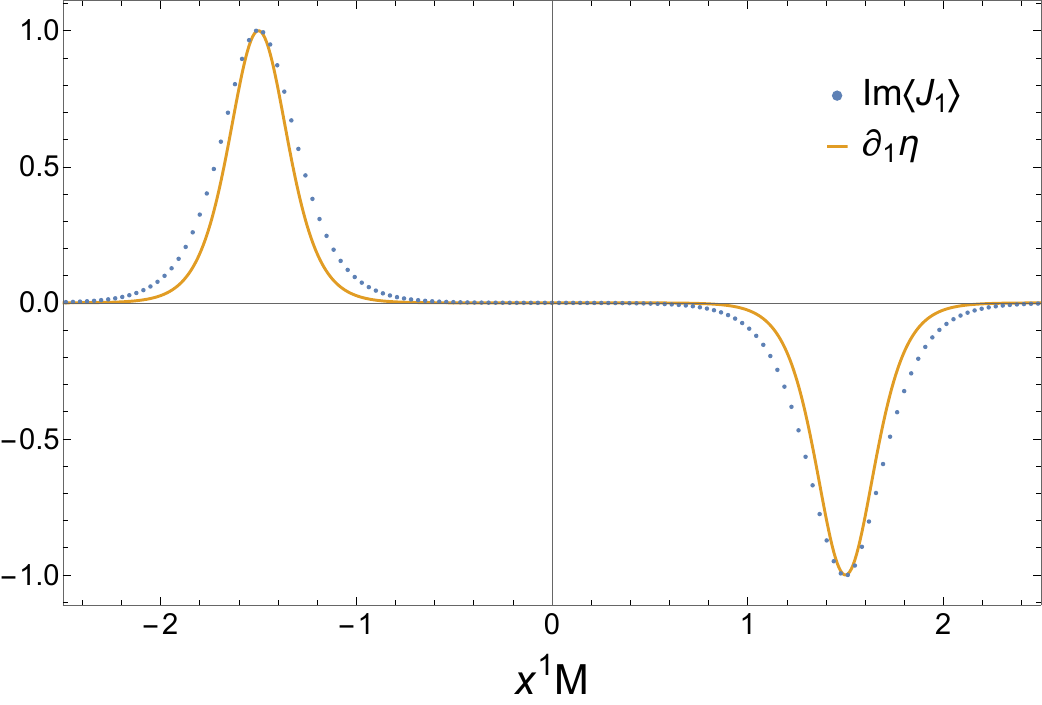}
    \caption{$\sigma M=0.2$}
    \label{fig:Currentvsetaprime_JJ_s0d2}
\end{subfigure}
\hfill
\begin{subfigure}{.49\textwidth}
    \includegraphics[width=\textwidth]{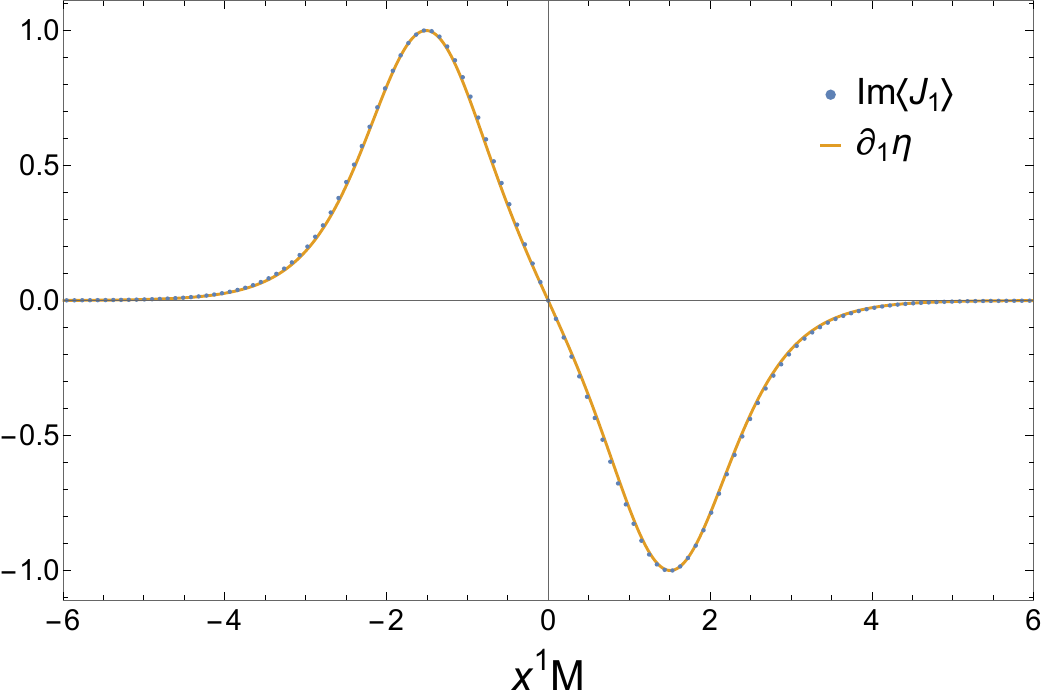}
    \caption{$\sigma M=1$}
    \label{fig:Currentvsetaprime_JJ_s1d0}
\end{subfigure}
\caption{Comparison between the current $\expval{J_1}$ and $\partial_1 \eta$. Both functions have been normalized so that their maximum is 1. Remarkably, the current appears only in the regions where $\partial_1 \eta\neq0$ and has the same sign as $\partial_1 \eta$.}
\label{fig:Currentvsetaprime_JJ}
\end{figure}

In figures \ref{fig:Os_JJ} and \ref{fig:TttandJ_JJ} we plot expectation values of the condensates $\expval{\mathcal{O}}$, $\expval{\mathcal{\bar O}}$, the current $\expval{J_1}$ and the energy density $\expval{T_{tt}}$. As expected, the current (which we have argued was a feature of the inhomogeneous setup) vanishes in the regions where $\eta\approx \text{const.}$ Moreover, we also find that the sign of the current matches the sign of $\partial_1\eta$ (see figure \ref{fig:Currentvsetaprime_JJ}). Hence, for $\mathcal{PT}$-invariant sources we expect the current to always be an odd function of $x^1$ so that $\mathcal{PT}$ is always preserved. Together, these observations seem to indicate that, in general, in a non-Hermitian junction the current appears localized at the interfaces and that its integral vanishes. In particular, in the limit $\sigma\rightarrow0$ we expect that the current becomes a pair of deltas of opposite sign localized at $x^1=\pm L/2$.

\section{Conclusions}\label{sec:Conclusions}

In this paper we have constructed $\mathcal{PT}$-symmetric inhomogeneous non-Hermitian models in holography. We have generalized the setup \cite{Arean:2019pom} to describe inhomogeneous systems and we have used it to study lattices where in each site there is some non-Hermiticity, as well as Hermitian/non-Hermitian/Hermitian junctions. We labelled the former setup \textit{non-Hermitian lattice} and the latter \textit{non-Hermitian junction}. 

Following the construction of \cite{Arean:2019pom}, we introduced the non-Hermiticity trough the sources $s$ and $\bar{s}$ of an operator charged under a $U(1)$ symmetry. We parameterized this non-Hermiticity using $\eta(x^1)$ which, for our models, was given by the ratio
\begin{equation}
    \eta=\frac{s-\bar{s}}{s+\bar{s}}\,.
\end{equation}
We also allowed for solutions with a complex external gauge field $a_\mu$. We distinguished between two cases, $a_1=iq^{-1}\partial_1\eta/(1-\eta^2)$ and $a_1=0$. For the former, we found that the Dyson map \eqref{eq:Dyson_Map} could map the model to a Hermitian one, and we explicitly checked that indeed the dynamics allowed for an equivalent Hermitian description. 
This equivalence is physically non-trivial as the systems described by the Hermitian and non-Hermitian models are fundamentally very different: one has a complex external gauge field and inflow/outflow of matter while the other does not \cite{Morales-Tejera:2022hyq}. 
This suggests that one should think of the Dyson map as a duality between different theories. Hence although the dynamics are mathematically equivalent the physical interpretation is fundamentally different.
An alternative interpretation where the Dyson map is considered a gauge symmetry can be found in \cite{Chernodub:2021waz,Chernodub:2024lkr}.

For models with $a_1=0$, 
we found that no Dyson map of the form \eqref{eq:Dyson_Map} could bring them into a Hermitian description. However, these models preserve $\mathcal{PT}$ despite presenting an emerging complex current localized in the region where $\partial_1\eta\neq0$. This indicates that they allow for unitary evolution and are dual to some unknown Hermitian theory accessible via some Dyson map, different from a complexified external $U(1)$ transformation, which we have not managed to identify.  
It could be interesting
to study the full spectrum of excitations of this phase with imaginary current. This could 
help one find the equivalent Hermitian theory.

It was observed in~\cite{Arean:2019pom,Xian:2023zgu} that in the case of constant $\eta$, the dual geometry becomes unstable for $|\eta|>1$. Nonetheless, here we have found stable
solutions with regions where $|\eta(x^1)|>1$.
However, near the AdS boundary, these solutions violate the null energy condition (NEC) and have an $a$-function that locally increases towards the IR. 
This seems to suggest that these solutions are problematic. In particular, as the $a$-function counts the number of degrees of freedom along the renormalization group flow, these solutions seemingly gain degrees of freedom after integrating out the UV. One should note however that this need not be the case. To prove that the number of degrees of freedom increases it is not sufficient to study the standard $a$-function, we would have to show that no monotonically decreasing function behaving as the central charge exists. 

We shall stress that despite the apparent issues, the 
solutions with $|\eta(x^1)|>1$ do not violate the boundary NEC. Moreover, our analysis of the quasinormal frequencies (QNFs) indicates that for $|\eta(x^1)|$ sufficiently small but larger than $1$, these solutions are linearly stable. However, this analysis is not fully conclusive as it can only find unstable QNFs
crossing to the upper half of the complex plane near $\Re \omega=0$. Hence, there could be unstable modes crossing
onto the upper half plane at large values of $|\Re\omega|$, which we would fail to observe. Furthermore, in our analysis we neglected the possibility of having momentum in the direction perpendicular to the inhomogeneities. Thus, there could also be instabilities arising from modes with non-zero momentum in that direction.

To further characterize the
solutions with $a_1=0$, we have also studied their low-temperature behavior. This analysis has been restricted to $|\eta(x^1)|<1$ due to limitations in the numerics. We have observed that the IR 
solution realizes a
Hermitian conformal fixed point rotated under a complexified $U(1)$.
This allowed us to establish the Dyson map between the IR of this non-Hermitian model and its equivalent Hermitian theory. 
Remarkably 
solutions with and without imaginary current
flow to the same IR fixed point.

It is worth noting that many of the above results have been tested primarily in the non-Hermitian lattice. However, we expect them to also hold for the non-Hermitian junction. Regrettably numerical difficulties have prevented us from a thorough exploration of the latter model. 

In this paper we have started the exploration of inhomogeneous non-Hermitian theories via holography. This work opens the way for several lines of research in the realm of strongly coupled non-Hermitian theories. We next list some of them.

\begin{itemize}

    \item Determine the phase diagram of non-Hermitian lattices and junctions. In \cite{Xian:2023zgu} the authors identified a rich phase structure for the homogeneous system and found a phase with complex metric that dominated at large $|\eta|$. It would be interesting to see if such phase exists for our inhomogeneous setup. This could provide     new insights on the nature of the low-temperature
    solutions with imaginary current, specially for $|\eta(x^1)|>1$.

    \item Study transport properties of non-Hermitian lattices and junctions. It would be interesting to extend the study of transport properties of \cite{Xian:2023zgu} to the inhomogeneous lattice.
    
    \item Study the non-Hermitian junction in greater detail. In this paper, we have limited our analysis of the junction to a proof-of-concept. It would be interesting to delve deeper into the phenomenology of these solutions and try to compare with realistic condensed matter systems. 

    \item Identify the Hermitian theory corresponding to the solutions with imaginary current. 
    This could be achieved by constructing the Dyson map, a potentially daunting task in a strongly coupled theory as this one. Alternatively, one could hope that the excitation spectrum at zero temperature could provide insights into the Hermitian counterpart.

    \item Construct holographic non-Hermitian Josephson junctions. In \cite{Li:2023sej,Cayao:2024pbd} non-Hermitian Josephson junctions were considered and their spectra and suppercurrents were studied in detail finding novel phenomena associated with non-Hermicity. It would be interesting to reproduce such results in holography constructing a model for a Josephson junction with a non-Hermitian deformation.
    
    \item An interesting new direction would be that of the study of non-Hermitian deformations around quantum critical points. In~\cite{Ashida_2017} it was observed that a non-Hermitian lattice can qualitatively modify the dynamics about a quantum critical point. Via holography we can explore these effects on strongly coupled systems.

\end{itemize}

\section*{Acknowledgements}
We thank J.L.F. Barbón, P.G. Romeu, and specially K. Landsteiner for valuable discussions. 
This work is supported through the grants CEX2020-001007-S and PID2021-123017NB-I00 funded by MCIN/AEI/10.13039/501100011033 and by ERDF ``A way of making Europe''.
The work of D.G.F. is supported by FPI grant PRE2022-101810. 
Some numerical simulations were carried out at the IFT Hydra cluster.

\appendix
\section{Holographic dictionary}\label{appendix:Holographic Dictionary}
In this appendix, we follow the standard holographic renormalization procedure \cite{Skenderis_2002}, to compute the one-point functions presented in equation \eqref{eq:VEVs}.

We commence by first defining the renormalized action $\mathcal{S}_{ren}$ associated with the bulk action \eqref{eq:Bulk_Grav_Action}
\begin{equation}\label{eq:Renormalized_Grav_Action}
    \mathcal{S}_{ren}=\int d^4y \sqrt{-g}\left(R-2\Lambda-\frac{1}{4}F_{MN}F^{MN}-\mathcal{D}_M\phi \mathcal{D}^M\bar{\phi}-m^2\phi\bar{\phi}-\frac{v}{2}\phi^2\bar{\phi}^2\right)+\mathcal{S}_{ct}\,,
\end{equation}
where $\mathcal{S}_{ct}$ is the standard counter-term (see e.g. \cite{Hartnoll_2008}) 
\begin{equation}\label{eq:Counterterm_Grav_Action}
    \mathcal{S}_{ct}=\int_{z=\epsilon} d^3x \sqrt{-\gamma}\left(2K-4-\phi\bar{\phi}\right)\,.
\end{equation}
Here, $\gamma_{\mu\nu}$ is the induced metric at the $z=\epsilon\rightarrow0$ hypersurface and $K$ is the trace of the extrinsic curvature $K_{\mu\nu}$

To compute one-point functions we need the variation of the renormalized action 
\begin{align}\label{eq:delta_Renormalized_Grav_Action_NoSeries}
    \delta S_{ren}  =   &-\int_{z=\epsilon} d^3x \sqrt{-\gamma } \left(\frac{1}{2} \gamma _{\mu \nu } \phi  \bar{\phi }+(K-2) \gamma _{\mu \nu }-K_{\mu \nu }\right)\delta \gamma ^{\mu \nu } \nonumber \\
   &-\int_{z=\epsilon} d^3x \sqrt{-\gamma } \left(n_{\alpha } \left(\delta  \phi   \mathcal{D}^{\alpha }\bar{\phi }+\delta \bar{\phi} \mathcal{D}^{\alpha } \phi  +\delta A_{\mu } F^{\alpha \mu }\right)+  \bar{\phi }\delta
   \phi +\phi \delta  \bar{\phi }\right)\,,
\end{align}
where $n_\alpha$ is the outward-pointing unit normal vector corresponding to the $z=\epsilon$ boundary. Plugging in the asymptotic expansion \eqref{eq:Asymptotic_Expansion} and redefining the fluctuations as
\begin{equation}
    \delta\gamma^{\mu\nu}=\delta\gamma^{\mu\nu}_{(b)}\,z^{-2}\,,\qquad  \delta A_\mu=\delta  A_\mu^{(b)} \,,\qquad  \delta\phi=\delta\phi^{(b)}\,z\,,\qquad \delta\bar{\phi}=\delta\bar{\phi}^{(b)}\,z\,,
\end{equation}
the above expression simplifies to
\begin{align}\label{eq:delta_Renormalized_Grav_Action_YesSeries}
    \delta S_{ren}  =&\int_{z=\epsilon} d^3x \left(-1-h_2^{(3)}-\frac{3}{2}(h_3^{(3)}+h_4^{(3)})-\frac{1}{2}\left(s\bar{\phi}^{(2)}+\bar{s}\phi^{(2)}\right) \right)\delta\gamma^{tt}_{(b)} \nonumber\\
   &+\int_{z=\epsilon} d^3x \left(-\frac{1}{2}+h_2^{(3)}+\frac{3}{2}(h_1^{(3)}+h_4^{(3)})+\frac{1}{2}\left(s\bar{\phi}^{(2)}+\bar{s}\phi^{(2)}\right) \right)\delta\gamma^{11}_{(b)} \nonumber\\
   &+\int_{z=\epsilon} d^3x \left(-\frac{1}{2}+h_2^{(3)}+\frac{3}{2}(h_1^{(3)}+h_3^{(3)})+\frac{1}{2}\left(s\bar{\phi}^{(2)}+\bar{s}\phi^{2)}\right) \right)\delta\gamma^{22}_{(b)} \nonumber\\
   &+\int_{z=\epsilon} d^3x \left( A_1^{(1)}\delta A_1^{(b)}+  \bar{\phi}^{(2)}\delta\phi^{(b)} + \phi^{(2)}\delta\bar{\phi}^{(b)}\right)+ \text{terms of order $\epsilon$}
\end{align}
where we have introduced $h_2^{(3)}=-( h_1^{(3)}+h_3^{(3)}+h_4^{(3)})/3$. Finally, upon considering the usual holographic prescription for flat AdS boundaries 
\begin{equation}
    \expval{T_{\mu\nu}}=-2\lim_{\epsilon\rightarrow0}\frac{\delta \mathcal{S}_{ren}}{\delta \gamma^{\mu\nu}_{(b)} }\,,\quad
    \expval{J^\mu}=\lim_{\epsilon\rightarrow0}\frac{\delta \mathcal{S}_{ren}}{\delta A_\mu^{(b)} }\,,\quad
    \expval{\mathcal{O}}=\lim_{\epsilon\rightarrow0}\frac{\delta \mathcal{S}_{ren}}{\delta \bar{\phi}^{(b)} }\,,\quad \expval{\bar{\mathcal{O}}}=\lim_{\epsilon\rightarrow0}\frac{\delta \mathcal{S}_{ren}}{\delta \phi^{(b)} }\,,
\end{equation}
we recover the one-point functions presented in equation \eqref{eq:VEVs}.

\section{Further details on the computation of the QNFs}\label{appendix:DetailsQNFs}

Here we describe in detail the setup used in the computation of the QNFs presented in section \ref{subsect:NHLattice_Stability}. We follow the approach of \cite{Yang:2021ssm}, generalizing it to a backreacted setup. The idea behind the method is to use coordinates regular at the horizon, so that ingoing boundary conditions become instead regularity conditions there; then write the equations of motion for the linearized fluctuations as a generalized eigenvalue problem
\begin{equation}\label{eq:Generalized_Eigenvalue_Problem_Operators}
    \hat{\mathcal{A}}\,\Phi=\omega \hat{\mathcal{B}}\,\Phi\,, 
\end{equation}
where $\Phi$ is a multi-component field and $\hat{\mathcal{A}}$, $\hat{\mathcal{B}}$ differential operators; and finally discretize the system in a Chebyshev grid to obtain the eigenvalues numerically. Notably, for the computation of the eigenvalues, it is convenient to rewrite the discretized version of the generalized eigenvalue problem \eqref{eq:Generalized_Eigenvalue_Problem_Operators}
\begin{equation}\label{eq:Generalized_Eigenvalue_Problem_Matrices}
    \mathcal{A}\,\Phi=\omega \mathcal{B}\,\Phi\,, 
\end{equation}
where now $\mathcal{A}$ and $\mathcal{B}$ are matrices, as a standard eigenvalue problem by inverting $\mathcal{B}$, namely 
\begin{equation}\label{eq:Standard_Eigenvalue_Problem_Matrices}
    \mathcal{Q}\,\Phi=\omega \Phi\,,\qquad \mathcal{Q}=\mathcal{B}^{-1}\mathcal{A}\,.
\end{equation}
This in turn allows us to determine the lowest lying QNFs using Arnoldi iteration, which significantly speeds up the computation (see e.g. chapter 28 of \cite{trefethen2005spectra}).

To better illustrate the numerical method, in subsection \ref{subsect:Info_QNFs_Probe} we consider the probe limit of the setup presented in section \ref{sec:NonHermitianLattice}. After discussing the approach in that toy model, in subsection \ref{subsect:Info_QNFs_Backracted} we explain how to generalize the method to the backreacted setup.

\subsection{Toy model: Probe limit}\label{subsect:Info_QNFs_Probe}
In the probe limit the Einstein Hilbert part of the action \eqref{eq:Bulk_Grav_Action} drops, and we consider the metric fixed to be SAdS$_{3+1}$ in regular coordinates~\cite{Warnick:2013hba,Arean:2023ejh} (these are the AdS equivalent of the hyperboloidal slicing frequently used in asymptotically flat spacetimes, see for instance \cite{Bizon:2020qnd,PanossoMacedo:2023qzp,PanossoMacedo:2024nkw} for a recent overview)
\begin{equation}
    ds^2=\frac{1}{z^2}\left[-(1-z^3)dt^2+(dx^1)^2+(dx^2)^2+(1+z^3)dz^2-2z^3dtdz\right]\,.
\end{equation}
Notably, with this choice of coordinates imposing infalling boundary conditions for the QNMs corresponds to demanding regularity at the horizon $z=1$. Alternatively, one could consider the metric written in ingoing Eddington-Finkelstein coordinates, which simplifies the computation of the QNFs significantly. However, when considering backreaction of the metric, we found that it was more convenient to implement regular-like coordinates as they allowed for an invertible $\mathcal{B}$ matrix in the discretized eigenvalue equation \eqref{eq:Generalized_Eigenvalue_Problem_Matrices}. Hence, in this section we consider regular coordinates to better connect with the upcoming discussion of the backreacted case. 

To study the QNMs we consider the following ansatz for the fields $\phi$ and $A_M$
\begin{subequations}
    \begin{align}
        \phi&=\phi^{(bg)}(x^1,z)+z\delta\phi(x^1,z)e^{-i\omega t +i k x^1}  \,,\\
        \bar{\phi}&=\bar{\phi}^{(bg)}(x^1,z)+z\delta\bar{\phi}(x^1,z)e^{-i\omega t +i k x^1}\,, \\
        A&=A_1^{(bg)}(x^1,z)dx^1+\delta A_\mu(x^1,z)e^{-i\omega t +i k x^1}  dx^\mu  \,,
    \end{align}
\end{subequations}
where all functions are regular at the horizon $z=1$. The terms with the superscript $(bg)$ correspond to the background solution
and thus satisfy the boundary conditions
\begin{equation}
    \partial_z\phi^{(bg)}(x^1,0)=s(x^1)\,,\qquad \partial_z\bar{\phi}^{(bg)}(x^1,0)=\bar{s}(x^1)\,,\qquad A_1^{(bg)}(x^1,0)=a_1(x^1)\,,
\end{equation}
and $\{\delta\phi,\delta\bar{\phi},A_\mu\}$ are linearized fluctuations that represent the quasinormal modes and thus satisfy Dirichlet boundary conditions at $z=0$
\begin{equation}
    \delta \phi(x^1,0)=\delta \bar{\phi}(x^1,0)=\delta A_\mu(x^1,0)=0\,.
\end{equation}
Here we focus on the non-Hermitian lattice introduced in section \ref{sec:NonHermitianLattice} in the phase with imaginary current (see section \ref{subsubsect:NonHermitianLattice_PTbroken}) with lattice constant $LM=6$ and scalar charge $q=1$.\footnote{Note that in the main text we have used $LM=3$ in the computation of the QNFs.} Hence, due to conformal invariance of the UV theory, solutions are characterized by only two background parameters $\{a,T/M\}$; and the momentum $k/M$.

As discussed in section \ref{subsect:NHLattice_Stability} fluctuations can be decoupled into two sectors according to their transformation under reflections along the $x^2$ axis
\begin{enumerate}
    \item[i.] Odd: $\{\delta A_2\}$.
    \item[ii.] Even: $\{\delta\phi,\delta\bar{\phi},\delta A_t,\delta A_1\}$.
\end{enumerate}
We consider only the even sector, as it is the more interesting for presentation of the method. Then, the equations of motion for the fluctuations are
  \begin{subequations}\label{eq:Probe_linearized_EoMs}
 \begin{align}
    &\delta\left[\nabla_M \tensor{F}{^M_t}+iq\left(\phi\, \mathcal{D}_t\bar{\phi}-\bar{\phi} \mathcal{D}_t\phi\right)\right]=0\,,\label{eq:Probe_linearized_EoMs_a}\\
    &\delta\left[\nabla_M \tensor{F}{^M_1}+iq\left(\phi\, \mathcal{D}_1\bar{\phi}-\bar{\phi} \mathcal{D}_1\phi\right)\right]=0\,,\label{eq:Probe_linearized_EoMs_b}\\
    &\delta\left[\left(\nabla_M-iqA_M\right) \mathcal{D}^M\phi+m^2\phi+v\phi^2\bar{\phi}\right]=0\,,\label{eq:Probe_linearized_EoMs_c}\\
    &\delta\left[\left(\nabla_M+iqA_M\right) \mathcal{D}^M\bar{\phi}+m^2\bar{\phi}+v\bar{\phi}^2\phi\right]=0\,,\label{eq:Probe_linearized_EoMs_d}\\
    &\delta\left[\nabla_M \tensor{F}{^M_z}+iq\left(\phi\, \mathcal{D}_z\bar{\phi}-\bar{\phi}\, \mathcal{D}_z\phi\right)\right]=0\,,\label{eq:Probe_linearized_EoMs_e}
\end{align}
\end{subequations}
were the $\delta$ indicates that in the equation we consider only the linear term in the fluctuations. We have a total of 5 equations for 4 fields. These equations however are linearly dependent and one can build a set of 4 linearly independent equations that, when satisfied, imply the original 5 equations. In the usual approach one would try to construct this set of 4 equations. However, there is also a more brute-forced alternative: we could choose any independent set of 4 equations not necessarily implying the remaining one, solve them, and then discard the solutions that do not satisfy the remaining equation. To illustrate this, consider the following example. Let us have a system of two differential equations
\begin{equation}
    y''(x)+\lambda y'(x)=0\,,\qquad y'(x)+\lambda y(x)=0\,.
\end{equation}
To solve this system, we note that satisfying the second equation implies satisfying the first as they are related by derivation. Hence the system can be reduced to solving only the second equation, finding that
\begin{equation}\label{eq:solution_toy_model_a}
    y(x)=c_1e^{-\lambda x}\,.
\end{equation}
Alternatively, we could simply choose one of the equations at random, solve it and then check whether the other one is satisfied. Let us choose the first one, the solutions are
\begin{equation}\label{eq:solution_toy_model_b}
    y(x)=c_2+c_1e^{-\lambda x}\,,
\end{equation}
then we see that in order to also be a solution to the second equation, all solutions with $c_2\neq0$ should be discarded.

The main takeaway here is any independent set of equations from \eqref{eq:Probe_linearized_EoMs} must necessarily contain all solutions to the system plus potentially some unphysical solutions that play the role of the  $c_2\neq0$ solutions in \eqref{eq:solution_toy_model_b}. Hence, instead of trying to find the minimal set of independent equations that imply \eqref{eq:Probe_linearized_EoMs}, we opt for the simpler, although less elegant, approach of selecting some set of equations; computing the eigenvalues and eigenfunctions and then discarding those eigenvalues that are unphysical as they do not satisfy the remaining equations.\footnote{Note that our approach could fail in the presence of degenerate eigenvalues. Say two eigenvectors $\varphi_1$ and $\varphi_2$ are solutions to the chosen equations with eigenvalue $\omega$. Then we could have that only some combination $\alpha\varphi_1+\beta\varphi_2$ satisfies the constraint while $\varphi_1$ and $\varphi_2$ independently do not satisfy it. In practice, within our setup, such possibility leads to no relevant issues. We make sure that indeed we are not losing information by checking that the QNFs we find continuously connect with those obtained in the homogeneous $\eta=0$ case; which we can compute without using the brute-forced approach.}

In practice, we do not choose the independent equations at random. Instead, we choose a set that is conveniently suited for our numerical implementation by demanding that the $\mathcal{B}$ matrix in \eqref{eq:Generalized_Eigenvalue_Problem_Matrices} is invertible and that only a few of the unphysical quasinormal frequencies appear near the origin of the complex plane. We found that an adequate choice was taking equations \eqref{eq:Probe_linearized_EoMs_a}-\eqref{eq:Probe_linearized_EoMs_d}. It is important to note that, as the aforementioned set of equations is quadratic in $\omega$, we need to introduce auxiliary fields 
\begin{equation}
    \{\delta \tilde{A}_1,\delta \tilde{\phi},\delta\tilde{\bar{\phi}}\}=-i\omega\{\delta A_1,\delta \phi,\delta\bar{\phi}\}\,,
\end{equation}
to be able to write a generalized eigenvalue problem.

Before moving on to the discussion of the numerical implementation, let us comment on two relevant issues. Firstly, the existence of more equations of motion than fields is a byproduct of fixing gauge. For any given sector, we always have as many equations as components before gauge fixing. Hence for the odd sector we need not worry about all the issues discussed above. Secondly, we want to stress that we can consistently impose Dirichlet boundary conditions instead of gauge-invariant boundary conditions. The key realization is that, after radial gauge fixing, imposing regularity on the horizon does not fix the remaining gauge invariance and hence we are free to impose Dirichlet boundary conditions on the boundary of AdS. Once we have fixed the radial gauge, the remaining residual gauge transformations cannot take us away from the regular solutions as they have to be $z$-independent.\footnote{After fixing radial gauge, the residual gauge transformations are large gauge transformations by construction.} Hence, we can use a gauge transformation to shift any solution to satisfy Dirichlet boundary conditions.

Regarding the numerics, we first place the system in a lattice, for the $z$-direction we consider a Chebyshev-Lobatto grid and for the $x^1$-direction, which is periodic with period $L$, we take an equispaced grid. This in turn, allows us to discretize the relevant differential operators using Chebyshev and Fourier differentiation matrices to replace $\partial_z$ and $\partial_1$, respectively. With this, we first solve for the background as indicated in section \ref{subsect:NonHermitianLattice_NumericalMethod} and then construct the $\mathcal{A}$ and $\mathcal{B}$ matrices corresponding to the discretization of the generalized eigenvalue problem arising from equations \eqref{eq:Probe_linearized_EoMs_a}-\eqref{eq:Probe_linearized_EoMs_d}. Once these matrices are constructed, we compute $\mathcal{Q}=\mathcal{B}^{-1}\mathcal{A}$ and obtain a small set of low-lying eigenvalues and eigenvectors using Arnoldi iteration. We then construct the discretized version of the remaining equation \eqref{eq:Probe_linearized_EoMs_e} and select only those eigenvalues whose eigenvectors satisfy it with a certain tolerance.

To give an explicit example, in table \ref{tab:SampleQNFsProbe_Teq1_aeq0d4} we present the QNFs computed for $\{a,T/M\}=\{0.4,1\}$ at momentum $k/M=\frac{4\pi}{3}\cdot10^{-4}$.\footnote{Notably, taking $k/M\neq0$ seems to improve the quality of the numerics specially when adding backreaction. For that reason we choose to always work at small but nonzero $k/M$.} We have employed a lattice with $N_1\times N_z=16\times20$ sites and a tolerance of $10^{-8}$. In order to locate the shown QNFs we have obtained a total of $20$ eigenvalues and eigenvectors of $\mathcal{Q}$. Remarkably, due to the spectral instability observed in \cite{Arean:2023ejh} the numerical round-off errors arising from the inversion of $\mathcal{B}$ caused the eigenvalues to migrate significantly. To avoid this issue, we found that it was sufficient to 
artificially\footnote{Background solutions are computed with MachinePrecision and then promoted to data with $5\times$MachinePrecision.}
increase the numerical precision to $5\times$MachinePrecision when computing the eigenvalues. Interestingly we found that the numerical error associated to the background was negligible for the low-lying QNFs.

\begin{table}[]
    \centering
    \begin{tabular}{|c|c|}
    \hline
    $\Re(\omega)$&$\Im(\omega)$\\
    \hline
    \hline
    0 & -0.0479634 \\
    \hline
 $\pm$0.996619 & -0.578461 \\
 \hline
 $\pm$0.996174 & -0.579758 \\
  \hline
 0 & -1.62071 \\
  \hline
 $\pm$1.27715 & -1.52475 \\
 \hline
 $\pm$1.35876 & -1.54718 \\
  \hline
$\pm$ 2.09013 & -0.467175 \\
  \hline
 $\pm$2.09033 & -0.467161 \\
 \hline
 $\pm$1.72459 & -1.46457 \\
 \hline
 $\pm$1.72859 & -1.4667 \\
  \hline
 $\pm$1.77647 & -1.47722 \\
 \hline
    \end{tabular}
    \caption{QNFs for $\{a,T/M\}=\{0.4,1\}$ at $k/M=\frac{4\pi}{3}\cdot 10^{-4}$.}
    \label{tab:SampleQNFsProbe_Teq1_aeq0d4}
\end{table}

\subsection{Adding backreaction}\label{subsect:Info_QNFs_Backracted} 
Now that the general approach has been introduced in the probe limit, we can proceed to add backreaction. However we wish to consider the background metric in regular-like coordinates as they were essential in the construction of the previous section. For that reason we take the following ansatz
\begin{subequations}
    \begin{align}
        ds^2&=g_{MN}^{(bg)}(x^1,z)dx^Mdx^N+\frac{1}{z^2}\delta g_{\mu\nu}(x^1,z)e^{-i\omega t +i k x^1}  dx^\mu dx^\nu\,,\\
        \phi&=\phi^{(bg)}(x^1,z)+z\delta\phi(x^1,z)e^{-i\omega t +i k x^1}  \,,\\
        \bar{\phi}&=\bar{\phi}^{(bg)}(x^1,z)+z\delta\bar{\phi}(x^1,z)e^{-i\omega t +i k x^1}\,, \\
        A&=A_1^{(bg)}(x^1,z)dx^1+\delta A_\mu(x^1,z)e^{-i\omega t +i k x^1}  dx^\mu \,, 
    \end{align}
\end{subequations}
where the background metric $g_{MN}^{(bg)}(x^1,z)$ is written in a regular-like ansatz
\begin{align}\label{eq:Ansatz_regular_metric}
    g_{MN}^{(bg)}dx^Mdx^N=\frac{1}{z^2}&\left[- \left(1-z^3\right)h_1 dt^2 +h_3(dx^1+2h_5\,dz)^2+h_4 (dx^2)^2 \right.\nonumber\\ &\left.+(1+z^3) h_2\, dz^2-2z^3 h_6\, dt\, dz+2 h_7\, dt\, dx \right]\,,
\end{align}
with $h_i=h_i(x^1,z)$. We demand regularity of all the functions at the horizon $z=1$ and at $z=0$ we require that the background satisfies
\begin{align}
    g_{MN}^{(bg)}&(x^1,0)dx^Mdx^N=\frac{1}{z^2}\left[- dt^2 +(dx^1)^2 + (dx^2)^2  + dz^2 \right]\,,\nonumber\\
    \partial_z\phi^{(bg)}(x^1,0)=&s(x^1)\,,\qquad \partial_z\bar{\phi}^{(bg)}(x^1,0)=\bar{s}(x^1)\,,\qquad A_1^{(bg)}(x^1,0)=a_1(x^1)\,,
\end{align}
while for the fluctuations we impose Dirichlet boundary conditions 
\begin{equation}
    \delta g_{\mu\nu}(x^1,0)=\delta A_{\mu}(x^1,0)=\delta \phi(x^1,0)=\delta \bar{\phi}(x^1,0)=0\,.
\end{equation}

Again, we focus on the phase with imaginary current of the non-Hermitian lattice of section \ref{sec:NonHermitianLattice}. Hence, after fixing $LM=6$ and $q=1$, the solutions are characterized by two background parameters $\{a,T/M\}$ and the dimensionless momentum $k/M$.

As indicated in section \ref{subsect:NHLattice_Stability} fluctuations can be decoupled into two sectors according to their transformation under reflections along the $x^2$ axis
\begin{enumerate}
    \item[i.] Odd: $\{\delta g_{t2},\delta g_{12},\delta A_2\}$.
    \item[ii.] Even: $\{\delta g_{tt},\delta g_{t1},\delta g_{11},\delta g_{22},\delta\phi,\delta\bar{\phi},\delta A_t,\delta A_1\}$.
\end{enumerate}
In the present discussion we consider only the even sector. We choose the following set of linearized equations for the fluctuations satisfying the requirements indicated in the previous subsection
\begin{equation}\label{eq:backreactedQNM_EoMs_Good}   \delta\mathcal{E}_{tz}=\delta\mathcal{E}_{11}=\delta\mathcal{E}_{22}=\delta\mathcal{E}_{1z}=\delta\mathcal{M}_t=\delta\mathcal{M}_1=\delta\mathcal{F}=\delta\mathcal{\bar{F}}=0\,,
\end{equation}
and we are left with 
\begin{equation}\label{eq:backreactedQNM_EoMs_Extra}
        \delta\mathcal{E}_{tt}=\delta\mathcal{E}_{t1}=\delta\mathcal{E}_{zz}=\delta\mathcal{M}_z=0\,,
\end{equation}
as the remaining unused equations. 
We do not present here the explicit form of these equations; instead we follow the notation used in equations \eqref{eq:EoMs_NoDeTurck} with the delta indicating that we consider the terms linear in the fluctuations. Similarly to what we found in the probe limit, the set of equations \eqref{eq:backreactedQNM_EoMs_Good} is quadratic in $\omega$ and to have a generalized eigenvalue problem we need to include the following auxiliary fields.
\begin{equation}
    \{\delta \tilde{g}_{11},\delta \tilde{g}_{22},\delta \tilde{A}_1,\delta \tilde{\phi},\delta\tilde{\bar{\phi}}\}=-i\omega\{\delta g_{11},\delta g_{22},\delta A_1,\delta \phi,\delta\bar{\phi}\}\,.
\end{equation}

With this, we have properly set up the problem and we proceed with the numerics as indicated in the toy model of the previous section. We discretize the system in a lattice, compute the background solution, construct the $\mathcal{Q}$ matrix, obtain a small set of low-lying eigenvalues and eigenvectors using Arnoldi iteration and finally select only those that satisfy the discretized version of equations \eqref{eq:backreactedQNM_EoMs_Extra} to a certain tolerance.

Regarding the background solution we proceed as described in section \ref{subsect:NonHermitianLattice_NumericalMethod}. The only relevant change being that now the reference background metric for the DeTurck trick is given by the SAdS$_{3+1}$ metric in regular coordinates 
\begin{equation}
   ds_{\text{ref}}^2= \frac{1}{z^2}\left[-(1-z^3)dt^2+(dx^1)^2+(dx^2)^2+(1+z^3)dz^2-2z^3dtdz\right]\,.
\end{equation}
It is also worth mentioning that the ansatz \eqref{eq:Ansatz_regular_metric} allows for non-stationary solutions \cite{Figueras:2012rb}. We however, want to focus solely on solutions which are stationary, i.e., solutions with constant temperature. For that reason, after discretizing the problem we choose to replace two of the equations of motion at $z=1$ by
\begin{equation}\label{eq:Regular_horizon_constraints}
    h_1(x^1,1)=h_6(x^1,1)\,,\qquad h_7(x^1,1)=0\,.
\end{equation}
This imposes stationarity and fixes the temperature to be $T=3/(4\pi)$.\footnote{The constraints \eqref{eq:Regular_horizon_constraints} are obtained from demanding that $\partial_t$ is the killing vector generating the event horizon at $z=1$ corresponding to a black brane with temperature $T=3/(4\pi)$.} For consistency, we check that the metric is regular and stationary throughout the bulk; the latter requirement corresponding to demanding that the time-like killing vector is surface orthogonal, i.e., that numerically
\begin{equation}
    \tilde{k}_t\wedge d\tilde{k}_t=0\,,
\end{equation}
where 
\begin{equation}
    \tilde{k}_t=g_{t\mu}dx^\mu\,.
\end{equation}

For completeness, in table \ref{tab:SampleQNFsBackreacted_Teq1_aeq0d4} we reproduce the QNFs corresponding to $\{a,T/M\}=\{0.4,1\}$ and $k/M=\frac{4\pi}{3}\cdot10^{-4}$. These are obtained using a lattice with $N_1\times N_z=16\times20$ sites and a tolerance of $10^{-8}$. In order to locate the shown QNFs, we have obtained a total of 70 eigenvalues and eigenvectors of $\mathcal{Q}$. As in the probe limit, we found that it was necessary to artificially increase the numerical precision to $5\times$MachinePrecision when computing the eigenvalues in order to avoid spectral instability.

\begin{table}[]
    \centering
    \begin{tabular}{|c|c|}
    \hline
    $\Re(\omega)$&$\Im(\omega)$\\
    \hline
    \hline
    0 & -0.0471596 \\
    \hline
     $\pm$ 0.99668 & -0.577632 \\
     \hline
      $\pm$ 0.996231 & -0.578915 \\
      \hline
    0 & -1.61755 \\
    \hline
      $\pm$ 1.2865 & -1.52129 \\
      \hline
      $\pm$ 2.09019 & -0.466238 \\
      \hline
    \end{tabular}
    \caption{QNFs for $\{a,T/M\}=\{0.4,1\}$ at $k/M=\frac{4\pi}{3}\cdot 10^{-4}$.}
    \label{tab:SampleQNFsBackreacted_Teq1_aeq0d4}
\end{table}

\bibliographystyle{JHEP}
\bibliography{biblio}

\providecommand{\href}[2]{#2}\begingroup\raggedright\begin{thebibliography}{10}

\bibitem{Bender:1998ke}
C.~M. Bender and S.~Boettcher, {\it {Real spectra in nonHermitian Hamiltonians having PT symmetry}},  {\em Phys. Rev. Lett.} {\bf 80} (1998) 5243--5246 [\href{http://arXiv.org/abs/physics/9712001}{{\tt physics/9712001}}].

\bibitem{Mannheim2018}
P.~D. Mannheim, {\it {Antilinearity rather than Hermiticity as a guiding principle for quantum theory}},  {\em Journal of Physics A: Mathematical and Theoretical} {\bf 51} (jun, 2018) [\href{http://arXiv.org/abs/1512.04915}{{\tt 1512.04915}}].

\bibitem{Bender_2005}
C.~M. Bender, {\it Introduction to pt-symmetric quantum theory},  {\em Contemporary Physics} {\bf 46} (July, 2005) 277–292.

\bibitem{Mostafazadeh:2001jk}
A.~Mostafazadeh, {\it {PseudoHermiticity versus PT symmetry. The necessary condition for the reality of the spectrum}},  {\em J. Math. Phys.} {\bf 43} (2002) 205--214 [\href{http://arXiv.org/abs/math-ph/0107001}{{\tt math-ph/0107001}}].

\bibitem{Mostafazadeh:2001nr}
A.~Mostafazadeh, {\it {PseudoHermiticity versus PT symmetry 2. A Complete characterization of nonHermitian Hamiltonians with a real spectrum}},  {\em J. Math. Phys.} {\bf 43} (2002) 2814--2816 [\href{http://arXiv.org/abs/math-ph/0110016}{{\tt math-ph/0110016}}].

\bibitem{Mostafazadeh:2002id}
A.~Mostafazadeh, {\it {PseudoHermiticity versus PT symmetry 3: Equivalence of pseudoHermiticity and the presence of antilinear symmetries}},  {\em J. Math. Phys.} {\bf 43} (2002) 3944--3951 [\href{http://arXiv.org/abs/math-ph/0203005}{{\tt math-ph/0203005}}].

\bibitem{Bender:2004sa}
C.~M. Bender, D.~C. Brody and H.~F. Jones, {\it {Extension of PT symmetric quantum mechanics to quantum field theory with cubic interaction}},  {\em Phys. Rev. D} {\bf 70} (2004) 025001 [\href{http://arXiv.org/abs/hep-th/0402183}{{\tt hep-th/0402183}}]. [Erratum: Phys.Rev.D 71, 049901 (2005)].

\bibitem{Alexandre:2022uns}
J.~Alexandre, J.~Ellis and P.~Millington, {\it {Discrete spacetime symmetries, second quantization, and inner products in a non-Hermitian Dirac fermionic field theory}},  {\em Phys. Rev. D} {\bf 106} (2022), no.~6 065003 [\href{http://arXiv.org/abs/2201.11061}{{\tt 2201.11061}}].

\bibitem{Li:2024xms}
Y.-D. Li and Q.~Wang, {\it {Isospectral local Hermitian theory for the PT-symmetric i\ensuremath{\phi}3 quantum field theory}},  {\em Phys. Rev. D} {\bf 111} (2025), no.~2 025016 [\href{http://arXiv.org/abs/2412.10732}{{\tt 2412.10732}}].

\bibitem{Arean:2024gks}
D.~Arean, D.~Garcia-Fari\~na and K.~Landsteiner, {\it {Strongly Coupled PT-Symmetric Models in Holography}},  11, 2024.
\newblock \href{http://arXiv.org/abs/2411.18471}{{\tt 2411.18471}}.

\bibitem{ElGanainy2018}
R.~El-Ganainy, K.~G. Makris, M.~Khajavikhan, Z.~H. Musslimani, S.~Rotter and D.~N. Christodoulides, {\it {Non-Hermitian physics and PT symmetry}},  {\em Nature Physics} {\bf 14} (Jan, 2018) 11--19.

\bibitem{Ashida_2017}
Y.~Ashida, S.~Furukawa and M.~Ueda, {\it Parity-time-symmetric quantum critical phenomena},  {\em Nature Communications} {\bf 8} (June, 2017).

\bibitem{Bender_2004}
C.~M. Bender, D.~C. Brody and H.~F. Jones, {\it Scalar quantum field theory with a complex cubic interaction},  {\em Physical Review Letters} {\bf 93} (Dec., 2004).

\bibitem{Chernodub:2024lkr}
M.~N. Chernodub and P.~Millington, {\it {Anomalous dispersion, superluminality, and instabilities in two-flavor theories with local non-Hermitian mass mixing}},  {\em Phys. Rev. D} {\bf 109} (2024), no.~10 105006 [\href{http://arXiv.org/abs/2401.06097}{{\tt 2401.06097}}].

\bibitem{Chernodub:2021waz}
M.~N. Chernodub and P.~Millington, {\it {IR/UV mixing from local similarity maps of scalar non-Hermitian field theories}},  {\em Phys. Rev. D} {\bf 105} (2022), no.~7 076020 [\href{http://arXiv.org/abs/2110.05289}{{\tt 2110.05289}}].

\bibitem{Maldacena:1997re}
J.~M. Maldacena, {\it {The Large N limit of superconformal field theories and supergravity}},  {\em Adv. Theor. Math. Phys.} {\bf 2} (1998) 231--252 [\href{http://arXiv.org/abs/hep-th/9711200}{{\tt hep-th/9711200}}].

\bibitem{Aharony:1999ti}
O.~Aharony, S.~S. Gubser, J.~M. Maldacena, H.~Ooguri and Y.~Oz, {\it {Large N field theories, string theory and gravity}},  {\em Phys. Rept.} {\bf 323} (2000) 183--386 [\href{http://arXiv.org/abs/hep-th/9905111}{{\tt hep-th/9905111}}].

\bibitem{BigJ}
M.~Ammon and J.~Erdmenger, {\em {Gauge/gravity duality}: {Foundations and applications}}.
\newblock Cambridge University Press, Cambridge, 4, 2015.

\bibitem{zaanen2015holographic}
J.~Zaanen, Y.~Liu, Y.~Sun and K.~Schalm, {\em Holographic Duality in Condensed Matter Physics}.
\newblock Cambridge University Press, 2015.

\bibitem{Hartnoll:2018xxg}
S.~A. Hartnoll, A.~Lucas and S.~Sachdev, {\em {Holographic Quantum Matter}}.
\newblock MIT Press, 2018.

\bibitem{Arean:2019pom}
D.~Are\'an, K.~Landsteiner and I.~Salazar~Landea, {\it {Non-hermitian holography}},  {\em SciPost Phys.} {\bf 9} (2020), no.~3 032 [\href{http://arXiv.org/abs/1912.06647}{{\tt 1912.06647}}].

\bibitem{Xian:2023zgu}
Z.-Y. Xian, D.~Rodr\'\i{}guez~Fern\'andez, Z.~Chen, Y.~Liu and R.~Meyer, {\it {Electric conductivity in non-Hermitian holography}},  {\em SciPost Phys.} {\bf 16} (2024) 004 [\href{http://arXiv.org/abs/2304.11183}{{\tt 2304.11183}}].

\bibitem{Morales-Tejera:2022hyq}
S.~Morales-Tejera and K.~Landsteiner, {\it {Non-Hermitian quantum quenches in holography}},  {\em SciPost Phys.} {\bf 14} (2023), no.~3 030 [\href{http://arXiv.org/abs/2203.02524}{{\tt 2203.02524}}].

\bibitem{Cai:2022onu}
W.~Cai, S.~Cao, X.-H. Ge, M.~Matsumoto and S.-J. Sin, {\it {Non-Hermitian quantum system generated from two coupled Sachdev-Ye-Kitaev models}},  {\em Phys. Rev. D} {\bf 106} (2022), no.~10 106010 [\href{http://arXiv.org/abs/2208.10800}{{\tt 2208.10800}}].

\bibitem{Cao:2024xaw}
S.~Cao and X.-H. Ge, {\it {Excitation Transmission through a non-Hermitian traversable wormhole}},  \href{http://arXiv.org/abs/2404.11436}{{\tt 2404.11436}}.

\bibitem{Donos:2012js}
A.~Donos and S.~A. Hartnoll, {\it {Interaction-driven localization in holography}},  {\em Nature Phys.} {\bf 9} (2013) 649--655 [\href{http://arXiv.org/abs/1212.2998}{{\tt 1212.2998}}].

\bibitem{Donos:2014uba}
A.~Donos and J.~P. Gauntlett, {\it {Novel metals and insulators from holography}},  {\em JHEP} {\bf 06} (2014) 007 [\href{http://arXiv.org/abs/1401.5077}{{\tt 1401.5077}}].

\bibitem{Donos:2014yya}
A.~Donos and J.~P. Gauntlett, {\it {The thermoelectric properties of inhomogeneous holographic lattices}},  {\em JHEP} {\bf 01} (2015) 035 [\href{http://arXiv.org/abs/1409.6875}{{\tt 1409.6875}}].

\bibitem{Freedman:1999gp}
D.~Z. Freedman, S.~S. Gubser, K.~Pilch and N.~P. Warner, {\it {Renormalization group flows from holography supersymmetry and a c theorem}},  {\em Adv. Theor. Math. Phys.} {\bf 3} (1999) 363--417 [\href{http://arXiv.org/abs/hep-th/9904017}{{\tt hep-th/9904017}}].

\bibitem{Myers:2010xs}
R.~C. Myers and A.~Sinha, {\it {Seeing a c-theorem with holography}},  {\em Phys. Rev. D} {\bf 82} (2010) 046006 [\href{http://arXiv.org/abs/1006.1263}{{\tt 1006.1263}}].

\bibitem{Myers:2010tj}
R.~C. Myers and A.~Sinha, {\it {Holographic c-theorems in arbitrary dimensions}},  {\em JHEP} {\bf 01} (2011) 125 [\href{http://arXiv.org/abs/1011.5819}{{\tt 1011.5819}}].

\bibitem{Hartnoll:2015faa}
S.~A. Hartnoll, D.~M. Ramirez and J.~E. Santos, {\it {Emergent scale invariance of disordered horizons}},  {\em JHEP} {\bf 09} (2015) 160 [\href{http://arXiv.org/abs/1504.03324}{{\tt 1504.03324}}].

\bibitem{Headrick:2009pv}
M.~Headrick, S.~Kitchen and T.~Wiseman, {\it {A New approach to static numerical relativity, and its application to Kaluza-Klein black holes}},  {\em Class. Quant. Grav.} {\bf 27} (2010) 035002 [\href{http://arXiv.org/abs/0905.1822}{{\tt 0905.1822}}].

\bibitem{Horowitz:2012ky}
G.~T. Horowitz, J.~E. Santos and D.~Tong, {\it {Optical Conductivity with Holographic Lattices}},  {\em JHEP} {\bf 07} (2012) 168 [\href{http://arXiv.org/abs/1204.0519}{{\tt 1204.0519}}].

\bibitem{Dias:2015nua}
O.~J.~C. Dias, J.~E. Santos and B.~Way, {\it {Numerical Methods for Finding Stationary Gravitational Solutions}},  {\em Class. Quant. Grav.} {\bf 33} (2016), no.~13 133001 [\href{http://arXiv.org/abs/1510.02804}{{\tt 1510.02804}}].

\bibitem{Gubser:2009cg}
S.~S. Gubser and A.~Nellore, {\it {Ground states of holographic superconductors}},  {\em Phys. Rev. D} {\bf 80} (2009) 105007 [\href{http://arXiv.org/abs/0908.1972}{{\tt 0908.1972}}].

\bibitem{Santos:T0}
S.~A. Hartnoll and J.~E. Santos, {\it Disordered horizons: Holography of randomly disordered fixed points},  {\em Phys. Rev. Lett.} {\bf 112} (Jun, 2014) 231601.

\bibitem{Li:2023sej}
C.-A. Li, H.-P. Sun and B.~Trauzettel, {\it {Anomalous Andreev spectrum and transport in non-Hermitian Josephson junctions}},  {\em Phys. Rev. B} {\bf 109} (2024), no.~21 214514 [\href{http://arXiv.org/abs/2307.04789}{{\tt 2307.04789}}].

\bibitem{Cayao:2024pbd}
J.~Cayao and M.~Sato, {\it {Non-Hermitian phase-biased Josephson junctions}},  {\em Phys. Rev. B} {\bf 110} (2024), no.~20 L201403 [\href{http://arXiv.org/abs/2408.17260}{{\tt 2408.17260}}].

\bibitem{Skenderis_2002}
K.~Skenderis, {\it Lecture notes on holographic renormalization},  {\em Classical and Quantum Gravity} {\bf 19} (Nov., 2002) 5849–5876.

\bibitem{Hartnoll_2008}
S.~A. Hartnoll, C.~P. Herzog and G.~T. Horowitz, {\it Holographic superconductors},  {\em Journal of High Energy Physics} {\bf 2008} (Dec., 2008) 015–015.

\bibitem{Yang:2021ssm}
P.~Yang, X.~Li and Y.~Tian, {\it {Instability of holographic superfluids in optical lattice}},  {\em JHEP} {\bf 11} (2021) 190 [\href{http://arXiv.org/abs/2109.09080}{{\tt 2109.09080}}].

\bibitem{trefethen2005spectra}
L.~N. Trefethen and M.~Embree, {\em Spectra and Pseudospectra: The Behavior of Nonnormal Matrices and Operators}.
\newblock Princeton University Press, 2005.

\bibitem{Warnick:2013hba}
C.~M. Warnick, {\it {On quasinormal modes of asymptotically anti-de Sitter black holes}},  {\em Commun. Math. Phys.} {\bf 333} (2015), no.~2 959--1035 [\href{http://arXiv.org/abs/1306.5760}{{\tt 1306.5760}}].

\bibitem{Arean:2023ejh}
D.~Are\'an, D.~Garcia-Fari\~na and K.~Landsteiner, {\it {Pseudospectra of holographic quasinormal modes}},  {\em JHEP} {\bf 12} (2023) 187 [\href{http://arXiv.org/abs/2307.08751}{{\tt 2307.08751}}].

\bibitem{Bizon:2020qnd}
P.~Bizo\'n, T.~Chmaj and P.~Mach, {\it {A toy model of hyperboloidal approach to quasinormal modes}},  {\em Acta Phys. Polon. B} {\bf 51} (2020) 1007 [\href{http://arXiv.org/abs/2002.01770}{{\tt 2002.01770}}].

\bibitem{PanossoMacedo:2023qzp}
R.~Panosso~Macedo, {\it {Hyperboloidal approach for static spherically symmetric spacetimes: a didactical introduction and applications in black-hole physics}},  \href{http://arXiv.org/abs/2307.15735}{{\tt 2307.15735}}.

\bibitem{PanossoMacedo:2024nkw}
R.~Panosso~Macedo and A.~Zenginoglu, {\it {Hyperboloidal Approach to Quasinormal Modes}},  \href{http://arXiv.org/abs/2409.11478}{{\tt 2409.11478}}.

\bibitem{Figueras:2012rb}
P.~Figueras and T.~Wiseman, {\it {Stationary holographic plasma quenches and numerical methods for non-Killing horizons}},  {\em Phys. Rev. Lett.} {\bf 110} (2013) 171602 [\href{http://arXiv.org/abs/1212.4498}{{\tt 1212.4498}}].

\end{thebibliography}\endgroup
\end{document}